\documentclass[traditabstract]{aa}
\usepackage{graphicx}
\usepackage{txfonts}
\usepackage{longtable}
\usepackage[authoryear]{natbib}
\bibpunct{(}{)}{;}{a}{}{,}

\def\ltsima{$\; \buildrel < \over \sim \;$}
\def\simlt{\lower.5ex\hbox{\ltsima}}
\def\gtsima{$\; \buildrel > \over \sim \;$}
\def\simgt{\lower.5ex\hbox{\gtsima}}
\def\kms{km~s$^{-1}$}
\newcommand{\HI}{\ion{H}{i}}

\begin{document}
   \title{The {\em StEllar Counterparts of COmpact high velocity clouds} (SECCO) survey. I.}
\subtitle{Photos of ghosts.\thanks{Based on data acquired using the Large Binocular Telescope (LBT).
The LBT is an international collaboration amongst institutions in the United
States, Italy, and Germany. LBT Corporation partners are The University of
Arizona on behalf of the Arizona university system; Istituto Nazionale di
Astrofisica, Italy; LBT Beteiligungsgesellschaft, Germany, representing the
Max-Planck Society, the Astrophysical Institute Potsdam, and Heidelberg
University; The Ohio State University; and The Research Corporation, on behalf
of The University of Notre Dame, University of Minnesota, and University of
Virginia.}}

   \author{ M. Bellazzini\inst{1}, G. Beccari\inst{2}, G. Battaglia\inst{3,9,1}, N. Martin\inst{4,5}, V. Testa\inst{6},  R. Ibata\inst{4}, M. Correnti\inst{7}, F. Cusano\inst{1}, \and E. Sani\inst{8}
           }
         
      \offprints{M. Bellazzini}

   \institute{INAF - Osservatorio Astronomico di Bologna,
              Via Ranzani 1, 40127 Bologna, Italy
             \email{michele.bellazzini@oabo.inaf.it} 
             \and
              European Southern Observatory, Alonso de Cordova 3107, Vitacura Santiago, Chile. 
             \and
             Instituto de Astrof'sica de Canarias, 38205 La Laguna, Tenerife, Spain
             \and
             Observatoire astronomique de Strasbourg, Universit\'e de Strasbourg, CNRS, UMR 7550, 11 rue 
             de l'Universit\'e, F-67000 Strasbourg, France 
             \and
             Max-Planck-Institut f\"ur Astronomie, K\"onigstuhl 17, D-69117 Heidelberg, Germany 
             \and
             INAF - Osservatorio Astronomico di Roma, via Frascati 33, 00040 Monteporzio, Italy
             \and
             Space Telescope Science Institute, 3700 San Martin Drive, Baltimore, MD 21218, USA
             \and 
             INAF - Osservatorio Astrofisico di Arcetri, Largo E. Fermi 5, I-50125, Firenze, Italy
              \and 
			Universidad de la Laguna, Dpto. Astrofisica, E-38206 La Laguna, Tenerife, Spain
          }

     \authorrunning{M. Bellazzini et al.}
   \titlerunning{The SECCO Survey. I.}

   \date{Submitted to A\&A }

\abstract{We present an imaging survey that searches for the stellar counterparts of recently discovered ultra-compact high-velocity \HI~ clouds (UCHVC). It has been proposed that these clouds are candidate mini-haloes in the Local Group and its surroundings within a distance range of 0.25-2.0~Mpc. Using the Large Binocular Telescope we obtained wide-field ($\simeq 23\arcmin \times 23\arcmin$) g- and r-band images of the twenty-five most promising and most compact clouds amongst the fifty-nine that have been identified. Careful visual inspection of all the images does not reveal any stellar counterpart that even slightly resembles Leo~P, the only local dwarf galaxy that was found as a counterpart to a previously detected high-velocity cloud. Only a possible distant ($D>3.0$~Mpc) counterpart to HVC274.68+74.70-123 has been identified in our images. The point source photometry in the central $17.3\arcmin\times 7.7\arcmin$ chips reaches $r\le 26.5$ and is expected to contain most of the stellar counterparts to the UCHVCs. However, no obvious stellar over-density is detected in any of our fields, in marked contrast to our comparison Leo~P field, in which the dwarf galaxy is detected at a $> 30\sigma$-significance level. Only HVC352.45+59.06+263 may be associated with a weak over-density, whose nature cannot be ascertained with our data. Sensitivity tests show that our survey would have detected any dwarf galaxy dominated by an old stellar population, with an integrated absolute magnitude of $M_V\le -8.0$ and  a half-light radius of $r_h\le 300$~pc that lies within 1.5~Mpc of us, thereby confirming that it is unlikely that the observed UCHVCs are associated with the stellar counterparts typical of known Local Group dwarf galaxies.}


\maketitle
%

\section{Introduction}
\label{intro}

The significant mismatch between the large number of low-mass sub-haloes predicted to be orbiting
Milky Way (MW)-sized galaxies in $\Lambda$-CDM N-body simulations and the observed
number of Local Group (LG) dwarf galaxies in the same mass range (the so-called  missing satellites problem;
\citealt{kauff,moore,klip}) has been the subject of much debate over the past decade in the field of local cosmology \citep[see, e.g.,][]{wyse}. 

Amounting to a factor larger than
10 in the realm of dwarf spheroidal (dSph) galaxies, this discrepancy has been mitigated by
advancements in both theory  and observations. On the theoretical
side, it is now generally believed that baryonic physics effects (such as
photo-ionisation, supernova feedback, etc., which were not included in the original
simulations) play a key role in preventing or quenching star formation in low-mass haloes \citep[see, e.g.,][and references therein]{kopo,sawa}. This leads to a significant
reduction in the number of observable satellites compared to the number of dark matter sub-haloes in simulations. On
the observational side, modern wide-area surveys, such as the Sloan Digital Sky Survey (SDSS, \citealt{dr9}) or the Pan-Andromeda Archaeological Survey \citep{pandas}, have more than doubled the number of known low-mass
satellites orbiting the MW and M31, revealing a wealth of
low-luminosity dwarfs ($M_V\ge -8$), usually referred to as ultra faint dwarfs (UFD), though they 
generally appear to be the simple extension of the dSph family to lower luminosities \citep[see][for a thorough discussion]{belo}. 
Since current surveys do not explore the whole sky and have limited depth, this implies the possibility 
that a significant number of faint members of the Local Group (LG) are still to be discovered
\citep[see, e.g.,][and references therein]{tolle}. 

All the new
systems have very low surface brightness \citep[SB hereafter, in the range $\mu_V\sim$
26.0-29.0 mag/arcsec$^2$,][]{martin} and have been detected as spatial
over-densities of resolved stars, which also
follow expectations for the location of stellar populations in the colour-magnitude
diagram \citep[CMD, see, e.g.,][]{belo07,richa,panda_ghosts}. 
Future, deeper surveys will greatly extend the exploration of the
relevant space of parameters, in search of faint distant satellites that cannot
be detected with the SDSS because of the relatively shallow limiting magnitude, $r\sim 22.5$ 
\citep[see, e.g.,][]{lsst} of that survey.

An alternative way to look for low-mass satellites that have so far escaped detection is to
search for low-mass compact HI clouds. Models of galaxy formation including
baryons suggest that  star formation is
dramatically reduced below a certain mass threshold. Consequently, it is expected that there should be a significant number of dark matter mini-haloes
containing $10^5-10^7~M_{\odot}$ of baryons in the form of neutral hydrogen, with or without an associated (small) stellar component  \citep{ricotti,kopo,sawa}. 

\citet[][A13 hereafter]{adams}
used the data from the Arecibo Legacy Fast ALFA (ALFALFA) \HI-line
survey to select a sample of 59 ultra compact high-velocity clouds (UCHVC) as
candidate dwarf galaxies, potentially hosted in mini-haloes within or just
outside the LG of galaxies \citep[see also][]{GALFA}. 
In particular, A13 selected clouds isolated from any known gas
complex in the halo (a) with an \HI~ major axis $<30\arcmin$,
corresponding to a physical diameter of  2~kpc at the distance of $D=250$~kpc, and (b)
with a Local Standard of Rest velocity far from the range covered by neutral gas in the
Galactic disc ($|V_{LSR}|>120$~\kms; see A13 for details). In addition, these authors identified a sub-sample of 17 top-quality
mini-halo candidates on the basis of the extreme isolation of the clouds { (i.e., clouds with no more than four neighbours within ten degrees of their centre, see A13)}; this is the ``most isolated sample'' (MIS). 

The A13 UCHVCs turn out to have small velocity widths ($W_{50}< 30$~\kms~
FWHM, with very few exceptions) and systemic velocities typical of known dwarf
galaxies members of the LG or lying just outside its zero-velocity boundary, floating in the Local Volume \citep[e.g., the NGC~3109 or the NGC~55 groups,][]{mcc,belfil}. 
As discussed in depth by A13, these systems are
 consistent with a population of LG dwarf galaxies located between 0.25 and 2~Mpc from us. 
A population of very faint, gas-rich, metal-poor, star-forming dwarf galaxies at
relatively short distances from us would not only play a crucial role in the 
missing satellites problem, but would also provide a direct probe of the
effects of feedback and re-ionization (as a function of baryonic and dark mass)
on the evolution of baryons within low mass haloes \citep[see][A13]{kopo,leop_1,brown}.   

The only way to directly confirm that the ALFALFA candidates are genuine dwarf galaxies
would be to identify a concomitant stellar counterpart whose distance could be estimated. This has indeed been the case for the most compact cloud identified in the survey: \citet{leop_1} and
\citet{leop_2} reveal that this cloud embeds a faint ($M_V\simeq -9.4$), low surface brightness
($\mu_V\simeq 24.5$~mag/arcsec$^2$), star-forming galaxy at a distance of $D\simeq 1.7$~Mpc, Leo~P \citep{leop_lbt}. 
Leo~P has a structure, as well as gas and stellar content, that is very similar to Leo~T, another recently discovered dwarf galaxy \citep{leoT}, 
which, quoting \citet{Ryan-Weber} ``$\dots $is not only the lowest luminosity galaxy with on-going star formation
discovered to date, but it is also the most dark matter-dominated, gas-rich
dwarf in the Local Group \dots''. Leo~T, which was initially discovered as a stellar
over-density in the SDSS at $D\sim 400$~kpc, lies just at the edge of the
completeness limit of the SDSS \citep[see][]{tolle}. Other similar galaxies could be
hidden at larger distances and the later discovery of Leo~P hints at the same conclusion. The ALFALFA UCHVCs, possibly together with the analogous
systems identified by the GALFA project based on the same observational material \citep{GALFA}, 
constitute the best available sample of candidates to follow-up on. 

\begin{figure}
   \centering
   \includegraphics[width=\columnwidth]{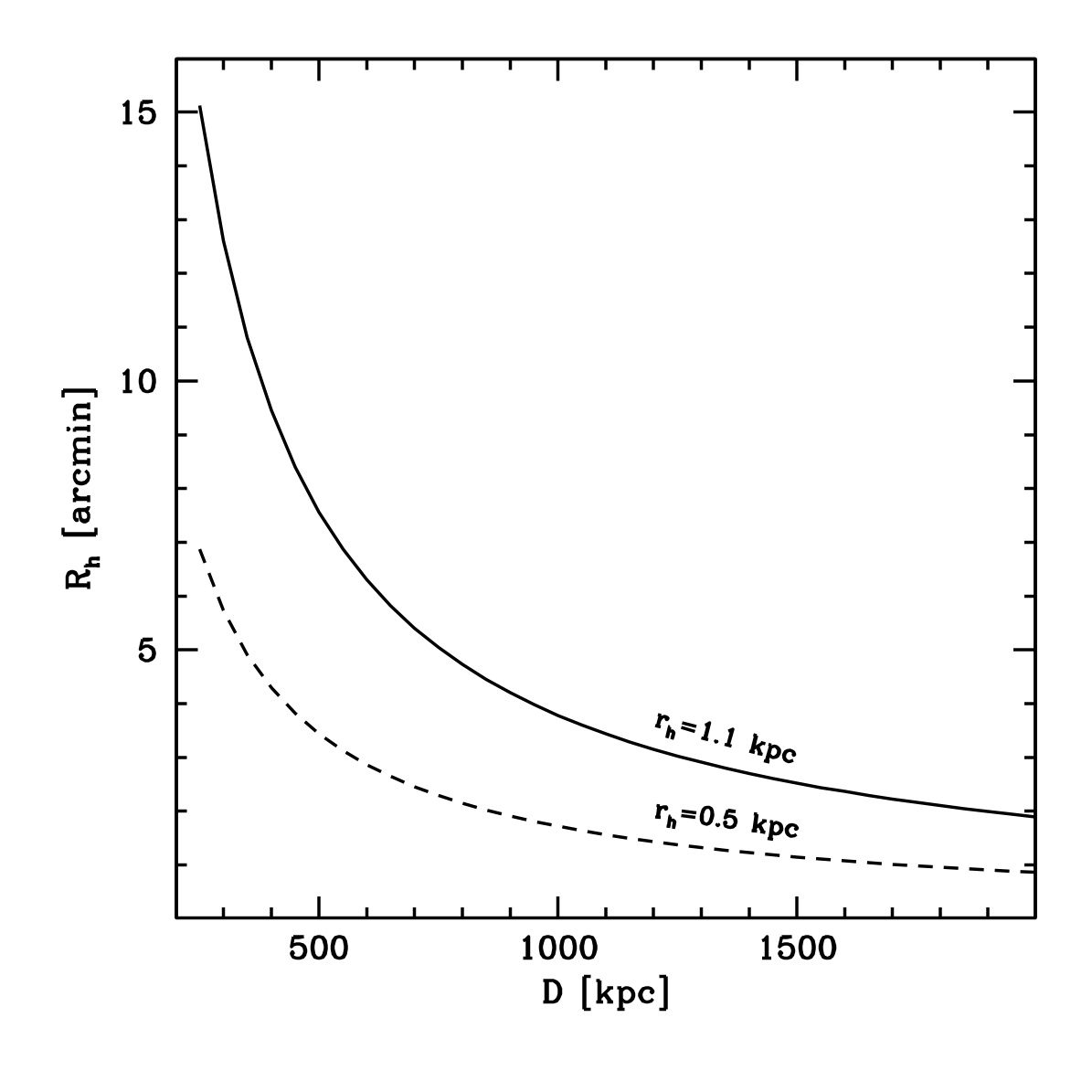}
     \caption{Angular half-light radius ($R_h$) as a function of distance for two assumptions on the physical half-light radius ($r_h$). A value of $r_h=1.1$~kpc corresponds to the upper limit of the distribution for dwarf galaxies in the Local Volume \citep{mcc} in the range of luminosities that is relevant to the present survey ($M_V\ge -12.0$; with the only exception of Andromeda~XIX, whose structure is likely strongly disturbed by the tidal interaction with M~31, see \citealt{XIX}). There are only nine dwarfs with $r_h\ge 0.5$~kpc amongst the fifty-four having $M_V>-12.0$; Leo~P has $r_h\sim 0.2$~kpc (N. Martin, private communication).}
        \label{disrad}
    \end{figure}

It is for these reasons that we undertook the study that we present in the present paper, the {\em StEllar Counterparts of COmpact high-velocity clouds} (SECCO) survey\footnote{{\tt www.bo.astro.it/secco} The acronym SECCO corresponds to the Italian word for ``dry;'' this appears appropriate since we are actually looking for the ``dry'' component (stars) of ``wet'' (gaseous) systems.}. The basic aim of the survey is to gather very deep wide-field imaging and photometry { (down to a magnitude level of $r\ga 26.0$)} in two passbands to search for the stellar counterparts of a significant fraction of the UCHVCs from the A13 sample. The analysis of the observational material will proceed in subsequent steps, from core to ancillary science, which will include progressive public data releases. In the present contribution, we describe the sample and present the first results of a search focused on dwarf galaxies similar to Leo~P, i.e. with typical angular sizes smaller than 3-4$\arcmin$ (half-light radius). 
For distances $D\ga 1.0$~Mpc this virtually encloses the whole range of sizes spanned by known Local Volume dwarf galaxies (see Fig.~\ref{disrad}).

The paper is organised as follows. In Sect.~\ref{secco} we provide a general description of the survey. In Sect.~\ref{visu} we present the results of the visual inspection of our images. Section~\ref{step1} presents the results of the deep stellar photometry performed in the central field of our mosaic images and the search for stellar over-densities by means of density maps. In Sect.~\ref{disc} we briefly summarise and discuss our main results.

\section{The SECCO survey}
\label{secco}

 The SECCO survey searches for stellar counterparts of a selected sub-sample of the A13 UCHVCs. From our observational experience \citep{vv124,sexAB}, we knew that very deep imaging and photometry for dwarf galaxies in the relevant range of distances (0.25~Mpc$\le D\le$~2.0~Mpc) could be obtained, under good seeing conditions, with a minimal observational effort using the LBC cameras \citep{lbc} mounted on the Large Binocular Telescope (LBT, Mount Graham, AZ) in binocular mode (see below). These observations provide the 
strongest constraints achievable with ground-based optical observations on the presence or lack of a stellar counterpart in a reasonable amount of observing time. Moreover, our intent is to put non-detections on a firm quantitative basis (see, e.g., Sect.~\ref{sens}), providing luminosity, size, and distance limits within which our observations would have detected a stellar system, if present. 

In the following sections we outline the main characteristics of the SECCO survey. 
Additional details on the various phases of the analysis will be provided in future contributions.

\subsection{The sample} 
\label{sample} 

We assembled our target list with the aim of sampling a significant fraction of the A13 sample, with a particular focus on the most promising candidates.
We included all the 17 MIS UCHVCs, classified as best mini-halo candidates by A13, to which we added the eight most compact candidates (i.e., those having the smallest mean diameter $d=\sqrt{ab}$) amongst the remaining 42 UCHVCs. { All of them have  $d<8.5'$  (where the mean and standard deviation of the whole sample are $\langle d\rangle = 10.7\arcmin$ and $\sigma_{d}=4.3\arcmin$)}, velocity width $W_{50}< 30$~\kms,  ($\langle W_{50}\rangle = 25.5$~\kms and $\sigma_{W_{50}}=8.9$~\kms, for the whole sample), and they are situated far away from the Magellanic Stream (A13). 
{ More compact clouds are better mini-halo candidates, since their physical sizes are
closer to those of known dwarf galaxies, in the relevant range of 
distances.}
The final sample tallies up to 25 fields that we observed with the large binocular cameras (LBC), i.e. 42\% of the whole A13 sample. In addition, to serve as a reference, we targeted a field centred on Leo~P with the same observing setup. Before submitting the proposal, we carefully inspected the images of all A13 UCHVC field in public archives (SDSS, DSS1, DSS2, 2MASS) to look for traces of an associated stellar system. While Leo~P is seen as a weak SB over-density in both SDSS and DSS2 images, none of the other targets shows a reliable sign of a counterpart in these images, thus confirming the results of previous inspections by \citet{giova07} and A13.

%
%
\begin{table*}
  \begin{center}
  \caption{List of surveyed fields.}
  \label{lista}
  \begin{tabular}{llrrccccc}
Field & CHVC & RA$_{J2000}$ & Dec$_{J2000}$ & $a \times b$ &$\langle E(B-V)\rangle$ & $\sigma_{E(B-V)}$ & $r_{90}$$^a$ & notes    \\
      &      &   [deg]      &    [deg]      &  [arcmin]    &     [mag]              &    [mag]          &  [mag]       &         \\
\hline
  A & HVC205.28+18.70+150$\star$ & 116.4995833	& +14.9769444 & $10\times 6$      & 0.029 &   0.004 &26.48  &  \\ 
  B & HVC204.88+44.86+147$\star$ & 142.5550000	& +24.2047222 & $ 8\times 6$      & 0.021 &   0.001 &26.37  &  \\ 
  C & HVC277.25+65.14-140$\star$ & 182.3333333	&  +4.3916667 & $ 7\times 4$      & 0.015 &   0.002 &26.31  &  \\ 
  D & HVC274.68+74.70-123$\star$ & 185.4779167	& +13.4694444 & $ 5\times 4$      & 0.048 &   0.004 &26.26  & Vis. cand. D1 \\ 
  E & HVC351.17+58.56+214$\star$ & 215.8383333	&  +4.5769444 & $ 7\times 5$      & 0.024 &   0.001 &26.20  &  \\ 
  F & HVC356.81+58.51+148$\star$ & 217.9950000	&  +6.5888889 & $ 6\times 5$      & 0.024 &   0.001 &26.11  &  \\ 
  G & HVC 13.59+54.52+169$\star$ & 226.8458333	& +11.5488889 & $10\times 5$      & 0.033 &   0.001 &26.38  &  \\ 
  H & HVC 13.60+54.23+179$\star$ & 227.1016667	& +11.4061111 & $15\times 7$      & 0.035 &   0.003 &25.63  &  \\ 
  I & HVC 13.63+53.78+222$\star$ & 227.5025000	& +11.1908333 & $ 9\times 6$      & 0.038 &   0.003 &25.59  &  \\ 
  J & Leo P		         & 155.4379167	& +18.0880556 & $ 3\times 1.6^{b}$& 0.025 &   0.008 &26.43  &  \\ 
  K & HVC196.09+24.74+166        & 119.0616667	& +25.1500000 & $10\times 5$      & 0.059 &   0.008 &26.35  &  \\ 
  L & HVC245.26+69.53+217$\star$ & 175.0337500	& +15.1122222 & $10\times 9$      & 0.026 &   0.001 &26.27  &  \\ 
  M & HVC298.95+68.17+270$\star$ & 191.3741667	&  +5.3397222 & $16\times 9$      & 0.019 &   0.002 &25.98  &  \\ 
  N & HVC326.91+65.25+316$\star$ & 202.6825000	&  +4.2272222 & $12\times 10$     & 0.025 &   0.001 &25.76  &  \\ 
  O & HVC 28.09+71.86-144$\star$ & 212.7420833	& +24.2011111 & $15\times 9$      & 0.019 &   0.002 &25.99  &  \\ 
  P & HVC353.41+61.07+257$\star$ & 214.9525000	&  +7.1875000 & $13\times 9$      & 0.027 &   0.002 &26.20  &  \\ 
  Q & HVC352.45+59.06+263$\star$ & 215.9904167	&  +5.3944444 & $16\times 11$     & 0.024 &   0.002 &26.16  & Overd. Q1 \\ 
  R & HVC  5.58+52.07+163$\star$ & 226.1720833	&  +6.2163889 & $11\times 10$     & 0.036 &   0.007 &26.35  &  \\ 
  S & HVC 27.86+38.25+124$\star$ & 246.1808333	& +12.7366667 & $11\times 9$      & 0.053 &   0.009 &25.76  &  \\ 
  T & HVC330.13+73.07+132        & 200.6733333	& +11.8752778 & $ 6\times 3$      & 0.027 &   0.005 &25.91  &  \\ 
  U & HVC250.16+57.45+139        & 167.3741667	&  +5.4336111 & $ 7\times 4$      & 0.045 &   0.009 &26.32  &  \\ 
  V & HVC324.03+75.51+135        & 198.1762500	& +13.5127778 & $ 7\times 5$      & 0.020 &   0.006 &25.64  &  \\ 
  W & HVC28.07+43.42+150         & 241.3858333	& +14.9888889 & $10\times 5$      & 0.035 &   0.006 &25.86  &  \\ 
  X & HVC290.19+70.86+204        & 188.6675000	&  +8.4022222 & $10\times 6$      & 0.019 &   0.005 &25.73  &  \\ 
  Y & HVC255.76+61.49+181        & 172.2316667	&  +6.4247222 & $11\times 6$      & 0.036 &   0.003 &26.36  &  \\ 
  Z & HVC26.01+45.52+161         & 238.7812500	& +14.4913889 & $ 8\times 6$      & 0.032 &   0.001 &25.89  &  \\ 
\hline
\end{tabular} 
\tablefoot{The listed coordinates are for the centroid of the UCHVC as reported by A13. { UCHVC major and minor diameters, $a$ and $b$, measured approximately at the level encircling half the total flux density, are also from A13.}
The mean colour excess ($\langle E(B-V\rangle$) and the associated standard deviation ($\sigma_{E(B-V}$) have been obtained from the \citet{ebv} reddening maps, read on a $15\times 15$ grid of $1.8\arcmin$ pixels and adopting the new calibration by \citet{schlaf}.\\
$\star$ Part of the {\em Most Isolated Sample} (MIS) as defined by A13. $^a$ r magnitude of the level including 90\% of the sources
of the DAOPHOT catalogues that passes the selections in CHI and SHARP and have $r_0>23.0$ and $-1.0<(g-r)_0\le 1.3$.
$^{b}$ From \citet{leop_1}.}
\end{center}
\end{table*}

The list of the surveyed fields is presented in Table~\ref{lista}. Each observed field is denoted by a capital letter (Column 1). The name of the associated UCHVC, its coordinates, and major and minor axes\footnote{As measured at the level encircling approximately half of the total \HI ~flux density (A13).}, taken from A13, are reported in Columns 2, 3-4, and 5, respectively. We interpolated the \citet{ebv} reddening maps, adopting the new calibration by \citet{schlaf}, on a $15\times 15$ grid centred on the UCHVC coordinates { with knots spaced by $1.8\arcmin$}. The mean and the standard deviation of these 225 values are reported in Columns 6 and 7, respectively. We note that the reddening towards the considered directions is from moderate to low, with small variations within each field. As a consequence, we only use the mean reddening values reported in Table~\ref{lista} to correct our photometry for the extinction in a given field. Throughout the paper (and the survey), we adopt the following reddening laws: $g_0 = g - 3.303 \, E(B-V)$ and $r_0 = r - 2.285 \, E(B-V)$, taken from \citet{schlaf}. In column 8 we report $r_{90}$, a parameter that will be defined in Sect.~\ref{obs},meant to characterise the limiting magnitude of the photometry in each field. Finally Column 9 lists any remarkable note stemming from our analysis of these fields.

\subsection{Observations}
\label{obs}

Observations were acquired using the wide-field LBC cameras  in binocular mode at the LBT, during the nights of January 29, March 7, and  March 29, 2014. For each field, two 300~s and two 20~s exposures were acquired in the two adopted passbands, SDSS g and r. The g and r images were obtained with the LBC-Blue and LBC-Red camera, respectively. 
The optics of each LBC feed a mosaic of four 4608~px $\times$ 2048~px CCDs with a pixel scale of 0.225 arcsec~px$^{-1}$. Each CCD chip covers a field of $17.3\arcmin\times 7.7\arcmin$. Chips 1, 2, and 3 flank one another, being adjacent along their long sides; Chip 4 is placed perpendicular to this array, with its long side adjacent to the short sides of the other chips \citep[see][]{lbc}. A small, simple dithering pattern was adopted to fill the gaps between Chips 1, 2 and 3. The pointings were
chosen so that the centre of the targeted UCHVC falls approximately in the middle of chip~2. 
All the images were acquired with the standard orientation of the camera except for field~N, where a rotation of 30$\degr$ from the north-south direction was performed to avoid a very bright foreground star.
The main details on the observations of each field are reported in Table~\ref{logobs}. Eighty-four percent of the $t_{exp}=300$~s images were gathered with seeing conditions compliant with our requirements, i.e. FWHM$_{PSF}\le 1.2\arcsec$, and 54\% of the images have FWHM$_{PSF}\le 1.0\arcsec$, while only 3\% have FWHM$_{PSF}\ge 1.5\arcsec$.

All the images are corrected for bias, dark, and flat field in a homogeneous way by the LBT-Italian Coordination Facility\footnote{\tt http://lbt.inaf.it}, which also provided sky-subtracted stacked mosaic images. 

\subsection{Plan of the survey: analysis}
\label{surv_ana}

The analysis of the whole dataset follows a series of steps, most of them related to the main scientific goal of the survey:

\begin{itemize}

\item STEP 0: visual inspection of the deep images. In the stacked $2\times 300$~s g and r images, we carefully search by eye for visible agglomeration of stars and/or partially resolved stellar systems that { could} be associated to the targeted UCHVCs. It is important to recall that Leo~P is readily visible on relatively deep optical images acquired with 4~m class telescopes \citep{leop_2}, and it stands out very clearly in typical LBC images \citep[see][and below]{leop_lbt}. It is very likely that stellar systems that are significantly less luminous than Leo~P can be visually identified in our images, so such an inspection is clearly worth doing. In the future we plan to quantify the sensitivity of the process by adding fake galaxies (of known luminosity, size, and distance) to our images and by repeating the visual inspection to find out the range of galaxy parameters that allow us to identify the system as a candidate counterpart in the images (see also Sect.~\ref{sens}).

\item STEP~1: very deep stellar photometry of Chip~2 \citep[with DAOPHOT,][]{daophot} and automatic search for over-densities. As mentioned above, the single $17.3\arcmin\times 7.7\arcmin$ field covered by the central chip is wide enough { to adequately
sample the stellar body of typical dwarf galaxies residing in the Local Volume \citep[see Fig.~\ref{disrad}, and][]{mcc,vv124,sexAB}}. 
In this context, it is important to recall that most of the stellar light in gas-rich dwarf galaxies is generally enclosed in a significantly smaller area than is spanned by the neutral gas distribution. For example the typical size (along the major axis) of the \HI ~cloud associated with Leo~P is $\simeq 3\arcmin$, while the half-light radius along the same direction is just $\simeq 0.7\arcmin$ (N. Martin, private communication). Over-densities of stellar sources are expected to be identified as significant peaks in density maps obtained from the stellar photometry catalogues. 
Experiments with synthetic galaxies will allow us to put robust constraints on the sensitivity of the whole process.
A detailed description of STEP~1 is provided in Sect.~\ref{step1}, below, where we present the main results of this part of the analysis.

\item STEP~2: star and galaxy photometry for all four chips in each field with Sextractor \citep{sex}. This  enables a search for stellar over-densities on larger angular scales  and allows us to test these against the over-densities of galaxies, which are (by far) the main source of false positives in our fields. 
The approach is analogous to
STEP~1 (i.e., search for peaks in density maps) but is more sensitive to closer low SB stellar systems, and less sensitive to compact and distant ones, which are best resolved by the exquisite PSF modelling of DAOPHOT.
Obtaining the photometric catalogues for all fields with Sextractor is much faster and requires much less human effort than the DAOPHOT photometry adopted in STEP~1, at the expense of some loss in photometric depth and accuracy at faint magnitudes; this appears to be a reasonable trade-off, since STEP~2 is focused on the search for more nearby stellar counterparts.

\item STEP~3: ancillary science, not related to the main goal of the survey. For example, our deep images of (mostly) high galactic latitude fields contain a large number of background galaxies that are not resolved into stars but are well resolved as extended objects. The catalogues from STEP~2 
will give us the opportunity to recover at least some of the information about these extended sources that is encoded in our images (e.g., half-light radii, ellipticity, position angles, etc.). This is not negligible since we cover a total of $\simeq 3.8$~deg$^2$ of high-latitude sky down to $r\simeq 26.5$. Moreover, these catalogues can be searched for galaxy clusters. Another example of ancillary science that is made possible by the survey data is the detailed study of several interacting galaxies that are serendipitously present in our images.

\end{itemize}

Finally, we will make publicly available most of our value added data in subsequent releases on the survey web site.

In the present contribution we present the results of STEP~0 and STEP~1 for all the fields, as well as an initial set of simulations to provide a first robust statement on the non-detection of significant over-densities in our STEP~1 catalogues. Extensive simulations should and will be performed in the course of the survey to obtain quantitative limits to the sensitivity of the search made in STEPs~0, 1 and 2. The first set of STEP~1 simulations is presented in this paper to provide a first robust statement about the non-detection of stellar systems in our fields. A fully detailed description of STEPs 2 and 3 are beyond the scope of the present paper, but will be provided in future contributions dedicated to these additional phases of the analysis.

\section{STEP 0: Visual inspection of the images}
\label{visu}

\begin{figure}
   \centering
   \includegraphics[width=\columnwidth]{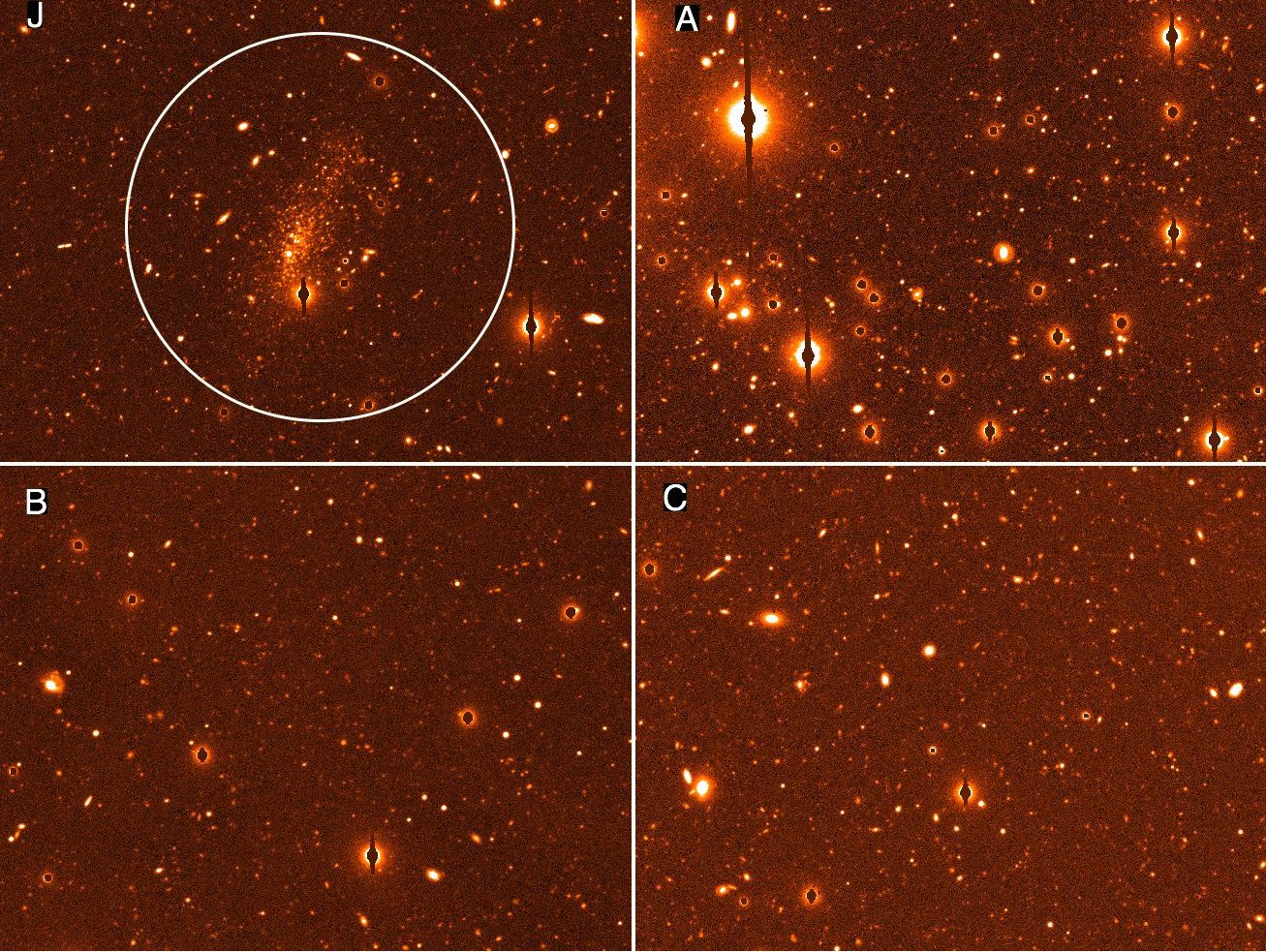}
     \caption{Portions of the r-band stacked images for fields J, A, B, and C, centred on the position of the associated UCHVCs. All the images shown here have the same size. In the upper left panel, a circle with radius $r=1.5\arcmin$ is plotted for reference. In all the images, north is up and east is to the left.}
        \label{quattro_ima}
    \end{figure}

In Fig.~\ref{quattro_ima} we show equal-size portions of the stacked r-band mosaics, centred on the location of the associated UCHVC for Leo~P and for the first three fields listed in Table~\ref{lista}. The aim of this plot is to illustrate how straightforward it would be to identify a stellar system similar to Leo~P in our images. 

We carefully inspected all our deep images, with  particular attention to the stacked g and r mosaics. In particular, two of us (M.B. and G.B.) independently perused all the images in search of groupings of stars and partially resolved stellar systems that could be plausibly identified as the counterparts of our target UCHVCs. For each potential candidate we searched the SIMBAD and NED databases to verify if it was a known system and, in this case, to ascertain its nature. The two independent searches were in excellent agreement. This reflects the very fundamental result of our STEP~0 analysis: there is no  stellar counterpart identifiable by visual inspection in our images, with the only exception the one discussed below. In particular, we can safely conclude that there is no galaxy similar to Leo~P in any of our 25 target fields. This is the main reason for the sub-title of this first paper of the series. However, while it is clearly advisable to model our expectations for the appearance of UCHVC stellar counterparts on the characteristics of known dwarf galaxies, { we cannot exclude that different kinds of stellar systems
have escaped detection until now.} 
Indeed the discoveries of the past decade have taught us that the range of parameter space populated by dwarf galaxies is much wider than previously thought and that  the classification boundaries between different classes of stellar systems can be quite blurred \citep[see, e.g.,][and references therein]{willstra,mic_pechino}. 

\begin{figure}
   \centering
   \includegraphics[width=\columnwidth]{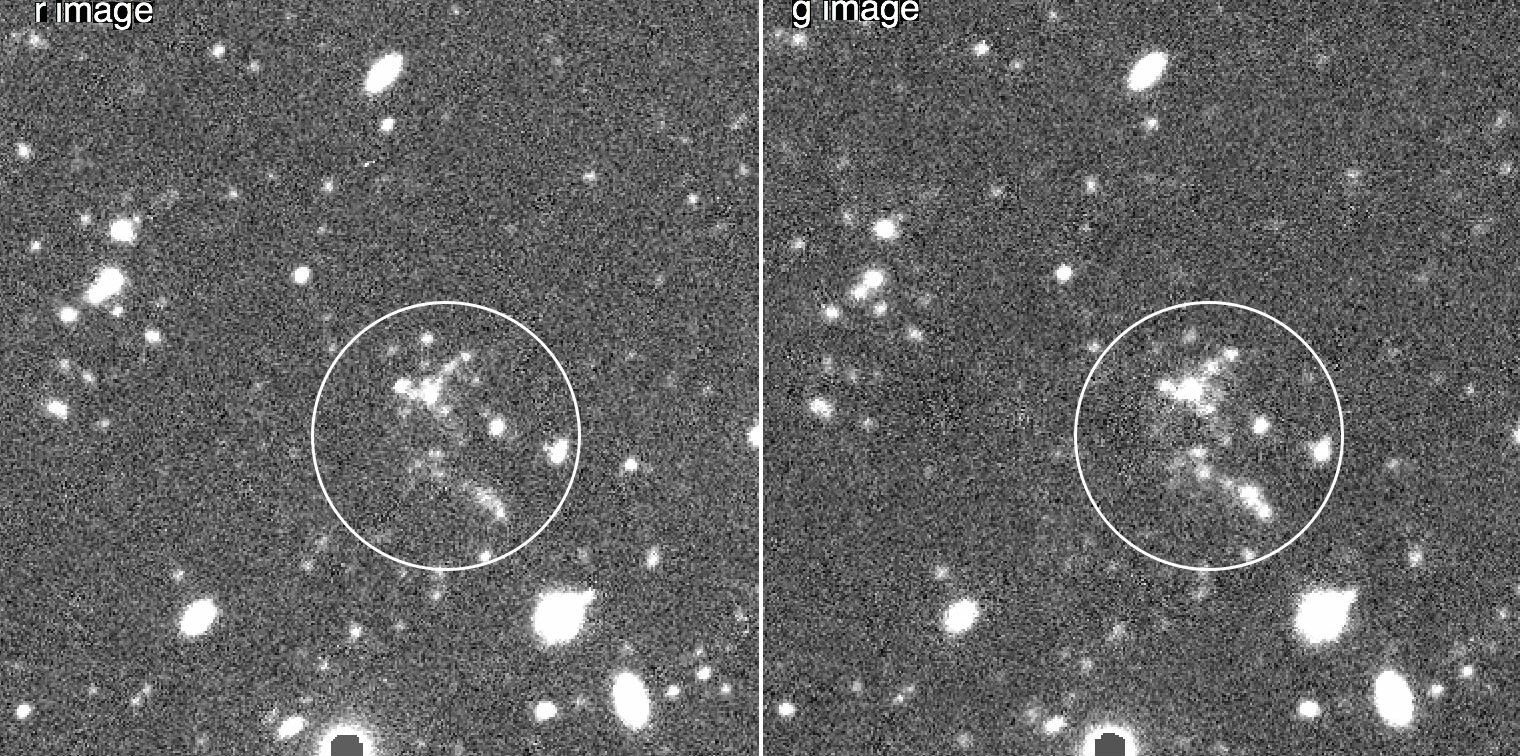}
     \caption{r-band (left panel) and g-band (right panel) stamp images zoomed-in on object D1. The superimposed circles have radii of $r=15\arcsec$. In all the images, north is up and east is to the left.}
        \label{D1_ima}
    \end{figure}

Inspecting the images from Field D with these caveats in mind, we noticed the $\sim15\arcsec$, irregular cluster of relatively bright sources surrounded by some light fuzziness that we show in Fig.~\ref{D1_ima} and  that we dub candidate D1. 
The approximate centre of the structure is located at $RA=185.47482$~deg and $Dec=13.46015$~deg, just $35\arcsec$ to the south-south-west of the centroid of the associated UCHVC (HVC274.68+74.70-123).

\begin{figure}
   \centering
   \includegraphics[width=\columnwidth]{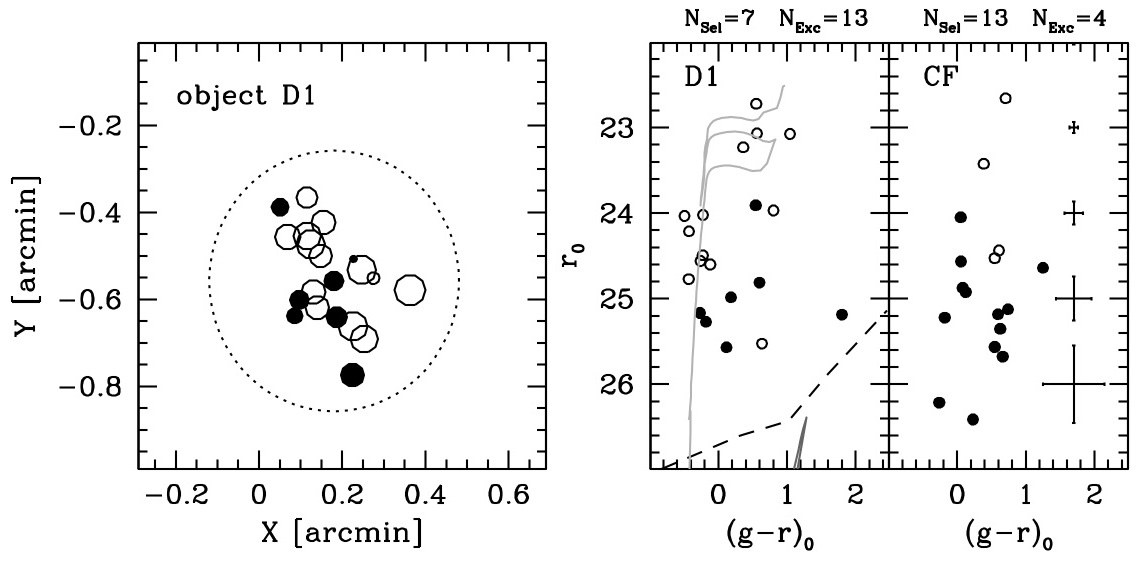}
     \caption{Left panel: maps of the sources in the catalogue of Field D lying within $18\arcsec$ (large dotted circle) of the centre of object D1. Filled circles are sources that pass our quality selection, and open circles are sources that would be excluded from the selection. The size of the circles is proportional to the magnitude of a source (larger circles corresponding to brighter sources). Right panels: the CMD of these sources compared to the CMD of a nearby control field of the same size. The meaning of the symbols is the same as for the right panel. { The number of selected ($N_{Sel}$) and excluded ($N_{Exc}$) stars is reported on top of each panel. The dashed line marks the level of the limiting magnitude as a function of colour for this field.} On the D1 CMD we have superimposed two theoretical isochrones from the \citet{marigo} set, with metallicity $Z= 0.001$ and age 28~Myr (light grey line) and 12.5~Gyr (dark grey line), shifted to a distance of D=7.2~Mpc.}
        \label{D1_cmd}
    \end{figure}

Given the overwhelming dominance of distant galaxies amongst faint sources in our catalogues the most likely hypothesis is that D1 is a galaxy group (or cluster). However, the grainy nature of the bright blobs and the fuzzy light around may also be compatible with a few clusters of young stars (and/or HII regions) within a very low SB envelope of unresolved older stars. In this case the system would be located at larger distances than the range targeted by our survey since in similar observing conditions, we resolve the RGB of Leo~P at $D\simeq1.7$~Mpc, down to $\sim 3.0$~mag below the RGB tip. A dwarf galaxy at $D=3.0$~Mpc would have the tip just 1.1 magnitudes fainter than Leo~P, so we would most likely still resolve its brightest RGB stars, while no RGB population is resolved in D1, as can also be appreciated from the CMD shown in Fig.~\ref{D1_cmd}.

Figure~\ref{D1_cmd} is aimed at checking if the CMD of the resolved sources is compatible with the hypothesis that D1 is a distant star-forming galaxy; the details of the stellar photometry are reported in Sect.~\ref{step1}, below. Only seven sources passing our selection criteria for stellar photometry ( see, again, Sect.~\ref{step1}, below) are enclosed within $18\arcsec$ of the centre of D1. The remaining fourteen sources are excluded because they are more extended than genuine single stars. This supports the idea that they are indeed distant galaxies but it is not incompatible with partially resolved star-forming knots. The comparison with the control field (CF) displayed in Fig.~\ref{D1_cmd} shows that the fraction of sources passing the selection is significantly lower in D1 than in the CF. This is a general property of the D1 set of sources due to the higher degree of crowding in the object with respect to the surrounding field.

Finally, Fig.~\ref{D1_cmd} shows that the CMD of D1 is compatible with the hypothesis of a distant stellar system  with a sparse population of young MS, { blue and red super giant stars}, and the old component (RGB) located below the detection threshold. { In the specific case shown in Fig.~\ref{D1_cmd}, we compare the observed CMD with two theoretical isochrones from the \citet{marigo} set, with metallicity $Z= 0.001$ and age 28~Myr and 12.5~Gyr, shifted to a distance of D=7.2~Mpc.}
The comparison is purely illustrative, since the brightest sources may not be single stars but, in fact, star clusters.
The comparison with a large set of CFs of the same area indicates that, when sources passing and not passing the selection are considered together, the D1 area does indeed have a higher-than-average number of sources bluer than $(g-r)_0=0.0$. In the specific case illustrated in Fig.~\ref{D1_cmd} the comparison yields 9 vs. 0. A spectroscopic follow-up of the brightest sources in D1 is currently underway to shed light on the nature of this detection.

\subsection{Chance superpositions}

In a few images the visual inspection reveal the presence of very bright sources near the position of the UVHVC. While these nearby stars or distant galaxies are clearly not related to the gas clouds, they may still be worth reporting for future researchers. We identified these sources using the NED and SIMBAD databases.

The very bright stars (r=11.2) SDSS~J124525.80+052116.6 lies at $\simeq 80\arcsec$ from the centre of  HVC298.95+68.17+270, in Field M.
A star of similar brightness, but lacking optical photometry in the consulted datasets, classified as a UV source, GALEXASC~J141051.96+241223.8, lies at $\simeq 90\arcsec$ from HVC28.09+71.86-144, in Field O.
The bright SA0/a galaxy IC~1174, with a magnitude of r=13.2 and a redshift $z\simeq 0.015$, is $\simeq 155\arcsec$ from the centre of HVC28.07+43.42+150, in Field W. A similarly bright face-on spiral located at redshift $z\simeq 0.037$ and with a remarkably disturbed morphology, lies at $\simeq 66\arcsec$ from HVC26.01+45.52+161 in field Z. It is SDSS~J155509.49, associated to the radio source AGC~258032, but lacks any optical magnitude estimate.

\section{STEP 1: deep PSF photometry of the central chips}
\label{step1}

   \begin{figure}
   \centering
   \includegraphics[width=\columnwidth]{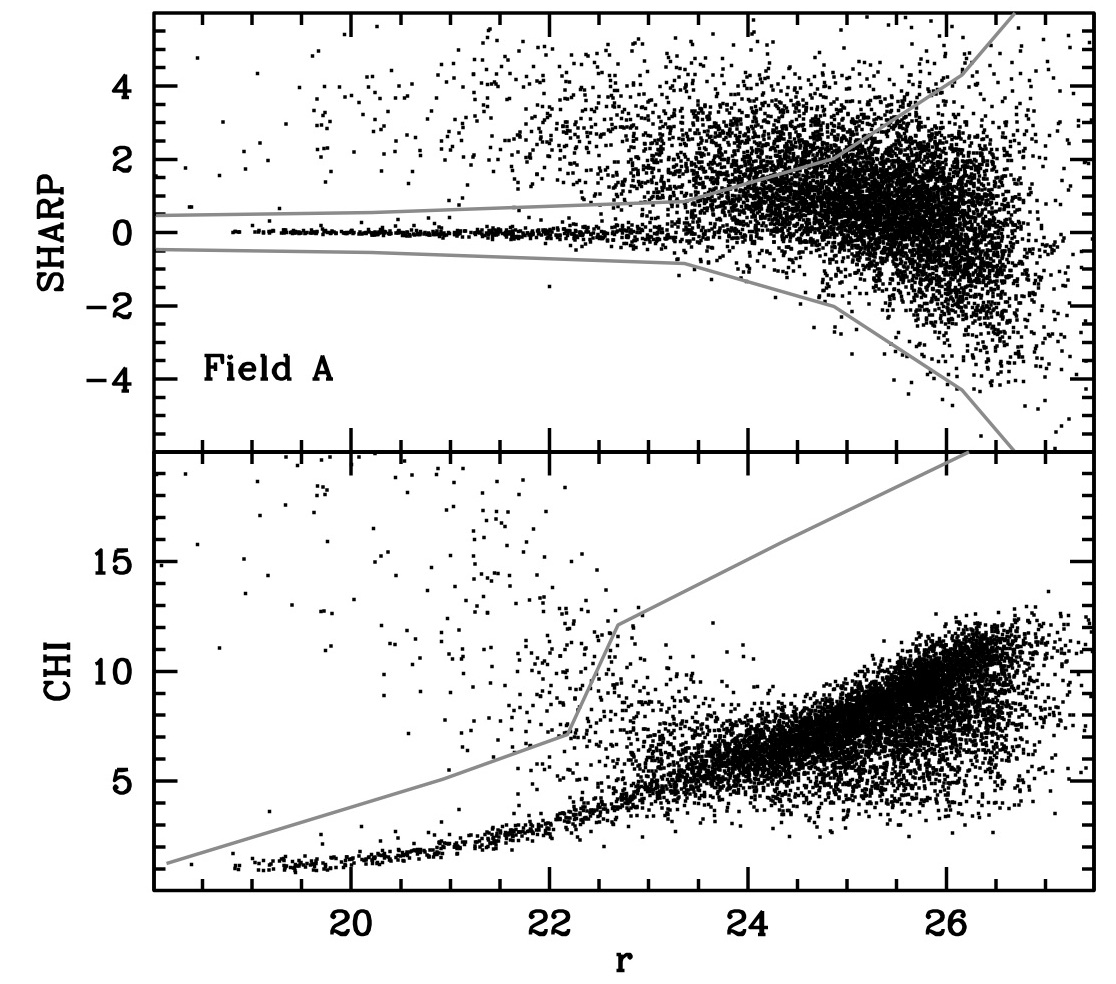}
     \caption{Quality selection of stellar sources based on the DAOPHOTII parameters SHARP (upper panel)
     and CHI (lower panel). The sources enclosed within (upper panel) and below (lower panel) the grey lines are selected as stars for further analysis. This illustrative example shows the photometric catalogue for Field A.}
        \label{seesel}
    \end{figure}

   \begin{figure*}
   \centering
   \includegraphics[width=\textwidth]{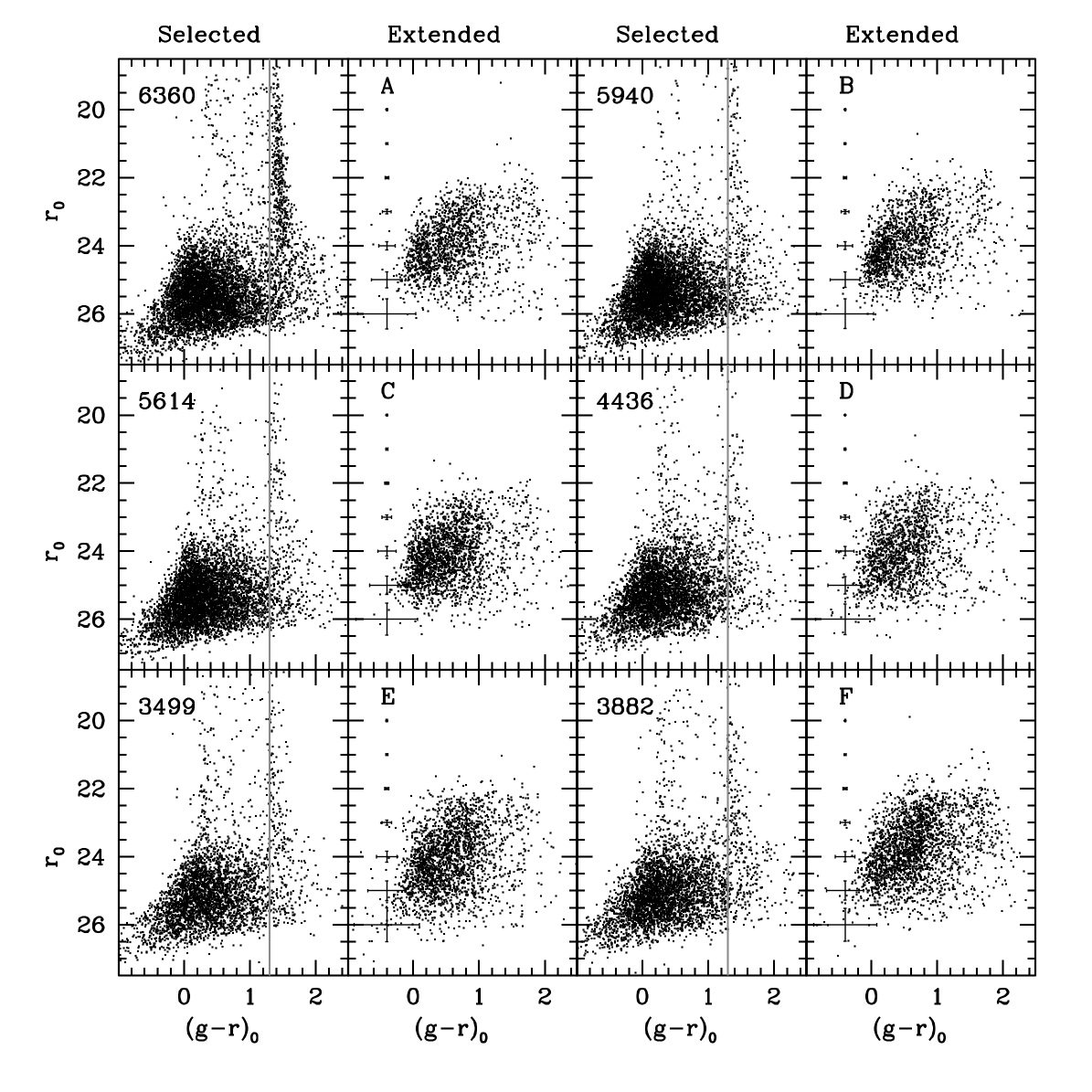}
     \caption{Reddening-corrected CMDs of six fields. For each field we present the CMD of the sources
     selected according to their CHI and SHARP parameters (left panels of each pair; ``Selected'' columns of the array of panels) and the CMD of those sources that are excluded from the selection because of their SHARP value larger which is higher than the (positive) limit; i.e., they are recognised by DAOPHOT as more extended than point sources (right panels of each couple;``Extended'' columns of the array of panels). The grey line in the CMDs of selected sources marks the colour threshold adopted to exclude local M dwarfs from the computation of density maps. The error bars are the average photometric errors in colour and magnitude as a function of magnitude, for {\em Selected} sources, computed over one magnitude wide bins. The number of {\em Selected} sources is also reported
     in the upper left corner of the respective CMD. }
        \label{cmd1}
    \end{figure*}

\subsection{Data reduction}
\label{reduz}

Since  we are focusing on STEP~1 in this section, the following refers to the Chip~2 images of each field.
Relative photometry was performed independently on each $t_{exp}=300$~s exposure using the point spread function (PSF)-fitting code DAOPHOTII/ALLSTAR \citep{daophot,allframe}. Intensity peaks more than 3$\sigma$ above the background were identified in a stacked image obtained by registering and co-adding all the images considered for the analysis. Then, each of these stars was re-identified and fitted on each image, when possible. 
Only sources found at least in two g and two r images were retained in the final catalogue. The average and the standard error of the mean of the independent measures obtained from the different images were adopted as the final values of the instrumental magnitude and of the uncertainty on the relative photometry.
It has been demonstrated \citep{vv124,sexAB} that this procedure works very well on similar datasets aimed at analysising of dwarf galaxies in the same range of distances as we are interested in here.

To clean the catalogues  from spurious and non-stellar sources, we also adopted cuts in the image-quality parameters $CHI$ and $SHARP$, provided by DAOPHOTII { \citep[as done, e.g., in][]{vv124,sexAB}}. The SHARP (sharpness) 
parameter provides a classification of detected sources according to the shape of their light distribution with respect to a point source, which should have SHARP=0.0. Negative values of SHARP indicate a source whose distribution is more peaked than a point source (e.g., cosmic spikes, hot pixels), whilst positive values indicate distributions that are less peaked than a point source, such as any kind of extended spurious sources (e.g., residuals from the subtraction of a bright star, erroneous decomposition of a resolved galaxy into several spurious point sources, unresolved blended stars) or genuinely extended objects (galaxies whose extended nature is resolved). The CHI ($\chi^2$) parameter provides an indication of the quality of the fit of a given source performed by DAOPHOT; a higher CHI value indicates a worse fit and may reveal, for example, badly identified sources or problems with the local estimates of the sky level.

After a careful inspection of the distribution of measured sources in the mag vs. $CHI$ and mag vs. $SHARP$ planes, we adopted the magnitude-dependent selection limits illustrated in Fig.~\ref{seesel}. This selection criterion is defined to be broad enough to avoid exceedingly severe selections for all the fields.  The upper panel of Fig.~\ref{seesel} clearly shows that our images are populated by two different families of genuine sources,
stars, tightly clustered around SHARP=0.0, and distant galaxies broadly spread at SHARP$\ga 1.0$.
The two kinds of sources are distinguished well down to $r\simeq 24.5$, at which point the two sequences merge and the ability to differentiate galaxies from stars is lost. The selection limits plotted in the upper panel of Fig.~\ref{seesel} imply that our selected samples necessarily include both genuine stars and unresolved galaxies below this magnitude limit. This choice is unavoidable if we want to take advantage of the full depth of our photometry. In Sect.~\ref{dens} we perform our search for over-densities on samples with different limits in magnitude that should be affected by significantly  different degrees of contamination from unresolved galaxies to explore the trade-off between depth and sample fidelity (see also Sect.~\ref{sens}, for discussion).

To provide a well-defined proxy for the limiting magnitude of each photometric catalogue, which is likely a key parameter to rank the sensitivity of the observations in the various fields, we define $r_{90}$ as the faint r magnitude level that encloses 90\% of the sources that passed the CHI and SHARP selection and have $r_0>23.0$ and $-1.0<(g-r)_0\le 1.3$, i.e. sources truly relevant to detecting distant dwarf galaxies. For $r\le 23.0$ and for $(g-r)_0> 1.3$, our samples are dominated by foreground Galactic stars (see below); no star is expected to have $(g-r)_0<-1.0$ \citep{fadely} even when taking photometric errors into account. The measured values of $r_{90}$ are listed in Table~\ref{lista}, and range from $r_{90}=26.48$ (Field A) to $r_{90}=25.59$ (Field I).

All the surveyed fields are within the SDSS Data Release 9 (DR9) footprint. Therefore, for each field we used the hundreds of stars in common between our catalogues and DR9 to transform the x,y pixel coordinates into Equatorial J2000 RA and Dec and to calibrate our relative photometry onto the SDSS system. The details of these processes are described in Appendix~\ref{app_obs}.

\subsection{Colour-magnitude diagrams}
\label{cmd}

In Fig.~\ref{cmd1} we present the reddening-corrected CMDs for the first six fields of the survey, which we use here to illustrate the main features that are common to all the fields. The same diagrams for the other fields are presented in Appendix~\ref{app_cmds}, in Figures \ref{cmd2} and \ref{cmd3}. For each field we present two CMDs: on the left-hand side, the diagram of the sources that pass our selection in CHI and SHARP (labelled as {\em Selected}) and, on the right-hand side, the sources that pass the selection in CHI but have higher positive SHARP values than the adopted limits (labelled as {\em Extended}).

In all cases the photometry of selected sources reaches $r_0\simeq 26.0$, albeit with different levels of completeness. There are two general features that are easily recognised as due to foreground Galactic stars, { in the CMDs of the {\em Selected} samples}: the vertical plume of local M dwarfs running along the whole diagram around $(g-r)_0\sim 1.5$, and the sparser halo main-sequence (MS) turn-off sequence at $(g-r)_0\sim 0.4$, which bends towards the M dwarfs at faint magnitudes. 
The grey line at $(g-r)_0=1.30$ shows that Galactic M-dwarfs can be selected out from further analysis with a simple colour cut.
The vast majority of sources with $r_0\ga 24.5$ are distant galaxies that are too faint to be reliably distinguished from stars based on their images. They are broadly distributed into a blue sequence $0.0\la (g-r)_0\la 0.6$ and a red sequence $0.6\la (g-r)_0\la 1.0$ \citep{bla}. For a more detailed discussion of { CMDs in fields at high galactic latitude} with comparable depth, see \citet{vv124} and references therein.

The CMDs of the sources recognised as {\em Extended} have a brighter limiting magnitude, very likely determined by a light distribution less peaked than point sources, and this frustrates their detection by DAOPHOTII.
Their colour distribution agrees with the blue sequence/red sequence scheme sketched above, but with an additional sparse population of  redder sources, whose colour [$(g-r)_0>1.0$] is consistent with passively evolving early type (red sequence) galaxies at $z\ga 0.4$, according to the \citet{buz} population synthesis models. It is important to remember that the {\em Extended} class sources, as presented in Fig.~\ref{cmd1}, are clearly dominated by background galaxies but may contain a 
non-negligible number of spurious sources. 

\subsection{Artificial stars experiments}
\label{comple}

To estimate the completeness of our photometry, we performed artificial star experiments on Field~B, taken as representative of 
top-quality observations (sub-arcsec seeing, amongst the fields with the deepest photometric limit), and on Field~I, taken as representative of the worst quality observations (bad seeing, the field with the shallowest photometric limit). This is sufficient, at the present stage, for characterising our photometry for the purposes of the present paper and producing the first set of sensitivity tests presented in Sect.~\ref{sens}, since the two considered cases bracket the whole range of survey sensitivities (see Sect.~\ref{sens}). 

   \begin{figure}
   \centering
   \includegraphics[width=\columnwidth]{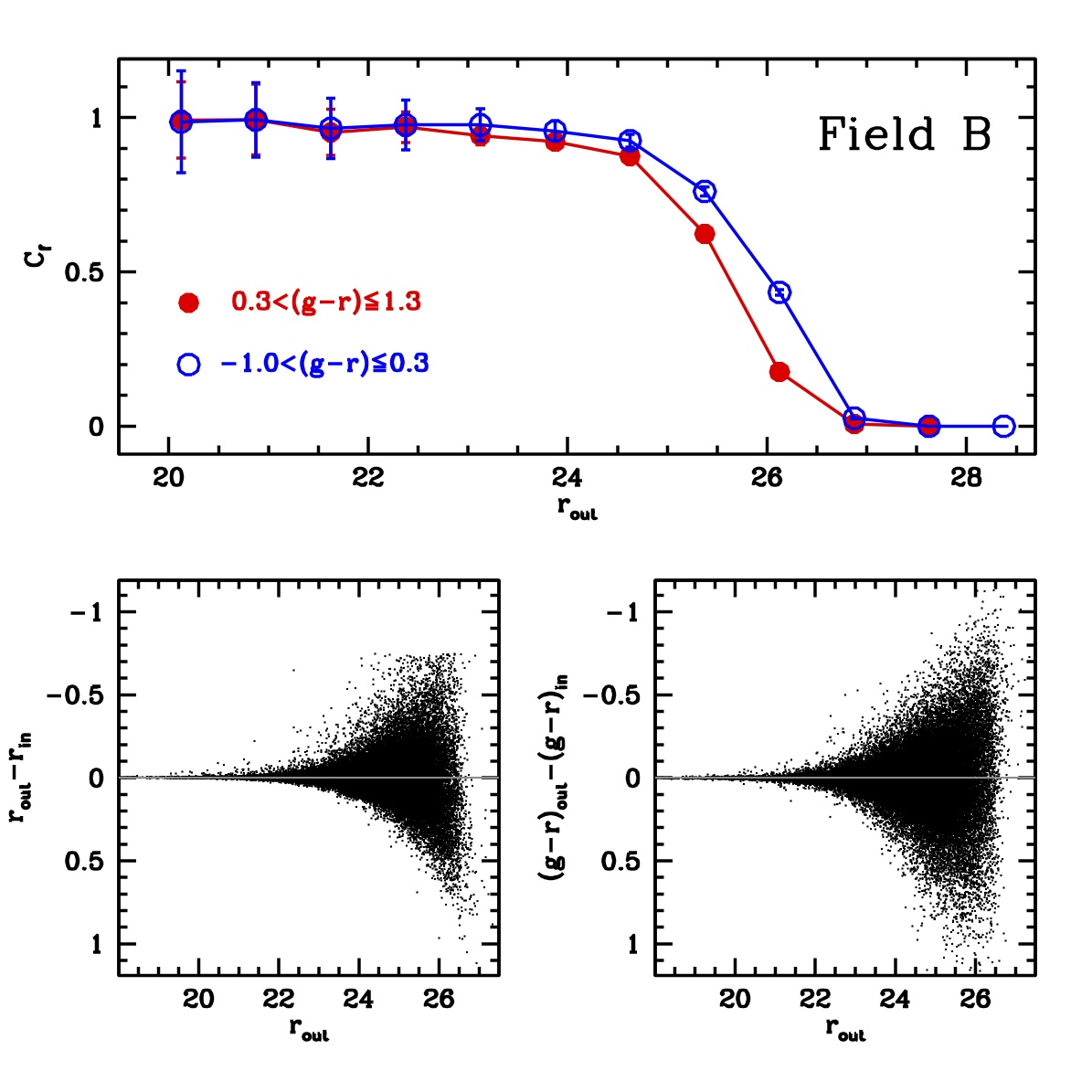}
   \includegraphics[width=\columnwidth]{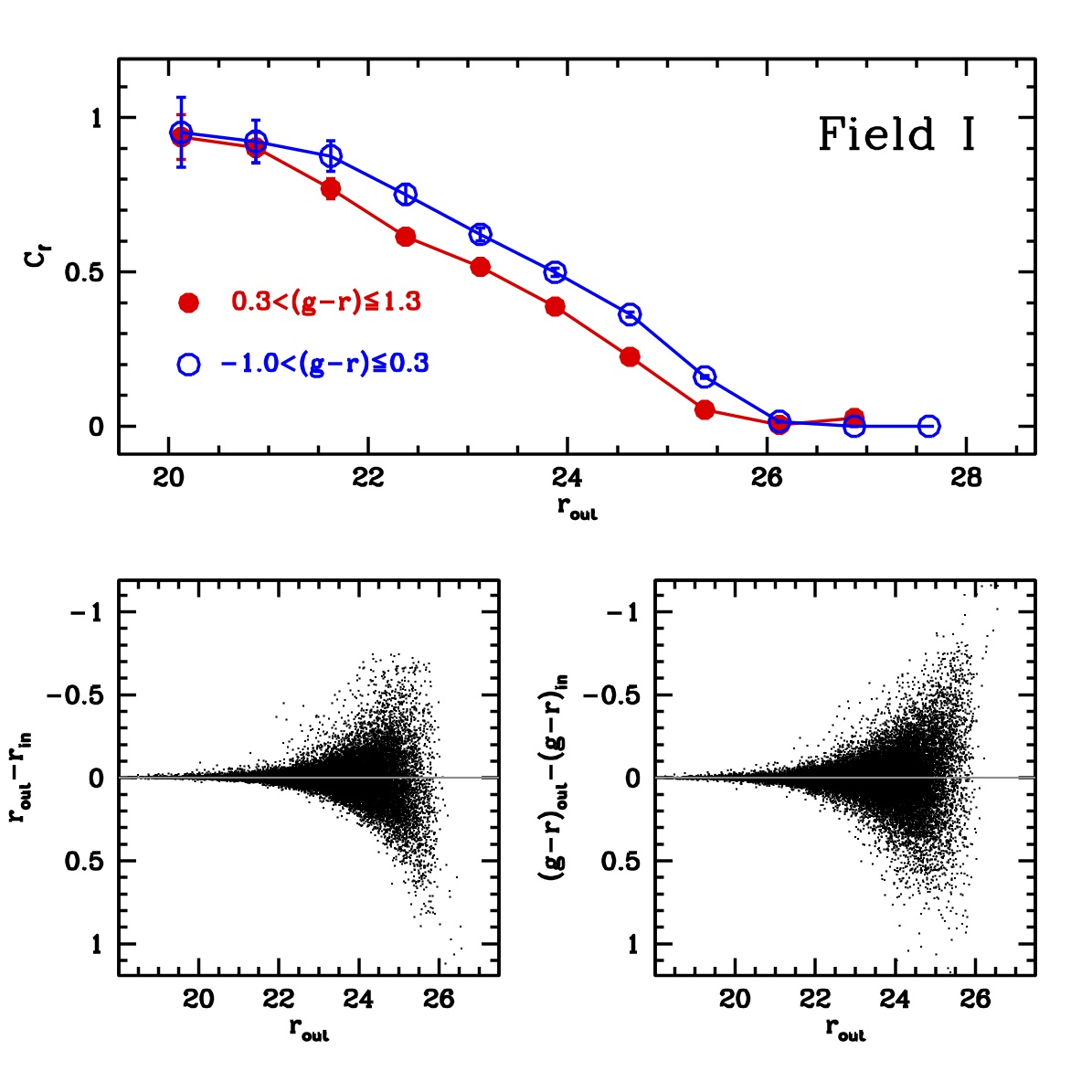}
     \caption{Results of artificial stars experiments for Field~B (upper set of panels), and for 
     Field~I (lower set of panels). The completeness factor ($C_f$) as a function of r magnitude is shown for two different colour ranges. Field~B is taken as representative of the fields observed under the best conditions; Field~I is representative of the worst observing conditions.}
        \label{comp_fig}
    \end{figure}


 A total of 100 000 artificial stars were added to the images \citep[following the recipe described in][]{lf}, and the entire data reduction process was repeated as in the real case, also adopting the same selection criteria described above. The PSF adopted as the best-fit model for photometry was also assumed as the model for the artificial stars. Artificial stars were distributed uniformly in position, over the entire extent of Chip~2, and in colour, over the range $-1.0\le g-r\le 2.0$. They were distributed in magnitude according to a luminosity function similar to the observed 
one but monotonically increasing beyond the limit of the photometry, down to $r\simeq 28.5$ \citep[see][for details and discussion]{lf}.

In Fig.~\ref{comp_fig}, we show the completeness fraction ($C_f$) as a function of r magnitude for two different colour ranges, $-1.0< g-r\le 0.3$ and $0.3< g-r\le 1.3$. Independent of the considered colour range, the completeness of the photometry in our best-seeing fields declines very gently, always remaining above 95\%, for $r\le 24.5$, where it begins to fade more abruptly, reaching $C_f\simeq 0.0$ at $r\simeq 27.0$. In this regime of the completeness curve, $C_f$ is always higher in the blue range than in the red one. The $C_f=0.5$ limit occurs at $r=25.6$ and $r=26.0$, in the red and blue ranges, respectively.  

The case of Field~I is clearly very different. The flat part of the completeness curve lies at $r\la 21.5$, and the $C_f=0.5$ limits occur at $r=23.2$ and $r=23.8$, respectively. This is not particularly surprising, since the observations for the I field were obtained far beyond our specifications. Still, we show that useful constraints on the presence or absence of stellar counterparts can be obtained also from such significantly  sub-optimal observational material.

The remarkable symmetry of the output-input magnitude and colour differences (in both fields) suggests that the effects of blending are quite limited in our relatively uncrowded fields. If this were not the case, a large number of artificial stars would have been recovered with an output magnitude significantly brighter than the input magnitude, owing to blending with a real source, resulting in an asymmetry on the negative side of the output-input magnitude differences.

   \begin{figure}
   \centering
   \includegraphics[width=\columnwidth]{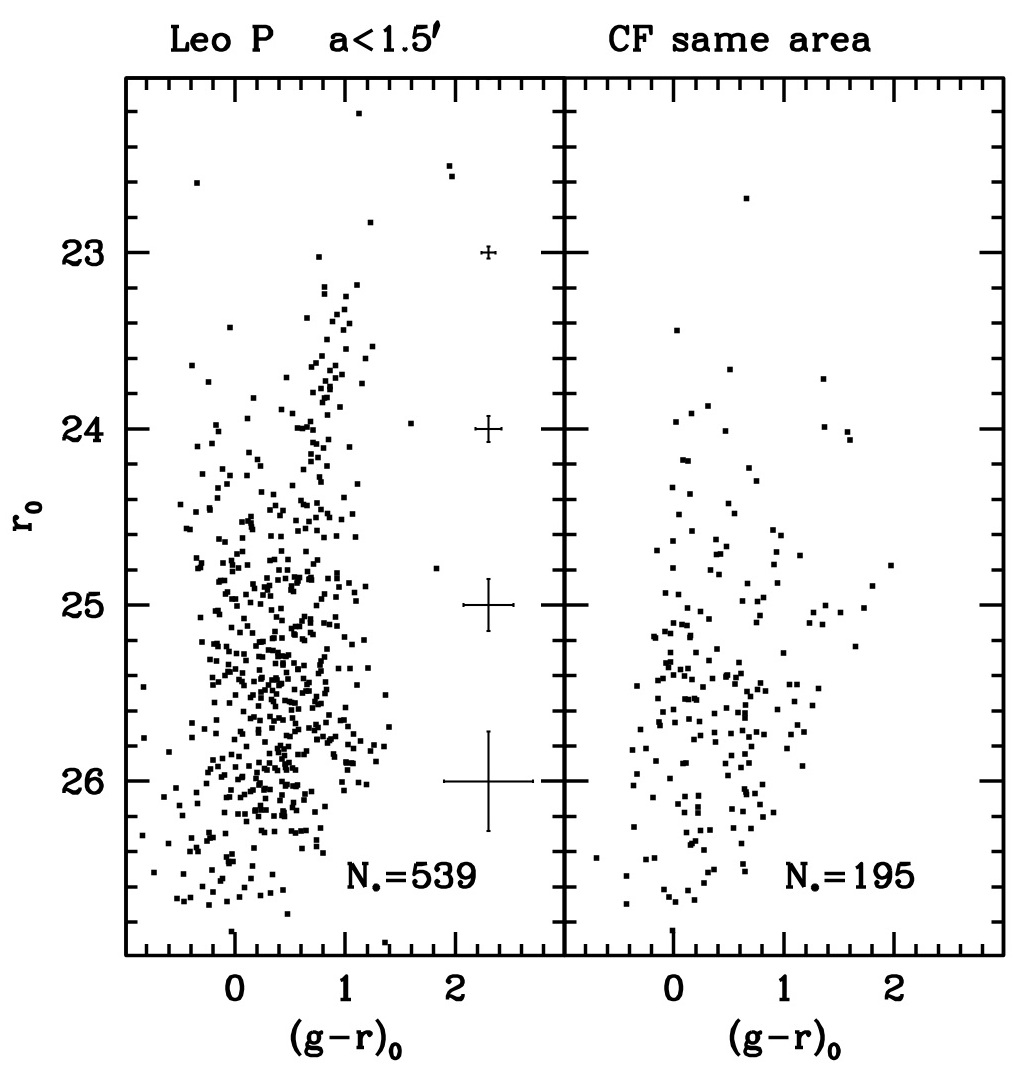}
     \caption{CMDs of stars within an ellipse of semi-major axis $1.5\arcmin$, position angle $PA=335\deg$, and ellipticity $\epsilon=0.52$ centred on the Leo~P dwarf galaxy (left panel), and for a control field within an ellipse of the same shape and dimension centred a few arcmin to the south of the dwarf galaxy. Average photometric errors are reported as in Fig.~\ref{cmd1}.}
        \label{fleop}
    \end{figure}


\subsection{The model case: Leo~P}
\label{leop}

As a basic quality control of the sensitivity of the survey, it is particularly interesting to check wether the presence of the only known counterpart of ALFALFA UCHVCs, Leo~P, is recognisable in the CMDs obtained from our data. In Fig.~\ref{fleop} we compare the CMD within an ellipse of semi-major axis $a=1.5\arcmin$, a position angle $PA=335\deg$, and an ellipticity $\epsilon=0.52$ \citep{leop_lbt} centred on Leo~P, with that of a CF of the same shape and area located $4.5\arcmin$ to the south of the dwarf galaxy. 

The excess of stars in the Leo~P diagram (539 vs. 195 stars) is obvious and the main features of the dwarf galaxy are easily recognised: the steep and wide red giant branch (RGB), tipping around $(g-r)_0\simeq 1.0$ and $r_0=23.2$, as well as the blue plume of young MS and He-burning stars reaching $r_0=24.0$ at $(g-r)_0\simeq 0.0$. Neither of these two features have a counterpart in the CMD of the CF. The overall CMD of Leo~P is very similar to the one obtained in the V and I passbands with the same instrument by \citet{leop_lbt} using $9\times 300$~s exposures per filter (thus suggesting that our observations are near the limit achievable with modern instrumentation on 8m class telescopes).
The severe crowding in the innermost region of the galaxy or in the vicinity of bright knots of young stars makes the Leo~P CMD much more affected by incompleteness than the CF one; for instance, the $C_f=0.50$ level occurs at $r\simeq 25.0$ within 1.0$\arcmin$ of the centre of LeoP and at $r\simeq 25.6$ in the region of the CF.

Figure~\ref{fleop} demonstrates that our data are completely adequate for detecting and characterising UCHVC stellar counterparts similar to Leo~P.

\subsection{Density maps}
\label{dens}

For each field studied in this paper we search for spatial over-densities within the photometric catalogues by producing density maps analogous to those presented in \citet{sexAB} from sources that passed identical SHARP and CHI selections. For each catalogue we proceed as follows:

\begin{itemize}

\item The relevant chip~2 field is covered by a uniform grid in X,Y coordinate space, with nodes spaced by 0.1$\arcmin$ in both directions.

\item For each node of the grid, we record the distance to the 40th nearest neighbour star, $d_{40}$, and estimate the local surface density as $\rho_{raw}=\frac{40}{\pi d_{40}^2}$. While the choice of the 40th neighbour may appear arbitrary, extensive experiments has shown that it is a good compromise between smoothing and sensitivity in the present case\footnote{In particular we explored sets of density maps for the synthetic galaxies discussed in Sect.~\ref{sens}, below, obtained by using various nearest-neighbour limits from 20 to 60. For instance, the maps using the 20th nearest neighbour recovered the same significant peaks found with the 40th nearest neighbour, but were much noisier.}.

\item The background surface density, $\rho_{bkg}$, and its standard deviation, $\sigma_{bkg}$, are estimated as above in a wide rectangular area in the southern part of the field ($-3.0\arcmin \le X\le +3.0\arcmin$ and $-9.0\arcmin \le Y\le -6.0\arcmin$).

\item Surface density maps are produced in terms of density excess above the background in units of $\sigma_{bkg}$.

\end{itemize}

   \begin{figure}
   \centering
   \includegraphics[width=\columnwidth]{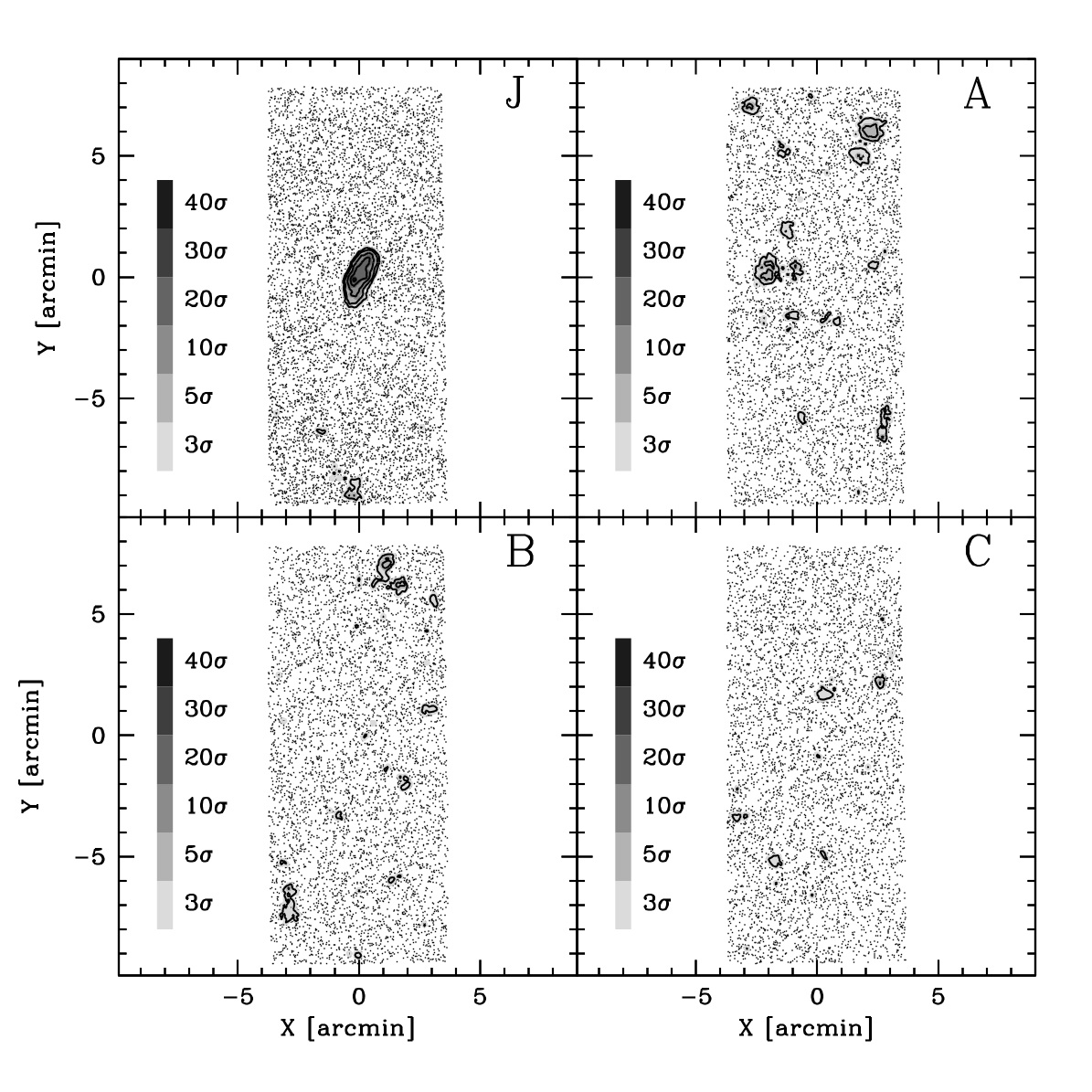}
   \includegraphics[width=\columnwidth]{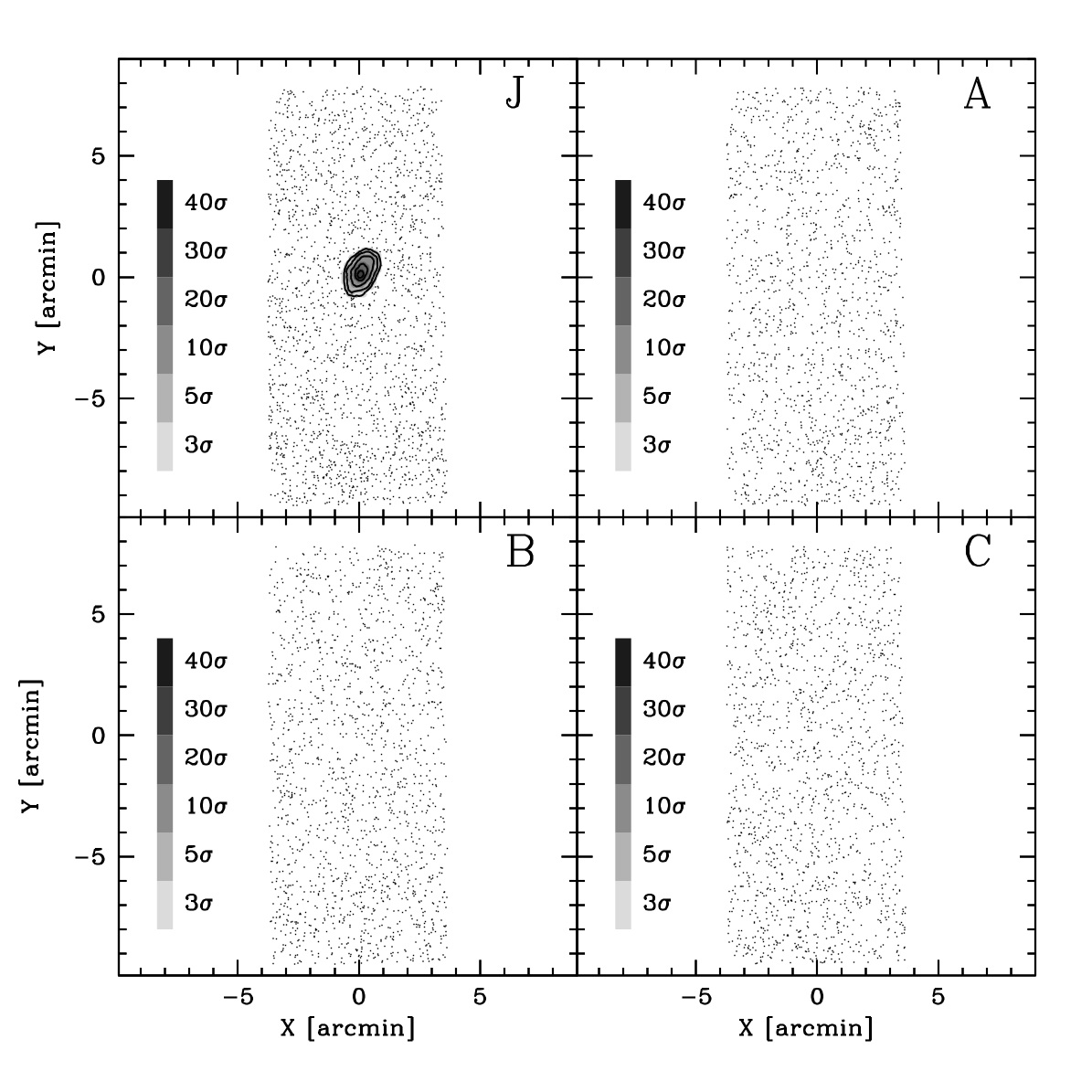}
     \caption{Upper array of four panels: r27 surface density maps for fields J (including Leo~P), A, B, and C. Dots correspond to the sources used to compute the density maps. Surface density levels are expressed in units of $\sigma_{bkg}$ above the background in different tones of grey,  from the lightest ($\rho\ge 3.0\sigma_{bkg}$ to the darkest $\rho\ge 40.0\sigma_{bkg}$, as illustrated by the vertical greyscale bar.
     Lower array of four panels: the same for the r25 maps.}
        \label{mappe1}
    \end{figure}


   \begin{figure}
   \centering
   \includegraphics[width=\columnwidth]{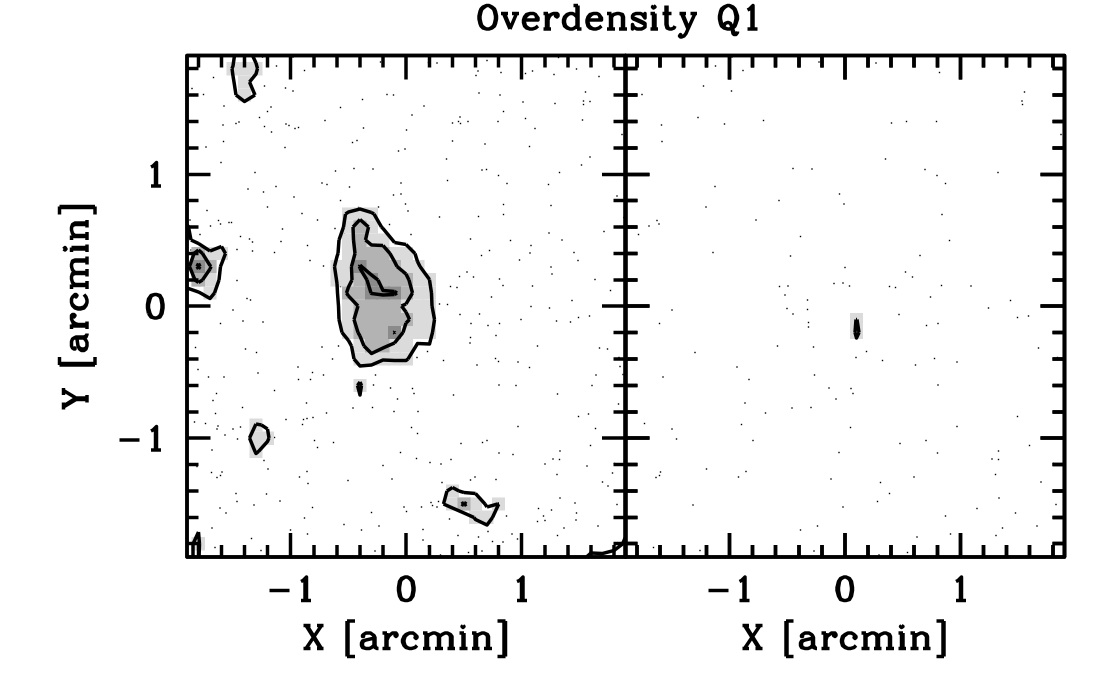}
     \caption{Zoomed-in view of the r27 (left panel) and r25 (right panel) density maps around over-density~Q1. The grey and contour levels correspond to 3, 5, and 7 $\sigma_{bkg}$.}
        \label{Q1map}
    \end{figure}


To avoid any bias induced by a priori assumptions on the nature of the stellar systems we are searching for,
we do not make any additional cuts in our catalogues, except to exclude the M~dwarfs with the simple colour cut illustrated in Fig.~\ref{cmd1} above, $(g-r)_0\le 1.3$. Dwarf galaxies that went undetected in the SDSS should be relatively faint, hence metal-poor, as is indeed the case for Leo~P \citep{leop_met}. For this reason it is extremely unlikely that these system would host a significant number of RGB stars redder than this limit, and, in any case, this would only happen for the coolest giant at the RGB tip { \citep{clem}.} 

Finally, for all the fields, we repeated the analysis for the whole sample (r27) and for the sub-sample with $r<25.0$ (r25). This second option focuses on a sample that is almost completely clean of extended sources and therefore serves as a control check for any over-densities produced by clusters and/or groups of galaxies; these should disappear in the r25 maps. It is important to note that this sanity check is only useful for relatively nearby galaxies ($D\la 1.5$), since the r=25.0 cut unavoidably limits the sensitivity for more distant systems, in general (see Sect.~\ref{sens}).  

In Fig.~\ref{mappe1} we show the r27 and r25 density maps for the field including Leo~P (Field J) and for 
Fields A, B, and C. This provides an illustration of the kind of analysis performed on the whole set of maps, deferred to Appendix~\ref{app_maps} in order to make the paper easier to read. In each map, we also used small dots to plot all the individual sources used to generate the density maps. They inform us of the level and uniformity of the underlying background.

The most important thing to note in Fig.~\ref{mappe1} is that {\em Leo~P is very clearly detected at more than $20\sigma$ in the r27 map and at more than $30\sigma$ in the r25 map}. The maps also recover the correct shape and position angle for the galaxy \citep{leop_lbt}. 

Moreover, while the signal-to-noise ratio of the Leo~P detection is enhanced passing from the r27 to the r25 map, the few sparse and weak additional over-densities
that are seen in the r27 maps of Fig.~\ref{mappe1} completely disappear in the r25 ones, thus suggesting that they very likely trace over-densities of extended sources. Indeed, the subsequent inspection of the images and the CMDs at the position of these over-densities does not reveal any plausible candidate stellar counterpart for any of UCHVC observed in our programme, with the only exception described below.

The general conclusions that can be drawn from the analysis of the density maps can be summarised as follows:

\begin{itemize}

\item No field contains a stellar over-density significantly above $5\sigma$, except for Leo~P.

\item The vast majority of 3-5$\sigma$ over-densities detected in the r27 maps are not present in the r25 maps.

\item The follow-up of the highest 3-5$\sigma$ over-densities on the images does not reveal any potentially resolved stellar system. In many cases groupings of distant galaxies are identified instead.

\end{itemize}

However, in the case of Field Q, a $>5\sigma$ over-density coincident with the position of the associated UCHVC in the r27 map also yields a $3\sigma$ over-density of very small angular scale in the r25 map (see Fig.~\ref{Q1map}, for a zoomed-in view of the maps). The over-density is also confirmed by the algorithm used in \citet{panda_ghosts}. This specific case, dubbed over-density Q1, is discussed in more detail below. 

   \begin{figure*}
   \centering
   \includegraphics[width=\textwidth]{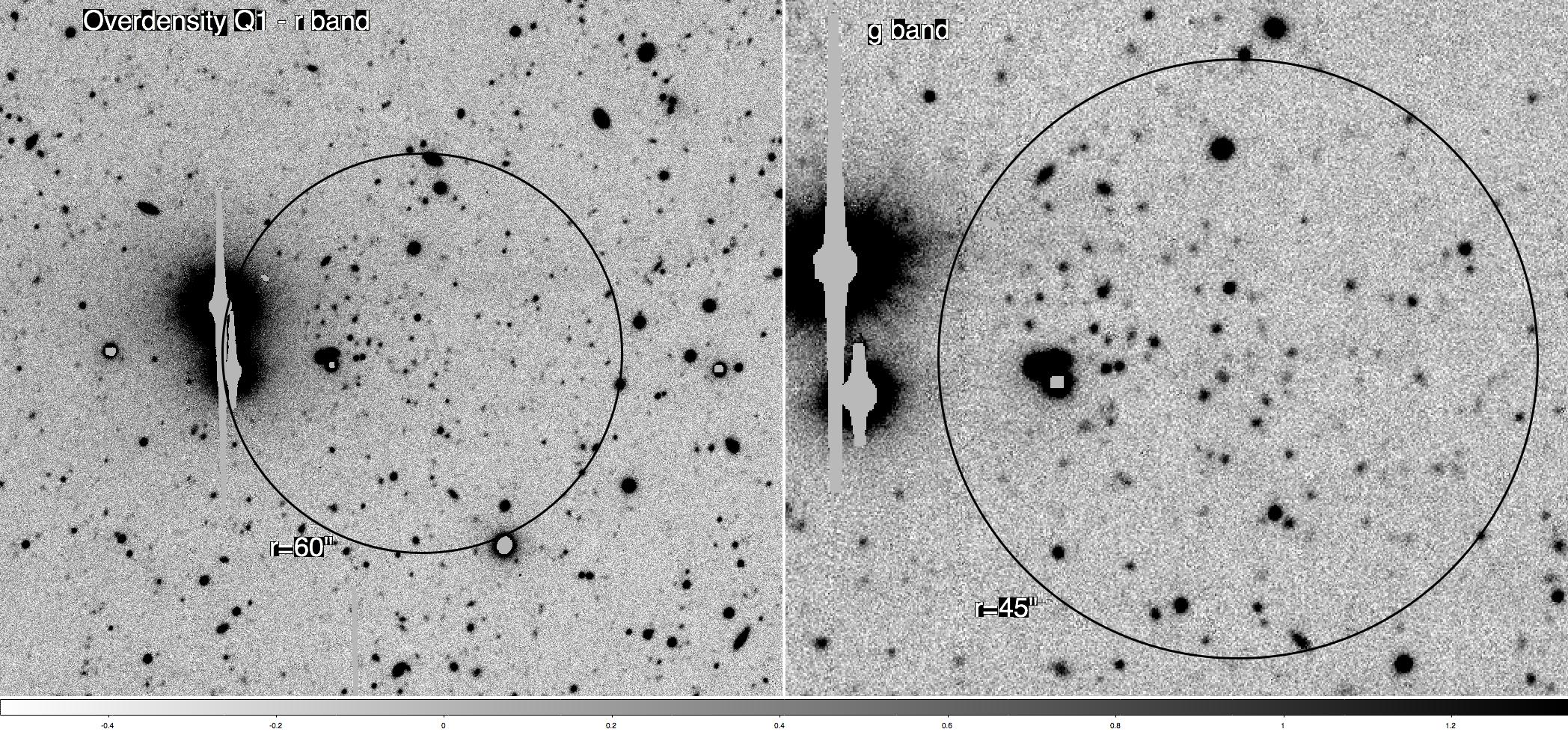}
     \caption{r-band (left panel) and g-band (right panel) images of over-density Q1 at different scales. The superimposed circles have radii of $r=60\arcsec$ (left panel) and $r=45\arcsec$ (right panel) and are centred on the position of the associated UCHVC.}
        \label{Q1_ima}
    \end{figure*}


\subsubsection{Over-density Q1}

Figure~\ref{Q1_ima} shows that the weak over-density near (X,Y)$\sim(0,0)$ detected in the density maps of Field Q can also be seen in the deep mosaic images. In particular, the zoom on the g-band image reveals a sparse swarm of very faint sources mostly concentrated in the upper left-hand quadrant of the superimposed circle, in agreement with the r27 map for which the over-density is approximately centred on (X,Y)$\sim(-0.5,0.0)$, as can be appreciated from Fig.~\ref{Q1map}. Several sources, especially amongst the brightest ones, are clearly identified as galaxies but most of them are too faint to be firmly classified.

\begin{figure}
   \centering
   \includegraphics[width=\columnwidth]{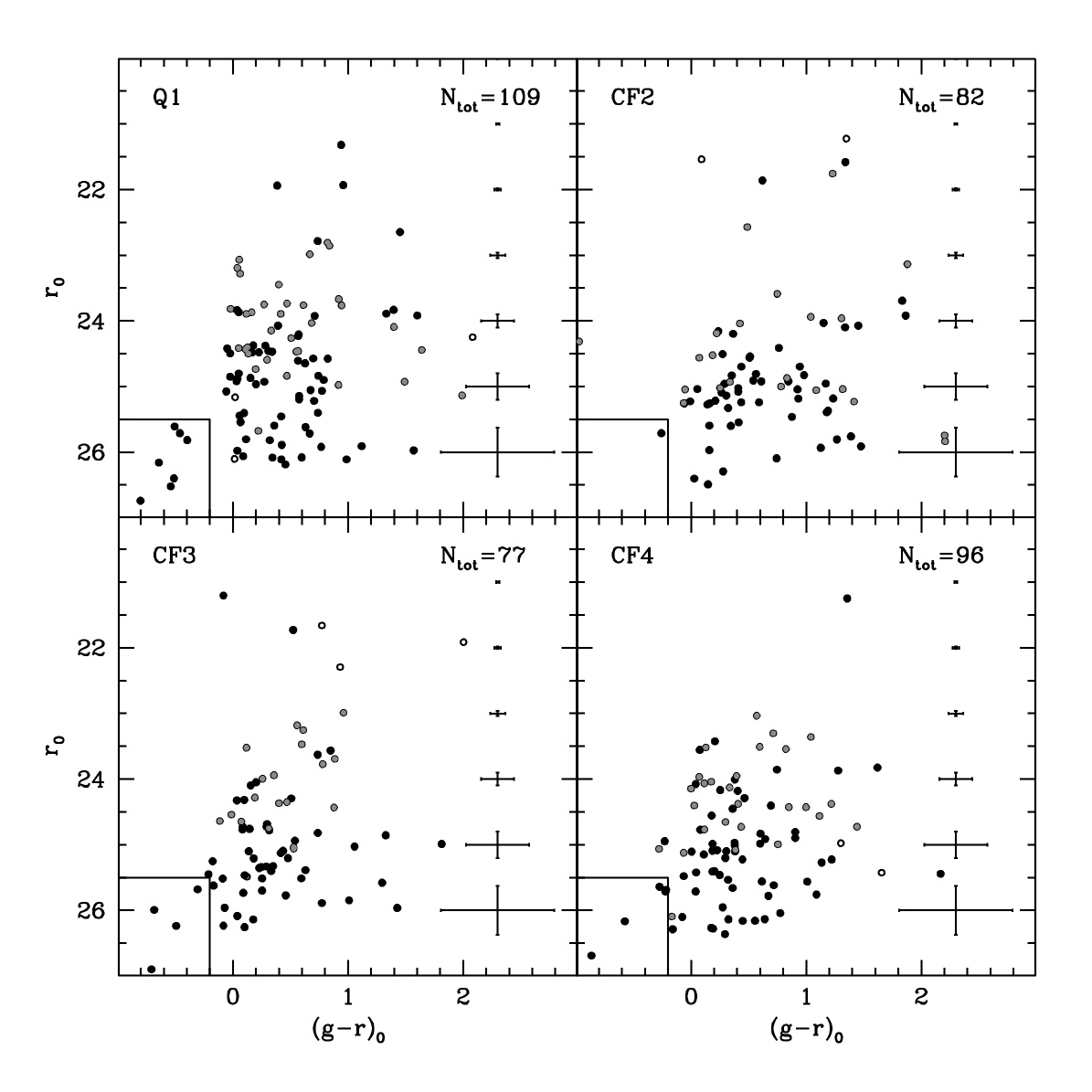}
     \caption{Comparison of the Q1 over-density CMD with three control fields with the same area. The Q1 CMD includes all sources within a circle of radius $r=45\arcsec$ of the centre of the UCHVC associated with Field Q. Black filled circles are sources that pass our CHI and SHARP selection, grey circles are sources that do not pass the selection because of positive SHARP values that are higher than the adopted limit (extended sources), and empty circles are sources that do not pass the selection for other reasons. { The total number of sources plotted in each CMD ($N_{tot}$) is reported. The faint blue sources discussed in the text are enclosed in a rectangle in the lower left corner of each CMD.}}
        \label{Q1_cmd}
    \end{figure}

The comparison between the CMD of the sources within the circle in the left-hand panel of Fig.~\ref{Q1_ima} and of three surrounding CFs of the same area, shown in Fig.~\ref{Q1_cmd}, does not help in ascertaining the nature of over-density Q1.
Apart for the overall excess in the total number of sources, the only noticeable difference is in the { slight excess of} faint and very blue sources [$r_0>25.5$ and $(g-r)_0<-0.2$] present in the Q1 CMD. These sources have colours that are compatible both with hot stars and with quasars \citep{fadely}, but the morphology of the over-density is not suggestive of a galaxy with recent star formation (like object D1, for example). Moreover, these sources lie just above the detection limit, where the photometric uncertainties are large.

The CMD of the Q1 region does not show any obvious feature that allows us to recognise it as a resolved stellar system. The ratio of the number of {\em extended} sources to the number of selected sources is only slightly larger in Q1 than in the considered CFs ($\sim 40$\% vs $\sim 30$\%). 
 
Given the available data, the most likely hypothesis is that over-density Q1 is a distant cluster of galaxies { (but see Sect.~\ref{sens}, for further discussion)}.
An easy spectroscopic follow-up is prevented by the extreme faintness of the sources that constitute the bulk of the over-density ($r\ga24.5$). Hubble Space Telescope (HST) imaging is probably the only viable solution to firmly determine the nature of this object. In the present context we must classify over-density Q1 as a weak candidate stellar counterpart to HVC352.45+59.06+263.

\subsection{Testing the sensitivity of the density maps}
\label{sens}

In analogy with the completeness of the photometry (Sect.~\ref{comple}), which is determined by means of artificial-star experiments, the sensitivity and the detection limit of our survey can be determined by using synthetic dwarf galaxies. 
For example, to test the sensitivity of density maps we inject a synthetic stellar population of given total luminosity, size and distance into one of our photometric catalogue, after having applied to its individual stars all the observational effects (incompleteness and photometric error) as estimated from the artificial-star experiments. Then the density map is derived as in Sect.~\ref{dens}, to verify that the synthetic galaxy is detected as a significant over-density. 

Here, to give a first quantitative insight into the kind of galaxies we would not have missed if they were present, we present the results of a set of tests of this kind.
A thorough exploration of the relevant parameter space is beyond the scope of the present paper and will be presented in a future contribution.

As a theoretical template we used a synthetic population (SP) that was obtained from the \citet{bressan} theoretical models. The SP stars span a metallicity range $-2.0\le[Fe/H]\le -1.6$, with a symmetric distribution with $\langle [Fe/H]\rangle=-1.8$ and $\sigma_{[Fe/H]}=0.1$~dex. The star formation history starts 13.0~Gyr ago and declines exponentially with an exponential scale of 0.5~Gyr. The initial mass function, in the form $N(m)\propto m^{-x}$, has $x=-2.35$ for $m>0.5~M_{\sun}$ and $x=-1.30$ for $m\le 0.5~M_{\sun}$. The details of the model are not particularly important for this application, what is relevant here is that the SP has a CMD that resembles that of a typical (prevalently) old and metal-poor dwarf galaxy like those we are looking for. The choice of a template completely dominated by old stars is conservative: the presence of a young MS or blue loop population would clearly favour detection. In the following we will (mainly) consider a synthetic population whose integrated V magnitude is $M_V=-8.0$ as a reference, i.e. a factor of $\sim 3.6$ fainter than Leo~P, but we also show a few experiments with a brighter synthetic galaxy ($M_V=-9.0$) with the same age and metallicity distribution.

   \begin{figure}
   \centering
   \includegraphics[width=\columnwidth]{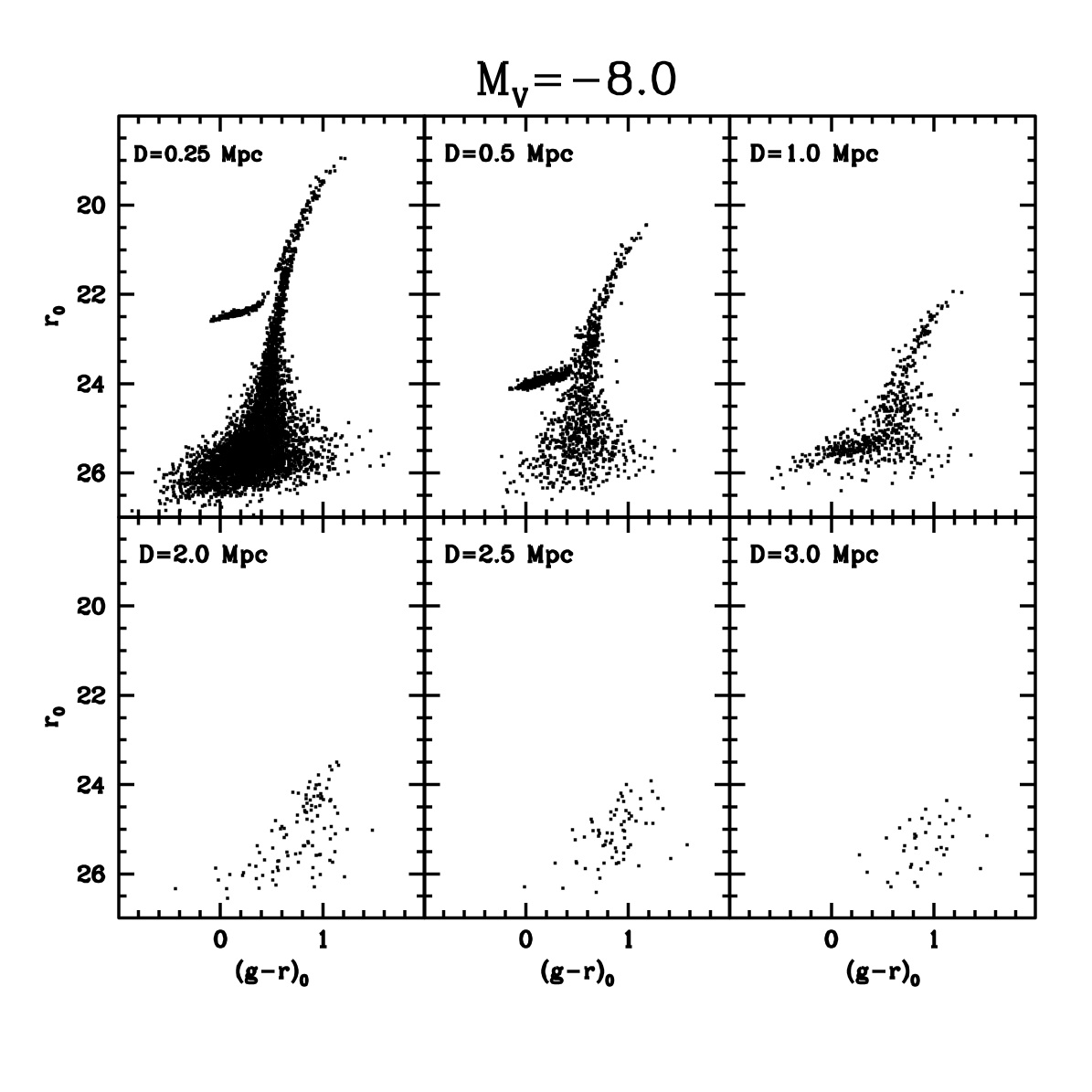}
     \caption{CMD of the synthetic population described in the text for a total luminosity corresponding to $M_V=-8.0$. The CMD of the population is shown for different assumptions on the distance. The effects of the Field~B incompleteness and observational errors are included.}
        \label{cm8}
    \end{figure}


In Fig.~\ref{cm8} we show the CMD of the $M_V=-8.0$ SP for different assumptions on its distance. The effects of incompleteness and observational errors are included as follows, after shifting all the magnitudes according to the adopted distance:

\begin{itemize}

\item For each SP star with magnitudes ($g_s$, $r_s$) we select from the artificial star catalogue described in Sect.~\ref{comple} the subset of artificial stars having r magnitudes within $\pm 0.5$~mag of $r_s$ and colours within $\pm 0.25$~mag of $g_s-r_s$;

\item We then extract one artificial star at random from this set. If the extracted artificial star is lost in the artificial-star experiments then the SP star is also considered lost, whereas if the artificial star is recovered, the magnitudes of the SP star are corrected by adding $\Delta_{mag}= mag_{out}-mag_{in}$ of the artificial star to both $g_s$ and $r_s$.

\end{itemize}

These corrected SPs (hereafter CSPs), produced with the set of artificial star catalogues from Field~B, probe the regime of optimal sensitivity ({\em best case}). There are ten fields with $r_{90}$ within $\pm 0.1$~mag from the value of Field~B, namely fields A, J, G, B, Y, R, K, U, C, L, in order of decreasing $r_{90}$, from $r_{90}=26.48$ to $r_{90}=26.27$. Field~B is representative of these 40\% of the surveyed fields that have been observed in optimal conditions. The CSPs produced with the set of artificial star catalogues from Field~I probe the regime of minimal sensitivity ({\em worst case}) because Field~I is the one with the brightest $r_{90}$. Therefore, Fields~B and I bracket the range of quality spanned by the observational material of the whole SECCO survey. 
 
We adopt an exponential profile for the spatial distribution model of our artificial galaxies. This is a very simple model, that only depends on a half-light radius, and it is known to provide a satisfactory description of the light distribution of low-SB dwarfs of all morphological types \citep{mateo}. All the other parameters being fixed, in particular the total luminosity, dwarfs of larger sizes should have lower SB and produce weaker over-densities in our maps. To be conservative, we therefore decided to simulate galaxies with the largest half-light radius for the considered luminosity \citep[see, e.g.,][]{mcc}. We adopt $r_h=500$~pc, an absolute upper limit for $M_V\simeq -8.0$ dwarfs. We also performed a set of simulations of models with a more typical half-light radius, $r_h=300$~pc \citep[very similar to the mean $r_h$ for galaxies of this luminosity, as derived by][]{brasseur}. We note that Leo~P has $r_h\simeq 200$~pc. With our simulations we explored the relevant range of distance suggested by A13 for the candidate mini-haloes ($0.25-2.0$~Mpc), extending also to slightly greater distances (up to 3.0~Mpc) for completeness.

\begin{figure}
   \centering
   \includegraphics[width=\columnwidth]{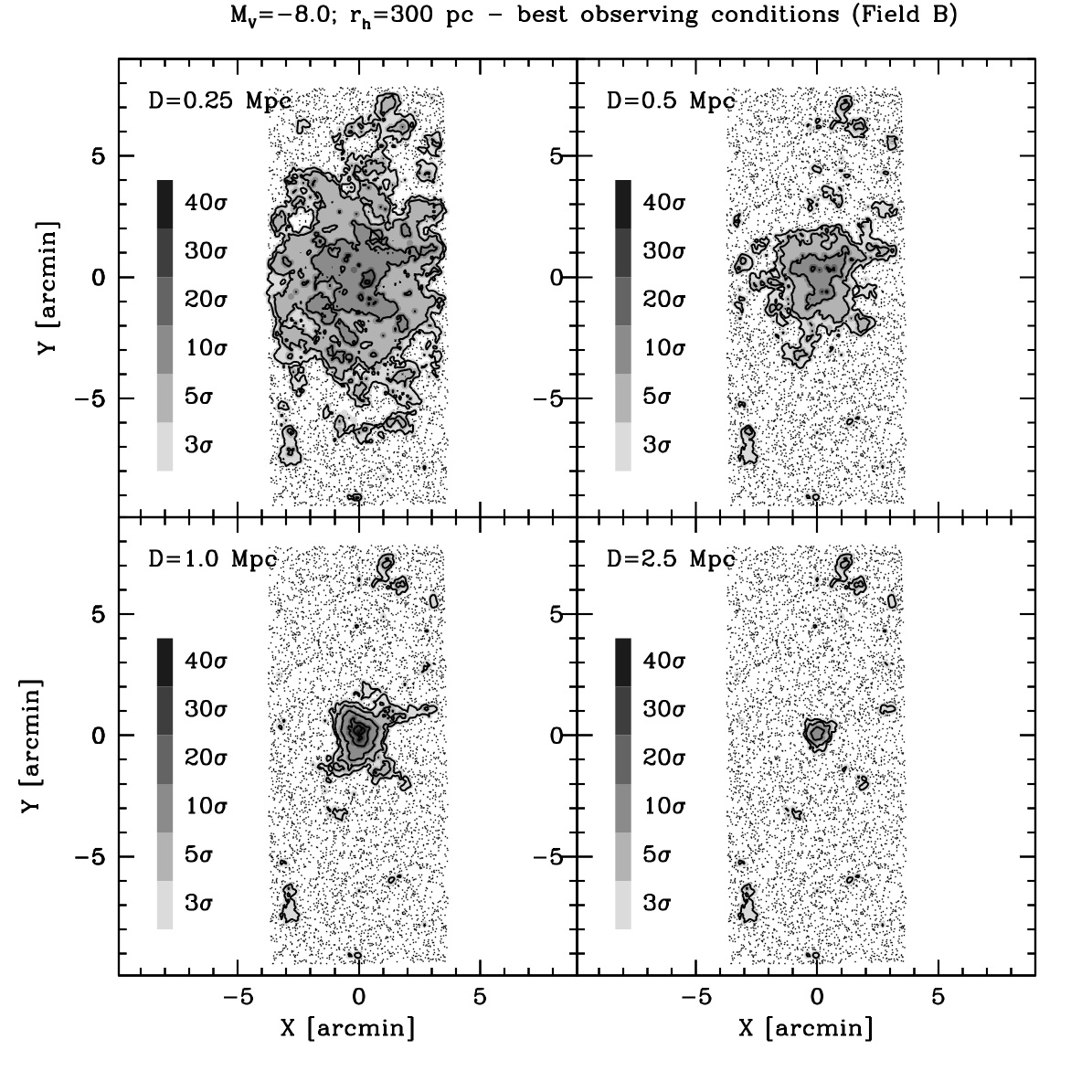}
     \caption{Density maps, derived as in Sect.~\ref{dens}, for a synthetic galaxy with $M_V=-8$ and $r_h=300$~pc, and different assumptions on the distance. The synthetic population has been corrected for observational incompleteness and photometric errors from the artificial-star experiments and the fore/background population of Field~B. This is intended to test the process of detection of over-densities in catalogues obtained from the images with the best quality.}
        \label{mapp8_300B}
    \end{figure}

\begin{figure}
   \centering
   \includegraphics[width=\columnwidth]{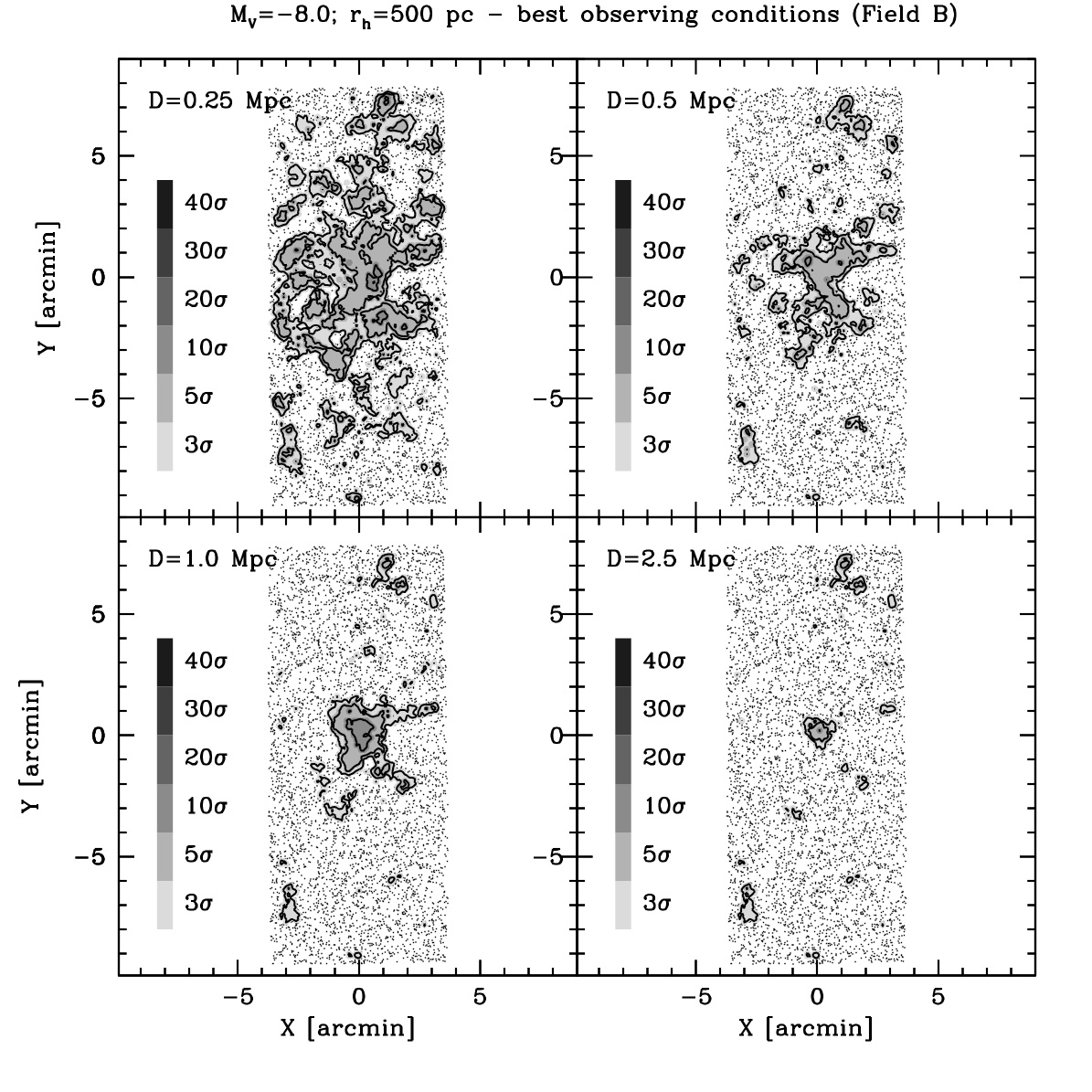}
     \caption{Same as Fig.~\ref{mapp8_300B} but for a synthetic galaxy with $r_h=500$~pc.}
        \label{mapp8_500B}
    \end{figure}

Given the catalogue of a CSP, we associate a set of X and Y coordinates
extracted from an exponential profile with $r_h=300$~pc (or 500~pc) to each synthetic star, centred on (X,Y)=(0,0), which we then convert into arcmin for the chosen distance. Conservatively, we assume spherical symmetry since more elliptical galaxies of the same total luminosity would lead to higher central density. We then add the synthetic stars to either Field~B or Field~I catalogues, for {\em best case} and {\em worst
case} CSPs, respectively. These provide the background population that would be present in the case of real observations of a dwarf galaxy. It is worth noting that both fields lack any reliable over-density around $(X,Y)=(0,0)$, even at a weak level, making Fields~B and~I optimally suited to our simulations. Finally, the density maps are generated as described above in Sect.~\ref{dens}.
In Figures~\ref{mapp8_300B} and \ref{mapp8_500B} we present the density maps for the {\em best case} synthetic galaxies, for different assumptions on the distance and for the two choices of $r_h$. 

The main results from these two plots can be summarised as follows:

\begin{enumerate}

\item The $r_h=300$~pc synthetic dwarf galaxy is detected as a density peak $\ge 10\sigma$ above the background for distances out to $D=2.5$~Mpc.  At $D=3.0$~Mpc, the density peak does not reach $10\sigma$ but is still above $5\sigma$, and the angular diameter is $\sim 1\arcmin$.

\item  The $r_h=500$~pc synthetic dwarf galaxy produces density peaks $\ge 5\sigma$ above the background for distances out to $D=2.5$~Mpc. Even though the peaks are not always strong and the morphology may appear quite irregular, it is clear that the galaxy is detected as an over-density of a significantly larger size than the typical $3-5\sigma$ blobs that are relatively common in our observed maps.

\item The adoption of r25 maps (not shown here) makes the signal cleaner from the synthetic dwarf galaxies in the distance range 0.25~Mpc$\la D\la$1.5~Mpc (depending also on $M_V$), but generally leads to a loss of sensitivity outside of this range. 

\end{enumerate}

\begin{figure}
   \centering
   \includegraphics[width=\columnwidth]{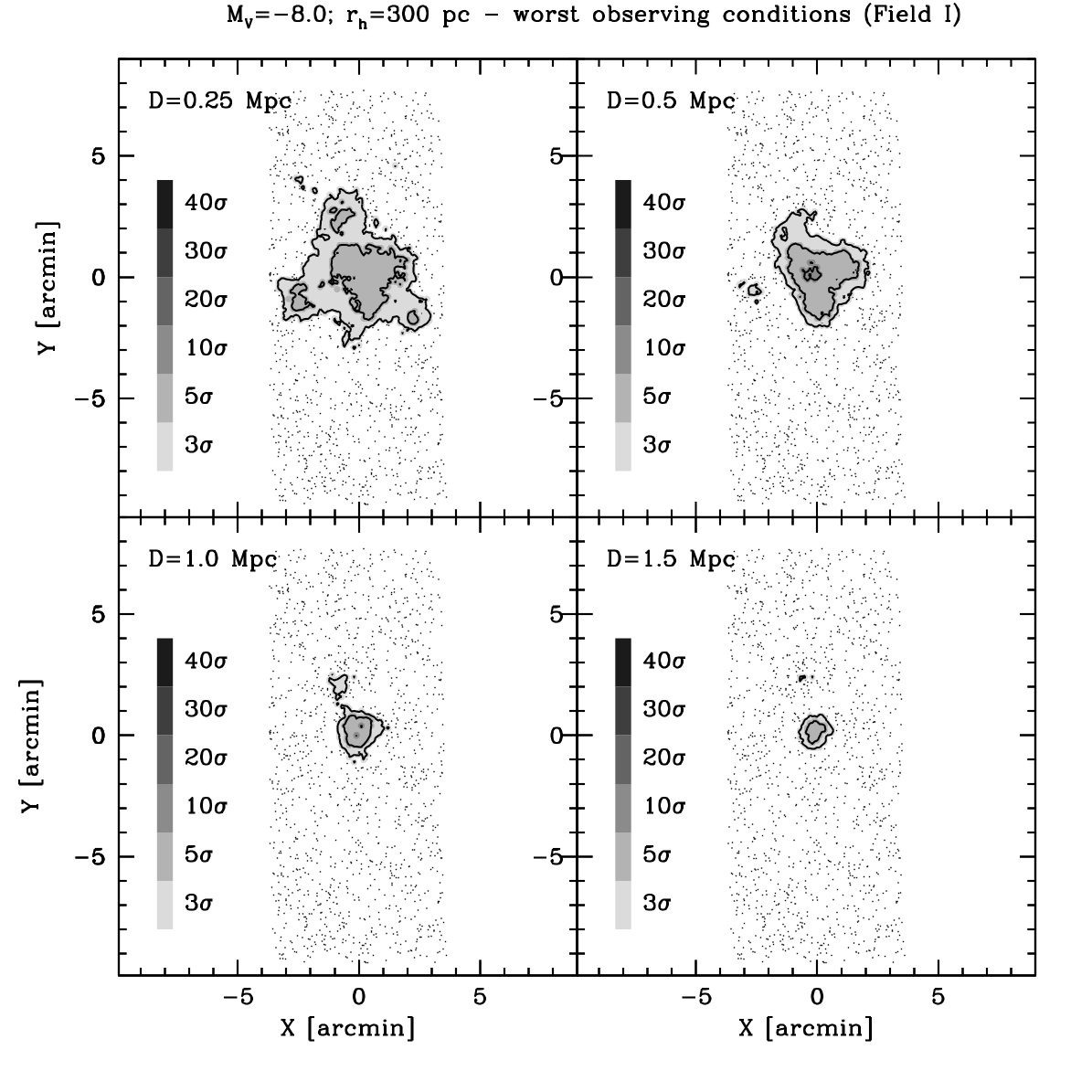}
     \caption{Same as Fig.~\ref{mapp8_300B}, but adopting the artificial-star experiments, as well as the fore/background population, of Field~I. This is intended to test the detection limits of over-densities in catalogues obtained from the observational material with the worst quality.}
        \label{mapp8_300I}
    \end{figure}

\begin{figure}
   \centering
   \includegraphics[width=\columnwidth]{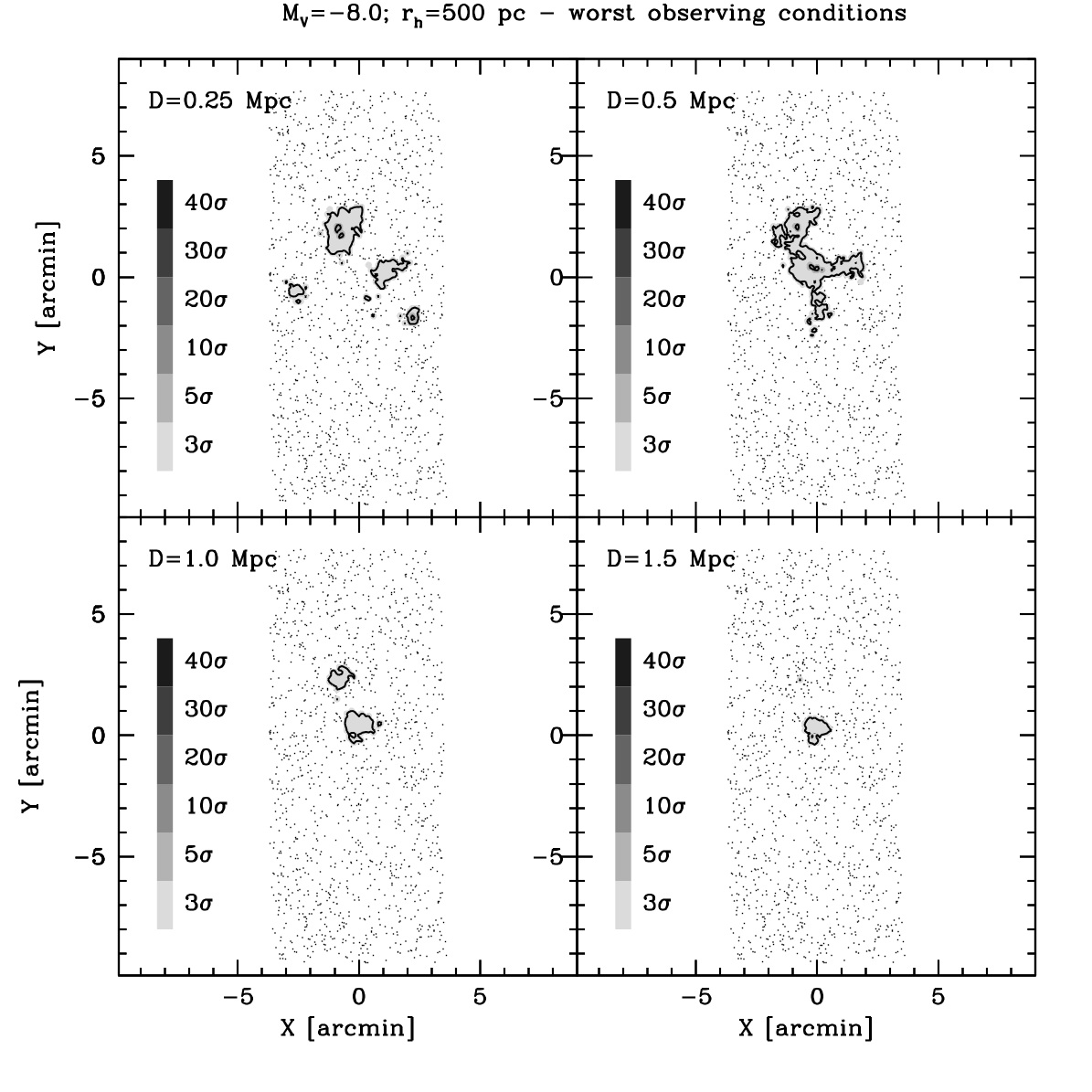}
      \caption{Same as Fig.~\ref{mapp8_300I} but for a synthetic galaxy with $r_h=500$~pc.}
        \label{mapp8_500I}
    \end{figure}

\begin{figure}
   \centering
   \includegraphics[width=\columnwidth]{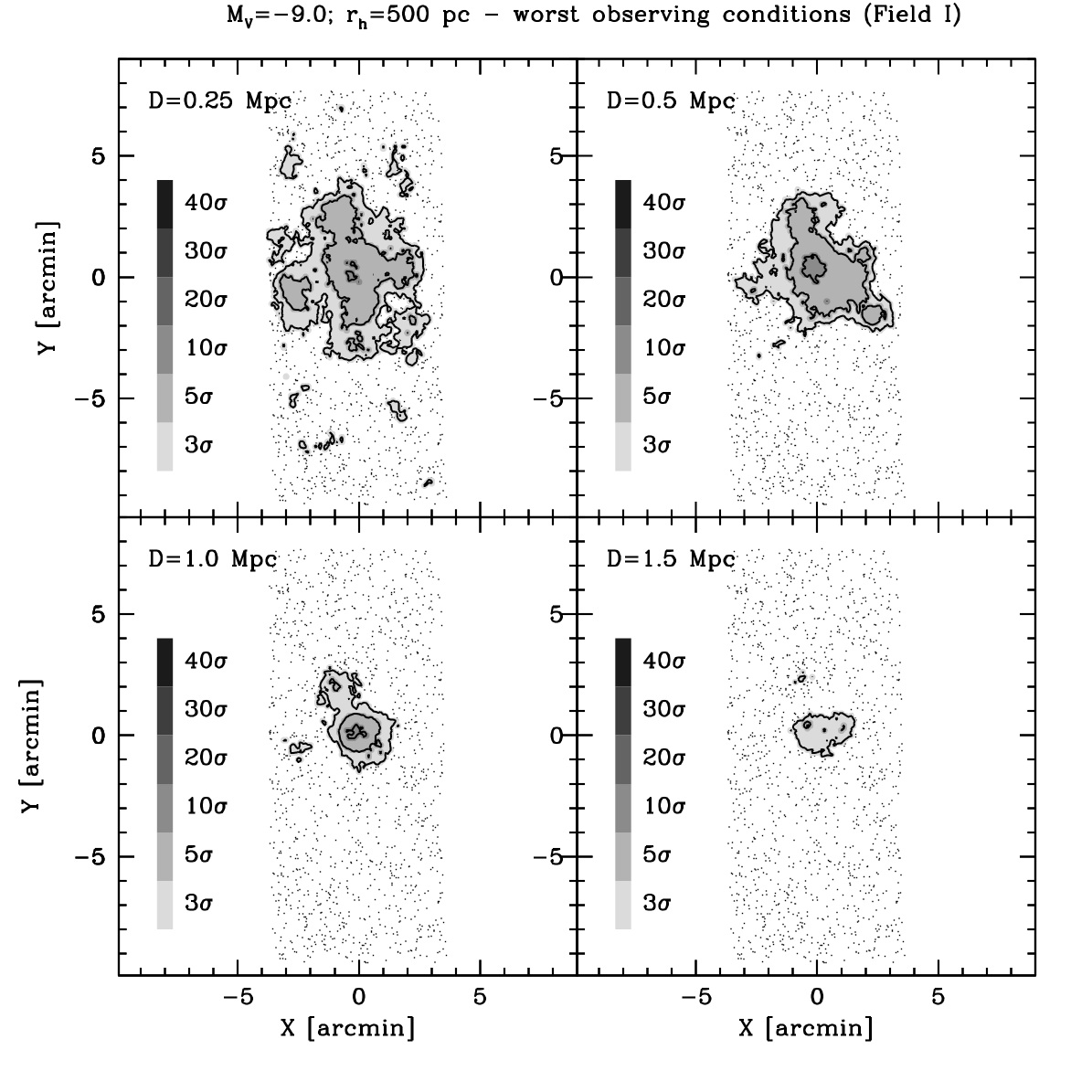}
      \caption{The same as Fig.~\ref{mapp8_500I}, but for a galaxy with $M_V=-9.0$.}
        \label{mapp9_500I}
    \end{figure}

Figures~\ref{mapp8_300I} and \ref{mapp8_500I}, which refer to {\em worst case} synthetic galaxies tell a different story. Synthetic galaxies with $r_h=300$~pc are detected as $\ge 5\sigma$ over-densities out to 1.5~Mpc, while $r_h=500$~pc galaxies produce only relatively small and irregular $\ge 3\sigma$ peaks at any of the considered distances. Figure~\ref{mapp9_500I} shows that, on the other hand, in {\em worst case} conditions, a $M_V=-9.0$  dwarf would be detected as a $\ga 10\sigma$ over-density out to $D=1.0$~Mpc even for $r_h=500$~pc.

\begin{figure*}
   \centering
   \includegraphics[width=\textwidth]{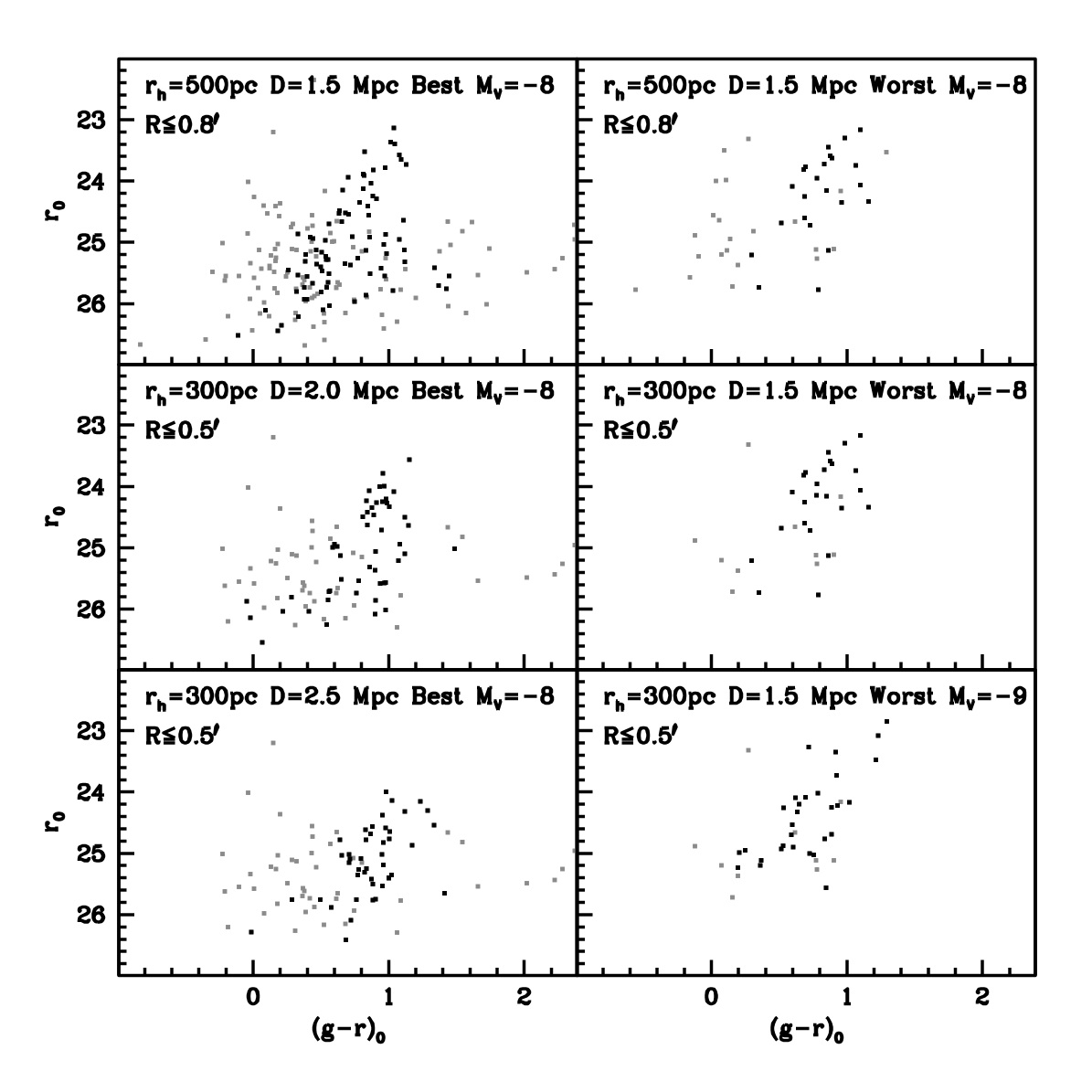}
     \caption{CMDs of the areas where synthetic galaxies are detected from density maps under various assumptions. They illustrate the morphology of the diagrams towards the detection limit of the survey. Grey dots are field sources (contaminants), black dots are stars of the synthetic dwarf galaxies.}
        \label{syncmd}
    \end{figure*}

It is interesting to look at the CMDs in areas enclosing the over-densities generated by synthetic dwarfs in the regime near the sensitivity limit. Figure~\ref{syncmd} illustrates a few remarkable cases, showing that interpreting the CMDs associated with these over-densities may not be straightforward, especially with {\em worst case} observations. For distances only slightly greater than the limits derived above, it would be virtually impossible to confirm the nature of a weak, small-size over-density from its CMD with our data. This may be particularly true if part of the total luminosity of the galaxy is made of young MS and/or He-burning stars; these would greatly help the detection by visual inspection and (probably) enhance the signal-to-noise of a detection in the density map, but it would also likely produce configurations in the CMD that are less clear in the low-completeness/high-error regime considered here. It is remarkable, in this context, that over-density Q1, as seen on the r27 density map (Figures~\ref{Q1map} and \ref{mappe5}), appears quite similar to the synthetic dwarf { at $D=2.5$~Mpc in Fig.~\ref{mapp8_500B} or that at $D=1.5$~Mpc shown in  Fig.~\ref{mapp8_300I}}. Figure~\ref{syncmd} shows that a fuzzy CMD is not an unexpected outcome for an actual { faint galaxy lying at $D\ge 1.5$~Mpc (see, e.g., the CMDs in the middle panels of the figure).} 

In conclusion, the overwhelming abundance of unresolved galaxies in the relevant range of magnitudes \citep[see][]{fadely}, strongly suggests that the vast majority of the weak ($3-5\sigma$) and small-size over-densities detected in our density maps are indeed clusters or groups of galaxies. Still, the simulations presented above demonstrate that the lack of recognisable features in the associated CMD that would qualify them as the stellar counterparts of UCHVCs does not imply that they cannot be dwarf galaxies just beyond the distance and luminosity limits considered here.

The experiments performed here however show that any dwarf galaxy brighter than $M_V=-8.0$ and with $r_h\le 300$~pc would have been detected in our density maps out to a distance of 1.5~Mpc. 
In ten of our twenty-five fields, dwarf galaxies of this luminosity would have been detected out to D=2.5~Mpc even if their size is as large as $r_h\le 500$~pc.
In some cases, there is evidence suggesting that tidal interactions may be at the origin of anomalously high $r_h$ values in local dwarfs (see, e.g., the cases of And~II, \citealt{amorisco}, and And~XIX, \citealt{XIX}). On the other hand, any galaxy associated with the surveyed UCHVCs would be isolated, so tidal interactions are unlikely to be a concern.

\begin{table*}
  \begin{center}
  \caption{Candidate stellar counterparts to the surveyed UCHVCs.}
  \label{cand}
  \begin{tabular}{lcccc}
Candidate & UCHVC & RA$_{J2000}$ & Dec$_{J2000}$ & approximate radius   \\
          &      &   [deg]      &    [deg]      &  [arcmin]           \\
\hline
  D1 & HVC274.68+74.70-123 & 185.4748235	& +13.4601493 & 0.3   \\ 
  Q1 & HVC352.45+59.06+263 & 215.9987500	&  +5.3944444 & 0.5    \\ 
\hline
\end{tabular} 

\end{center}
\end{table*}

\section{Summary and discussion}
\label{disc}

We have presented the general scheme and the first results of the SECCO survey, a project aimed at searching for stellar counterparts in twenty-five UCHVCs proposed by A13 to be potential mini-haloes within the Local Group and its surroundings, in the distance range 0.25~Mpc$\le D\le 2.0$~Mpc. 

From the visual inspection of deep and wide-field images for all our targets, which are the most compact amongst those identified by A13, we did not find any obvious stellar counterpart (e.g., similar to the recently discovered Leo~P) associated with the surveyed UCHVCs, except maybe for a distant, very low SB counterpart to cloud HVC274.68+74.70-123 (candidate D1). 
We were able to identify only one possible (weak) candidate from density maps based on the photometric catalogues extracted from the images, over-density~Q1. The main properties of these candidates, which can be inferred from the data currently available, are summarised in Table~\ref{cand}.

We also presented an initial investigation of the sensitivity of our survey to faint and low-SB stellar systems that are expected to be the counterparts of the surveyed UCHVC. This is the first step in a set of simulations that will also quantitatively assess the limits of the visual (non-)detections. We demonstrate that any galaxy with $r_h\le 300 (500)$~pc and $M_V\le -8.0$, lying within 1.5 (1.0)~Mpc, would have appeared as a $\ge5\sigma$ over-density in our density maps. This specific limit is fixed by the performance of our worst set of observations (Field~I), which can be considered as representative of only a few other fields observed under similar conditions (e.g., only fields H and V have $r_{90}$ within $\pm 0.1$~mag of the value from Field~I). On the other hand, best case simulations show that in the ten fields with observing material of quality very similar to Field~B, we would have detected any galaxy with $r_h\le 500$~pc and $M_V\le -8.0$ out to $D=2.5$~Mpc.

The full set of sensitivity tests will explore the relevant parameter space more thoroughly (e.g., simulating dwarfs fainter than $M_V=-8$) and will also probe the the visual inspection process, which may be more sensitive than density maps in some range of distances and luminosities. These tests will push the sensitivity of the survey
beyond the simple limit provided here, tracing the manifold where we would have detected any stellar counterpart in luminosity - distance - radius space.
While we must wait for these tests to obtain the most stringent limits from our survey,  we can already conclude from this first part of the analysis that local ($D\la 1.5$~Mpc) dwarf galaxies brighter than $M_V\simeq -8.0$ and similar to the dwarfs we already know are not associated with the A13 UCHVCs that we explored. 

Since our sample is composed of the best mini-halo candidates selected by A13 and includes a significant fraction of the whole sample of UCHVC identified by these authors (42\%), it can also be concluded by extension that UCHVCs are very rarely associated with the kind of stellar counterparts our optical survey is sensitive to. This does not exclude, however, that all or some of the UCHVCs lie within a DM mini-halo. They could be ``unborn'' galaxies that were never able to switch on their star formation. Still, the lack of a stellar counterpart (or, at least, of a detectable one with the present survey) makes it exceedingly difficult to gain any further insights into their actual nature at present.

\begin{acknowledgements}

We acknowledge the support from the LBT-Italian Coordination Facility for the execution of observations, data distribution, and reduction. We are grateful to Michele Cignoni for providing synthetic populations for our tests on the survey sensitivity. 

M.B acknowledge the financial support from the PRIN MIUR 2010-2011 project ``The
Chemical and Dynamical Evolution of the Milky Way and Local Group Galaxies'',
prot. 2010LY5N2T. 

G. Battaglia gratefully acknowledges support through a Marie-Curie action Intra European Fellowship, funded by the European Union Seventh Framework Program (FP7/2007-2013) under Grant agreement number PIEF-GA-2010-274151, as well as the financial support by the Spanish Ministry of Economy 
and Competitiveness (MINECO) under the Ram\'on y Cajal Programme 
(RYC-2012-11537).

This research made use of the SIMBAD database, operated at the CDS, Strasbourg, France.
This research made use of the NASA/IPAC Extragalactic Database (NED) which is operated by the Jet Propulsion Laboratory, California Institute of Technology, under contract with the National Aeronautics and Space Administration. 
This research has made use of NASA's Astrophysics Data System.

This research made use of SDSS data. Funding for the SDSS and SDSS- II has been provided by the Alfred P. Sloan Foundation, the Participating Institutions, the National Science Foundation, the US Department of Energy, the National Aeronautics and Space Administration, the Japanese Monbukagakusho, the Max Planck Society, and the Higher Education Funding Council for England. The SDSS Web Site is http:www.sdss.org. The SDSS is managed by the Astrophysical Research Consortium for the Participating Institutions. The Participating Institutions are the American Museum of Natural History, Astrophysical Institute Potsdam, University of Basel, University of Cambridge, Case Western Reserve University, University of Chicago, Drexel University, Fermilab, the Institute for Advanced Study, the Japan Participation Group, Johns Hopkins University, the Joint Institute for Nuclear Astrophysics, the Kavli Institute for Particle Astrophysics and Cosmology, the Korean Scientist Group, the Chinese Academy of Sciences (LAMOST), Los Alamos National Laboratory, the Max-Planck-Institute for Astronomy (MPIA), the Max-Planck- Institute for Astrophysics (MPA), New Mexico State University, Ohio State University, University of Pittsburgh, University of Portsmouth, Princeton University, the United States Naval Observatory, and the University of Washington.
\end{acknowledgements}

\bibliographystyle{apj}

\begin{thebibliography}{999}
\bibitem[Adams et al.(2013)]{adams}
         Adams, E.K., Giovanelli, R., Haynes, M.P., 2013, \apj, 768, 77 (A13)
\bibitem[Ahn et al.(2012)]{dr9} 
         Ahn, C.P., Alexandroff, R., Allende Prieto, C., et al. 2012, \apjs, 
	 203, 21
\bibitem[Amorisco, Evans \& van de Ven(2014)]{amorisco}
         Amorisco, N.C., Evans, N.W., van de Ven, G., 2014, Nature, 507, 335
\bibitem[Bellazzini et al.(2002)]{lf}
         Bellazzini, M., Fusi Pecci, F., Messineo, M., Monaco, L., 
	 \& Rood, R. T. 2002, \aj, 123, 1509 
\bibitem[Bellazzini et al.(2011a)]{vv124}
         Bellazzini, M., Beccari, G., Oosterloo, T.A., et al., 2011a, \aap, 527,
	 A58 (Pap~I)	 
\bibitem[Bellazzini(2012)]{mic_pechino}
         Bellazzini, M., 2012, in Origin and Complexity of Massive Star
	 Clusters, IAU SpS1, XXVIIth IAU General Assembly, Higlights of
	 Astronomy, in press (arXiv:1210.6169)
\bibitem[Bellazzini et al.(2013)]{belfil}
         Bellazzini, M., Oosterloo, T., Fraternali, F., Beccari, G., 2013, \aap,
	 559, L11
\bibitem[Bellazzini et al.(2014)]{sexAB}
         Bellazzini, M., Beccari, G., Fraternali, F., et al., 2014, \aap, 566, 44
\bibitem[Belokurov et al.(2006)]{belo06}
         Belokurov, V., Zucker, D.B., Evans, N.W., et al., 2006, \apj, 642, L137
\bibitem[Belokurov et al.(2007)]{belo07}
         Belokurov, V., Zucker, D.B., Evans, N.W., et al., 2007, \apj, 654, 897
\bibitem[Belokurov(2013)]{belo}
         Belokurov, V., 2013, New Astronomy Rev., 57, 100
\bibitem[Bertin \& Arnouts(1996)]{sex}
         Bertin, E., Arnouts, S., 1996, \aaps, 117, 393
\bibitem[Blanton \& Moustakas(2009)]{bla}
         Blanton, M.R., Moustakas, J., 2009, \araa, 47, 159 
\bibitem[Brasseur et al.(2011)]{brasseur}
         Brasseur, C.M., Martin, N.F., Maccio', A.V., Rix, H.-W., Kanf, X., 2011, \apj, 743, 179
\bibitem[Bressan et al.(2012)]{bressan}
         Bressan, A., Marigo, P., Girardi, L., Salasnich, B., Dal Cero, C., Rubele, S., Nanni, A., 2012,  
         \mnras, 427, 127
\bibitem[Brown et al.(2014)]{brown}
         Brown, T.M., Tumlinson, J., Geha, M., et al., 2014, \apj, in press (arXiv:1410.0681)
\bibitem[Buzzoni(2005)]{buz}
         Buzzoni, A., 2005, \mnras, 362, 725
\bibitem[Cannon et al.(2011)]{cannon}
         Cannon, J.M., Giovanelli, R., Haynes, M.P., et al., 2011, \apj, 739, L22
\bibitem[Clem et al.(2008)]{clem}
         Clem, J. L., Vandenberg, D. A., Stetson, P. B. 2008, \aj, 135, 628
\bibitem[Fadely et al.(2012)]{fadely}
         Fadely, R., Hogg, D.W., Willman, B., 2012, \apj, 760, 15
\bibitem[Giallongo et al.(2008)]{lbc} 
         Giallongo, E., Ragazzoni, R., Grazian, A., 2008, \aap 482, 349	
\bibitem[Giovanelli et al.(2007)]{giova07}
         Giovanelli, R., Haynes, M.P., Kent, B.R., et al., 2007, \aj, 133, 2583
\bibitem[Giovanelli et al.(2013)]{leop_1}
         Giovanelli, R., Haynes, M.P., Adams, E., 2013, \aj, 146, 15
\bibitem[Irwin et al.(2007)]{leoT}
         Irwin, M.J., Belokurov, V., Evans, N.W., rt al., 2007, \apj, 656, L13
\bibitem[Ivezi\'c et al.(2008)]{lsst}
         Ivezi\'c, Z., Axelrod, T., Becker, A.C., et al., 2008, in Classification and 
	 discovery in large astronomical surveys, AIP Conf. Proc., 1082, 359
\bibitem[Kauffmann, White \& Guiderdoni(1993)]{kauff}
         Kauffman, G., White, S.D.M., Guiderdoni, B., 1993, \mnras, 264, 201
\bibitem[Klypin et al.(1999)]{klip}
         Klypin, A., Kravtsov, A.V., Valenzuela, O., Prada, F., 1999, \apj, 522,
	 82
\bibitem[Koposov et al.(2009)]{kopo}
         Koposov, S.E., Yoo, J., Rix, H.-W., Weinberg, D.H., Macci\`o, A.V., 
	 Miralda-Escud\`e, J. 2009, \apj, 696, 2179
\bibitem[Marigo et al.(2008)]{marigo}
         Marigo, P., Girardi, L., Bressan, A., Groenewegen, M.A.T., Silva, L., Granato, G.L., 
	 2008, \aap, 482, 883
\bibitem[Martin, de Jong \& Rix(2008)]{martin}
         Martin, N.F., de Jong, J.T.A, Rix, H.-W., 2008, \apj, 684, 1075
\bibitem[Martin et al.(2013)]{panda_ghosts}
         Martin, N.F., Ibata, R.A., McConnachie, A.W., Mackey, A.D., Ferguson,
	 A.M.N., Iriwin, M.J., Lewis, G.F., Fardal, M.A., \apj, 776, 80
\bibitem[Martin et al.(2014)]{martin14}
         Martin, N.F., Ibata, R.A., Rich, R.M., et al., 2014, \aj, 787, 19
\bibitem[Mateo(1998)]{mateo}
         Mateo, M., 1998, \araa, 36, 435
\bibitem[McConnachie(2008)]{XIX}
         McConnachie, A., Huxor, A., Martin, N.F., et al., 2008, \apj, 688, 1009
\bibitem[McConnachie(2009)]{pandas}
         McConnachie, A.W., Irwin, M.J., Ibata, R.A., 2009, Nature, 461, 66
\bibitem[McConnachie(2012)]{mcc}
         McConnachie, A., 2012, \aj, 144, 4 
\bibitem[McQuinn et al.(2013)]{leop_lbt}
         McQuinn, K.D.W., Skillman, E.D., Berg, D.A., et al., 2013, \aj, 146, 145
\bibitem[Moore et al.(1999)]{moore}
         Moore, B., Ghigna, S., Governato, F., et al., 1999, \apj, 524, L19
\bibitem[Rhode et al.(2013)]{leop_2}
         Rhode, K.L., Salzer, J.J., Haurberg, N.C., et al., 2013, \aj, 145, 149
\bibitem[Richardson et al.(2011)]{richa}
         Richardson, J.C., Irwin, M.J., McConnachie, A.W., et al., 2011, \apj, 732, 76
\bibitem[Ricotti(2009)]{ricotti}
         Ricotti, M., 2009, \mnras, 392, L45
\bibitem[Ryan-Weber et al.(2008)]{Ryan-Weber}
         Ryan-Weber, E.V., Begum, A., Oosterloo, T., Pal, S., Irwin, M.J., Belokurov, V., Evans, N.W., Zucker, D.B.
         2008, MNRAS, 384, 535
\bibitem[Saul et al.(2013)]{GALFA}
         Saul, D.R., Peek, J.E.G., Grchevich, J., etal., 2012, \apj, 758, 44
\bibitem[Sawala et al. (2013)]{sawa}
         Sawala, T., Frenk, C.S., Crain, R.A., Jenkins, A., Schaye, J.,
	 Theus, T., Zavala, J., 2013, MNRAS, 431, 1366
\bibitem[Schlafly \& Finkbeiner(2011)]{schlaf}
         Schlafly, E., \& Finkbeiner, D.P., 2011, \apj, 737, 103
\bibitem[Schlegel et~al.(1998)]{ebv}
         Schlegel, D.~J., Finkbeiner, D.~P., \& Davis, M. 1998, \apj, 500, 525
\bibitem[Skillman et al.(2013)]{leop_met}
         Skillman, E.D., Salzer, J.J., Berg, D.A., et al., 2013, \aj, 146, 3
\bibitem[Stetson(1987)]{daophot}
         Stetson, P.B., 1987, \pasp, 99, 191
\bibitem[Stetson(1994)]{allframe}
         Stetson, P.B., 1994, \pasp, 106, 250
\bibitem[Tollerud et al.(2008)]{tolle}
         Tollerud, E.J., Bullock, J.S., Strigari, L.E, Willman, B., 2008, \apj, 688, 277
\bibitem[Tolstoy, Hill \& Tosi et al.(2009)]{tht}
         Tolstoy, E., Hill, V., \& Tosi, M., 2009, \araa, 47, 371
\bibitem[Willman \& Strader(2012)]{willstra}
         Willman, B., Strader, J., 2012, \aj, 144, 76
\bibitem[Wyse(2012)]{wyse}
         Wyse, R.F.G., 2012, in Galactic Archaeology: Near-Field Cosmology and the Formation 
	 of the Milky Way, W. Aoki, M. Ishigaki, T. Suda, T. Tsujimoto, and 
	 N. Arimoto Eds., ASP Conf. Ser., 458, 251	 	 

\end{thebibliography}

\appendix

\section{Observational material and astrometric and photometric calibrations}
\label{app_obs}


\begin{table*}
  \begin{center}
\caption{Observational material}
 \label{logobs}
   \begin{tabular}{ccccccc}
Field & Filter & Date     & UT & Airmass &  FWHM$_{PSF}$  & Bkg    \\
      &        &yyyy-mm-dd&  hh:mm:ss  &         &[arcsec]&  [ADU]\\
\hline
 A&g &  2014-01-29 &  04:04:34 & 1.27 & 0.86 &  7183 \\
 A&g &  2014-01-29 &  04:11:20 & 1.25 & 0.83 &  7120 \\
 A&r &  2014-01-29 &  04:04:16 & 1.27 & 0.79 & 11170 \\
 A&r &  2014-01-29 &  04:11:06 & 1.25 & 0.78 & 11113 \\
 B&g &  2014-01-29 &  06:05:26 & 1.16 & 0.89 &  7758 \\
 B&g &  2014-01-29 &  06:12:12 & 1.14 & 0.91 &  7752 \\
 B&r &  2014-01-29 &  06:05:07 & 1.16 & 0.91 & 10544 \\
 B&r &  2014-01-29 &  06:11:55 & 1.14 & 0.96 & 10694 \\
 C&g &  2014-01-29 &  09:01:47 & 1.28 & 0.95 &  9299 \\
 C&g &  2014-01-29 &  09:08:34 & 1.27 & 0.88 &  9129 \\
 C&r &  2014-01-29 &  09:01:29 & 1.28 & 0.91 & 12979 \\
 C&r &  2014-01-29 &  09:08:18 & 1.27 & 0.88 & 12636 \\
 D&g &  2014-01-29 &  09:16:53 & 1.18 & 0.92 &  8381 \\
 D&g &  2014-01-29 &  09:23:39 & 1.16 & 0.98 &  8290 \\
 D&r &  2014-01-29 &  09:16:48 & 1.18 & 0.92 & 11662 \\
 D&r &  2014-01-29 &  09:23:25 & 1.16 & 0.98 & 11801 \\
 E&g &  2014-01-29 &  11:15:42 & 1.28 & 1.07 &  8643 \\
 E&g &  2014-01-29 &  11:22:33 & 1.26 & 1.18 &  8722 \\
 E&r &  2014-01-29 &  11:15:37 & 1.28 & 0.97 & 13949 \\
 E&r &  2014-01-29 &  11:22:18 & 1.26 & 0.97 & 13914 \\
 F&g &  2014-01-29 &  12:03:26 & 1.17 & 1.02 &  7690 \\
 F&g &  2014-01-29 &  12:10:10 & 1.16 & 1.02 &  7620 \\
 F&r &  2014-01-29 &  12:03:19 & 1.17 & 1.00 & 13012 \\
 F&r &  2014-01-29 &  12:09:56 & 1.16 & 0.99 & 12997 \\
 G&g &  2014-01-29 &  12:50:37 & 1.11 & 0.83 &  6631 \\
 G&g &  2014-01-29 &  12:57:23 & 1.10 & 1.01 &  7616 \\
 G&r &  2014-01-29 &  12:50:21 & 1.11 & 0.74 & 11599 \\
 G&r &  2014-01-29 &  12:57:09 & 1.10 & 0.81 & 12126 \\
 H&g &  2014-03-07 &  09:01:48 & 1.30 & 1.07 &  6406 \\
 H&g &  2014-03-07 &  09:10:43 & 1.27 & 1.77 &  6312 \\
 H&r &  2014-03-07 &  09:01:31 & 1.30 & 1.05 & 10043 \\
 H&r &  2014-03-07 &  09:10:41 & 1.27 & 1.68 &  9994 \\
 I&g &  2014-03-07 &  09:53:59 & 1.17 & 1.31 &  6135 \\
 I&g &  2014-03-07 &  10:00:47 & 1.15 & 1.47 &  6100 \\
 I&r &  2014-03-07 &  09:53:43 & 1.17 & 1.35 & 10384 \\
 I&r &  2014-03-07 &  10:00:32 & 1.15 & 1.49 & 10362 \\
 J&g &  2014-01-29 &  07:30:40 & 1.12 & 0.77 &  8243 \\
 J&g &  2014-01-29 &  07:37:27 & 1.10 & 0.76 &  8215 \\
 J&r &  2014-01-29 &  07:30:23 & 1.12 & 0.74 & 11043 \\
 J&r &  2014-01-29 &  07:37:11 & 1.10 & 0.76 & 11046 \\
 K&g &  2014-01-29 &  03:38:40 & 1.32 & 0.90 &  7959 \\
 K&g &  2014-01-29 &  03:45:25 & 1.29 & 0.94 &  7792 \\
 K&r &  2014-01-29 &  03:38:23 & 1.32 & 0.81 & 12465 \\
 K&r &  2014-01-29 &  03:45:10 & 1.29 & 0.86 & 12113 \\
 L&g &  2014-01-29 &  08:37:15 & 1.16 & 1.04 &  8846 \\
 L&g &  2014-01-29 &  08:44:03 & 1.15 & 1.08 &  8822 \\
 L&r &  2014-01-29 &  08:37:14 & 1.16 & 0.85 & 11962 \\
 L&r &  2014-01-29 &  08:43:48 & 1.15 & 0.88 & 12062 \\
 M&g &  2014-01-29 &  09:50:19 & 1.24 & 1.20 &  8724 \\
 M&g &  2014-01-29 &  09:57:05 & 1.22 & 1.19 &  8602 \\
 M&r &  2014-01-29 &  09:50:02 & 1.24 & 1.07 & 12908 \\
 M&r &  2014-01-29 &  09:56:53 & 1.22 & 1.14 & 12885 \\
 ... continue&&&&&&
\end{tabular} 
\end{center}
\end{table*}
%


\setcounter{table}{0}
\begin{table*}
  \begin{center}
\caption{Observational material - continued}
   \begin{tabular}{ccccccc}
Field & Filter & Date     & UT & Airmass &  FWHM$_{PSF}$  & Bkg    \\
      &        &yyyy-mm-dd&  hh:mm:ss  &         &[arcsec]&  [ADU]\\
\hline
 N&g &  2014-01-29 &  10:05:21 & 1.34 & 1.47 &  9751 \\
 N&g &  2014-01-29 &  10:12:07 & 1.32 & 1.23 &  9421 \\
 N&r &  2014-01-29 &  10:05:17 & 1.34 & 1.33 & 15022 \\
 N&r &  2014-01-29 &  10:11:53 & 1.32 & 1.14 & 14557 \\
 O&g &  2014-01-29 &  10:31:00 & 1.19 & 1.34 &  8509 \\
 O&g &  2014-01-29 &  10:37:47 & 1.17 & 1.54 &  8138 \\
 O&r &  2014-01-29 &  10:30:42 & 1.19 & 1.10 & 13364 \\
 O&r &  2014-01-29 &  10:37:31 & 1.17 & 1.04 & 13131 \\
 P&g &  2014-01-29 &  11:00:54 & 1.28 & 0.96 &  9296 \\
 P&g &  2014-01-29 &  11:07:43 & 1.26 & 0.90 &  8809 \\
 P&r &  2014-01-29 &  11:00:39 & 1.28 & 0.98 & 14351 \\
 P&r &  2014-01-29 &  11:07:26 & 1.26 & 0.93 & 13804 \\
 Q&g &  2014-01-29 &  11:42:37 & 1.21 & 0.94 &  8732 \\
 Q&g &  2014-01-29 &  11:49:25 & 1.19 & 1.09 &  8558 \\
 Q&r &  2014-01-29 &  11:42:22 & 1.21 & 1.02 & 13859 \\
 Q&r &  2014-01-29 &  11:49:11 & 1.19 & 1.11 & 13745 \\
 R&g &  2014-01-29 &  12:24:58 & 1.19 & 0.84 &  7682 \\
 R&g &  2014-01-29 &  12:31:44 & 1.18 & 0.81 &  7404 \\
 R&r &  2014-01-29 &  12:24:50 & 1.19 & 0.83 & 13522 \\
 R&r &  2014-01-29 &  12:31:29 & 1.18 & 0.80 & 13296 \\
 S&g &  2014-03-28 &  11:32:42 & 1.06 & 1.22 &  5091 \\
 S&g &  2014-03-28 &  11:53:23 & 1.07 & 1.45 &  5871 \\
 S&r &  2014-03-28 &  11:23:09 & 1.06 & 1.12 &  8851 \\
 S&r &  2014-03-28 &  11:32:32 & 1.06 & 1.11 &  8905 \\
 S&r &  2014-03-28 &  11:53:22 & 1.07 & 1.31 &  9183 \\
 T&g &  2014-03-07 &  08:35:28 & 1.11 & 1.03 &  6458 \\
 T&g &  2014-03-07 &  08:42:19 & 1.10 & 1.08 &  6493 \\
 T&r &  2014-03-07 &  08:35:12 & 1.11 & 0.90 & 10039 \\
 T&r &  2014-03-07 &  08:42:03 & 1.10 & 0.95 & 10078 \\
 U&g &  2014-01-29 &  08:21:53 & 1.22 & 0.92 &  8783 \\
 U&g &  2014-01-29 &  08:28:40 & 1.21 & 1.00 &  8835 \\
 U&r &  2014-01-29 &  08:21:36 & 1.22 & 0.85 & 12038 \\
 U&r &  2014-01-29 &  08:28:25 & 1.21 & 0.92 & 12080 \\
 V&g &  2014-03-07 &  08:11:25 & 1.12 & 1.39 &  6611 \\
 V&g &  2014-03-07 &  08:18:13 & 1.11 & 1.22 &  6577 \\
 V&r &  2014-03-07 &  08:11:20 & 1.12 & 1.25 & 10486 \\
 V&r &  2014-03-07 &  08:17:58 & 1.11 & 1.08 & 10509 \\
 W&g &  2014-06-20 &  05:44:41 & 1.05 & 0.98 &  5284 \\
 W&g &  2014-06-20 &  05:51:32 & 1.05 & 1.11 &  5347 \\
 W&r &  2014-06-20 &  05:44:29 & 1.05 & 0.92 &  8391 \\
 W&r &  2014-06-20 &  05:51:28 & 1.05 & 1.04 &  8364 \\
 X&g &  2014-03-07 &  06:08:14 & 1.43 & 1.16 & 15700 \\
 X&g &  2014-03-07 &  06:15:00 & 1.40 & 1.23 & 14305 \\
 X&r &  2014-03-07 &  06:07:58 & 1.43 & 1.01 & 16745 \\
 X&r &  2014-03-07 &  06:14:45 & 1.40 & 1.08 & 15221 \\
 Y&g &  2014-01-29 &  07:54:58 & 1.34 & 0.89 &  9155 \\
 Y&g &  2014-01-29 &  08:01:42 & 1.32 & 0.95 &  9252 \\
 Y&r &  2014-01-29 &  07:54:40 & 1.34 & 0.85 & 13042 \\
 Y&r &  2014-01-29 &  08:01:28 & 1.32 & 0.93 & 12810 \\
 Z&g &  2014-06-20 &  05:11:42 & 1.05 & 1.04 &  5438 \\
 Z&g &  2014-06-20 &  05:18:34 & 1.05 & 1.11 &  5471 \\
 Z&r &  2014-06-20 &  04:55:13 & 1.06 & 0.87 &  9576 \\
 Z&r &  2014-06-20 &  05:18:18 & 1.05 & 0.91 &  8953 \\
\end{tabular} 
\tablefoot{List of all $t_{exp}=300$~s images gathered for the survey. 
FWHM$_{PSF}$ is Full Width at Half Maximum of the PSF as measured 
on the image. Bkg is the mean level of the sky background, in ADU.}
\end{center}
\end{table*}

\subsection{Astrometry}

We transform the coordinates of our catalogues from x,y in pixels to Equatorial J2000 RA and Dec using stars in common between our catalogue and SDSS-DR9 (hereafter DR9, for brevity), using the latter catalogue as our astrometric standard. The transformation is performed with the dedicated code CataXcorr\footnote{Part of the CataPack package developed by P. Montegriffo at INAF - Osservatorio Astronomico di Bologna: {\tt http://davide2.bo.astro.it/~paolo/}} in two steps. First, a third-degree polynomial solution is fitted using DR9 sources classified both as stars or as galaxies as astrometric standard. The derived solution is based on a few hundred shared sources per field (from a minimum of 323 to a maximum of 612) and have a median rms of $0.11\arcsec$ in both RA and Dec (from a minimum of $0.07\arcsec$ to a maximum of $0.14\arcsec$). Second, a first-degree polynomial solution for the derived coordinates is fitted using only DR9 sources classified as stars. The derived solutions are based, for each field, on several tens to a few hundred shared stars (from a minimum of 65 to a maximum of 330) and have a median rms of $0.08\arcsec$ in both RA and Dec (from a minimum of $0.05\arcsec$ to a maximum of $0.12\arcsec$).

A second set of X,Y coordinates projected on the sky is also calculated for each field. The origin of these coordinates is located at the position of the centroid of the UCHVC associated with the considered field; X increases towards the west and Y increases towards the North.

\subsection{Photometric calibration}

The absolute photometric calibration is determined using stars in common with DR9 as secondary calibrators, as done in
\citet{vv124,sexAB}. Instrumental g,r magnitudes for each field have been transformed into the SDSS DR9 photometric system
individually, using in-field calibrators.
Here we adopted a two step procedure. We use Sextractor to reduce the 20s images of four fields 
(A, J, K, R) that have many stars in common with DR9. Since our deep images saturate around $r\sim 19.0$, and DR9 g,r photometry has photometric errors $\le 0.03$~mag only for $r<20.0$, the catalogues from these short exposures have a much larger (useful) magnitude range that overlaps more with DR9 than those from the long exposures. This leads to better comparisons between instrumental ($g_i,r_i$) magnitudes and standard magnitudes ($g_{DR9},r_{DR9}$) since they rely on the large numbers of DR9 stars observed at high signal-to-noise, i.e. with more accurate photometry \citep[see][for a discussion]{vv124}.  We use the hundreds of stars in common for each field within the relevant colour range to fit the coefficient of a first-degree colour term to transform instrumental magnitudes into natural magnitudes ($g_{nat},r_{nat}$):

\begin{equation}
g_{nat} = g_i +a_g(g_i-r_i)~~~~~~r_{nat} = r_i +a_r(g_i-r_i)
\end{equation}

The equations are found to provide an adequate description of the data over the colour range of interest for our scientific purposes.
We adopt the mean of the four values as our final coefficients to be used for all our photometry:
$a_g=0.084$ with standard deviation $\sigma_{a_g}=0.002$ and $a_r=0.040$ with standard deviation $\sigma_{a_r}=0.012$. 

These equations are used to transform all of our DAOPHOTII photometry into the natural system. 
We then use the stars in common between our deep photometry catalogues and DR9 to determine a single constant for each field and passband. To estimate these photometric zero points, $ZP_r$, $ZP_g$, we carefully inspect, for each field, the $mag_{DR9}-mag_{nat}$ vs. $mag_{DR9}$ diagrams to select the set of unsaturated common stars showing low scatter in $mag_{DR9}-mag_{nat}$ that were finally used to estimate the zero points, with typical rms $\le 0.02$ mag. All the natural magnitudes were transformed into the standard systems with the equations:

\begin{equation}
g_{DR9} = g_{nat} + ZP_g~~~~~~r_{DR9} = r_{nat} + ZP_r 
\end{equation}

\section{Colour-magnitude diagrams: the complete data set}
\label{app_cmds}

We present here the CMDs for the fields that are not shown in the main text.
The arrangement and the symbols are the same as in Fig.~\ref{cmd1}.

  \begin{figure*}
   \centering
   \includegraphics[width=\textwidth]{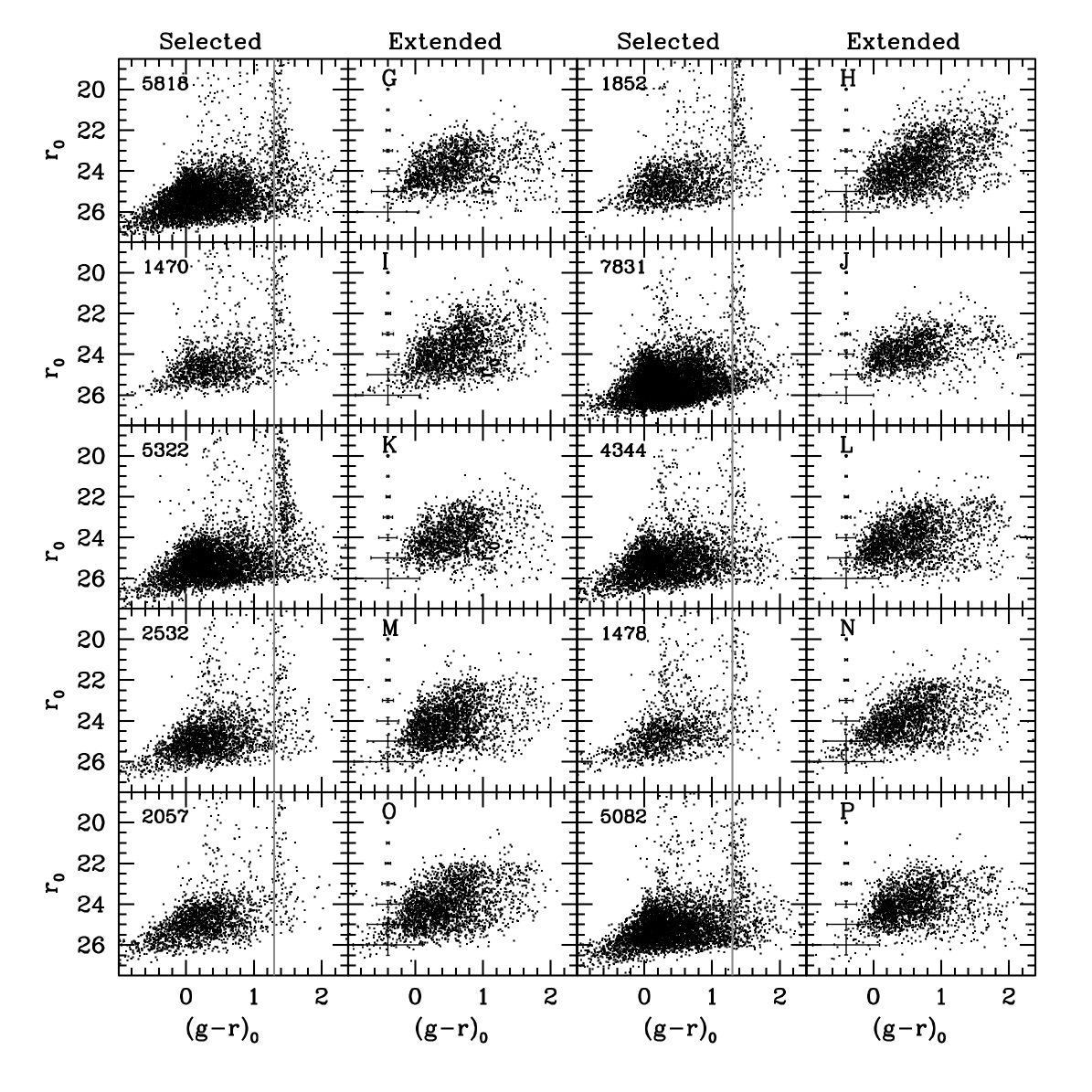}
     \caption{Same as Fig.~\ref{cmd1} but for Fields~G to~P.}
        \label{cmd2}
    \end{figure*}

  \begin{figure*}
   \centering
   \includegraphics[width=\textwidth]{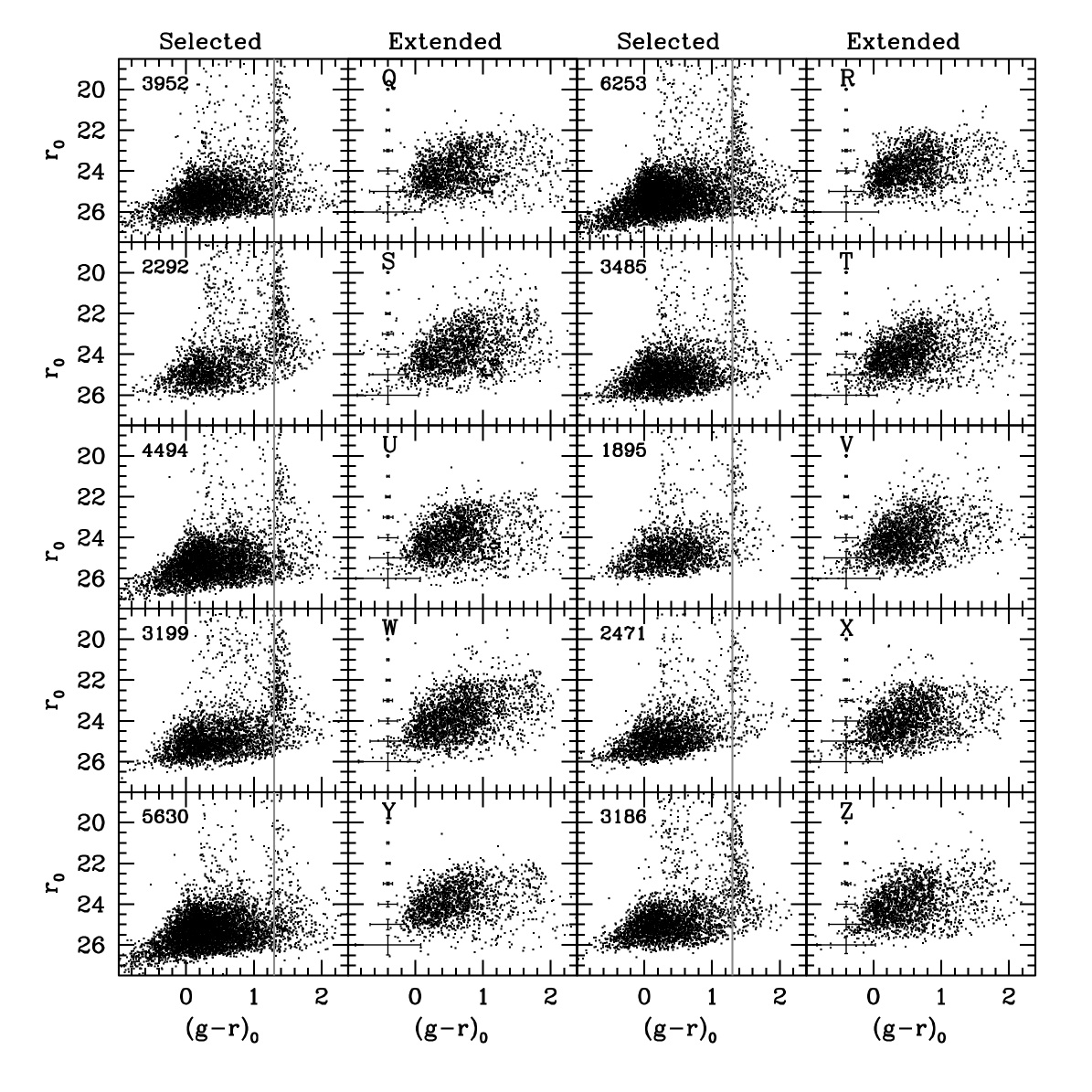}
     \caption{Same as Fig.~\ref{cmd1} but for Fields~Q to~Z.}
        \label{cmd3}
    \end{figure*}


\section{Density maps: the complete data set}
\label{app_maps}

We present here the density maps for the fields not shown in the main text.
The arrangement and the symbols are the same as in Fig.~\ref{mappe1}.
We remind the reader that the r27 maps are obtained using the entire, colour-selected sample, while r25 stars are obtained from the sub-sample of stars with $r<25.0$.

   \begin{figure}
   \centering
   \includegraphics[width=\columnwidth]{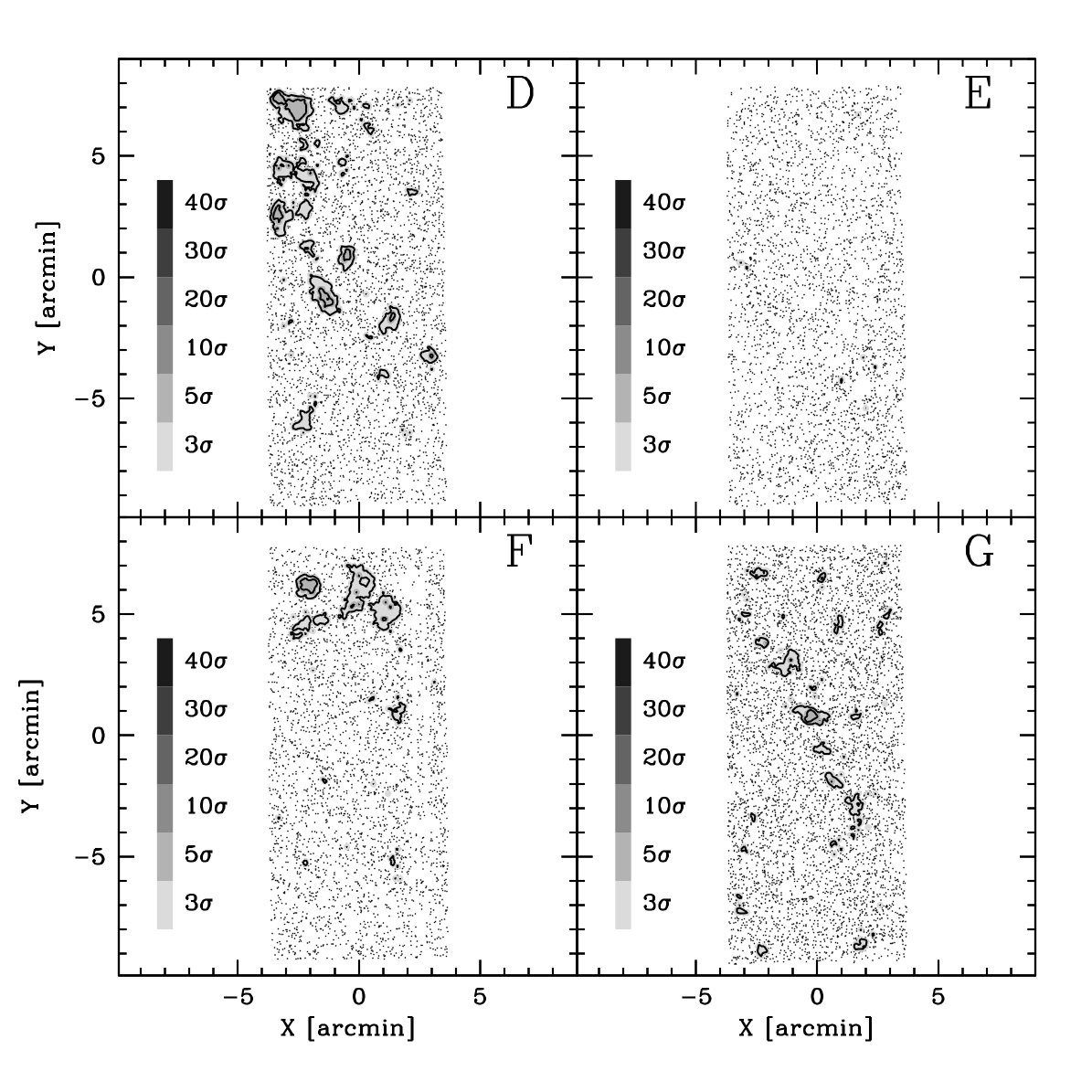}
   \includegraphics[width=\columnwidth]{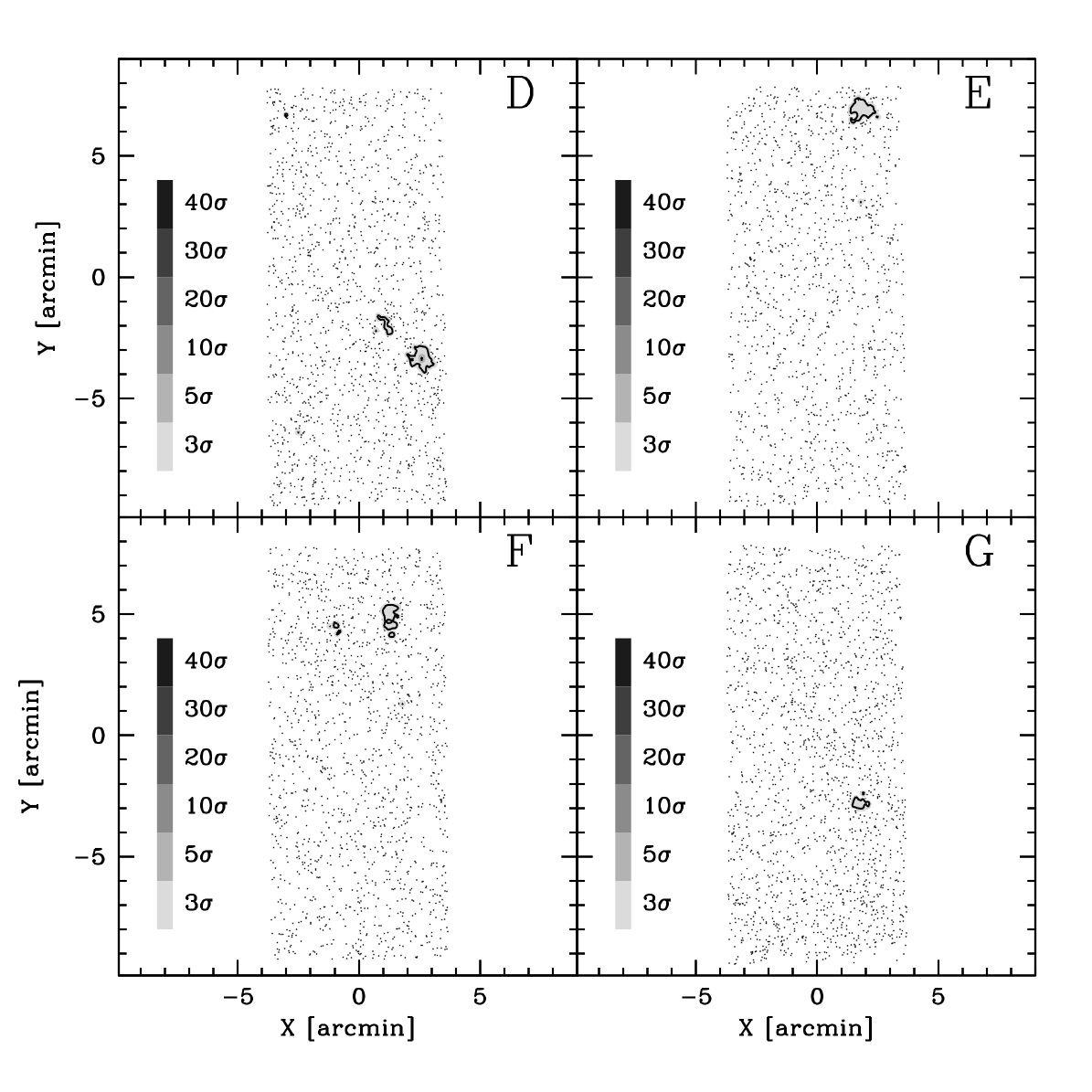}
     \caption{r27 maps and r25 maps for fields D, E, F, and G.
     The arrangement and the meaning of the symbols are the same as in Fig.~\ref{mappe1}.}
        \label{mappe2}
    \end{figure}

   \begin{figure}
   \centering
   \includegraphics[width=\columnwidth]{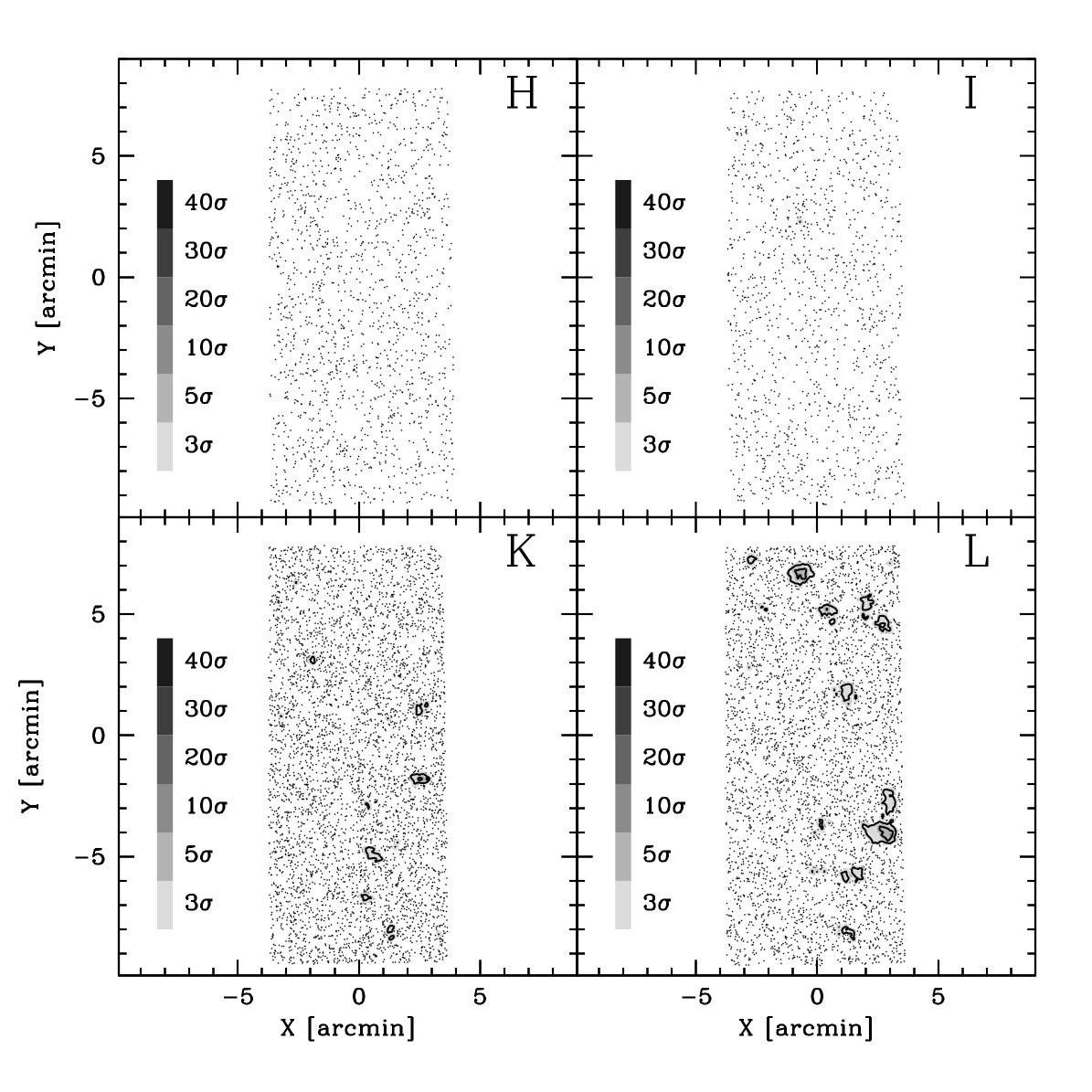}
   \includegraphics[width=\columnwidth]{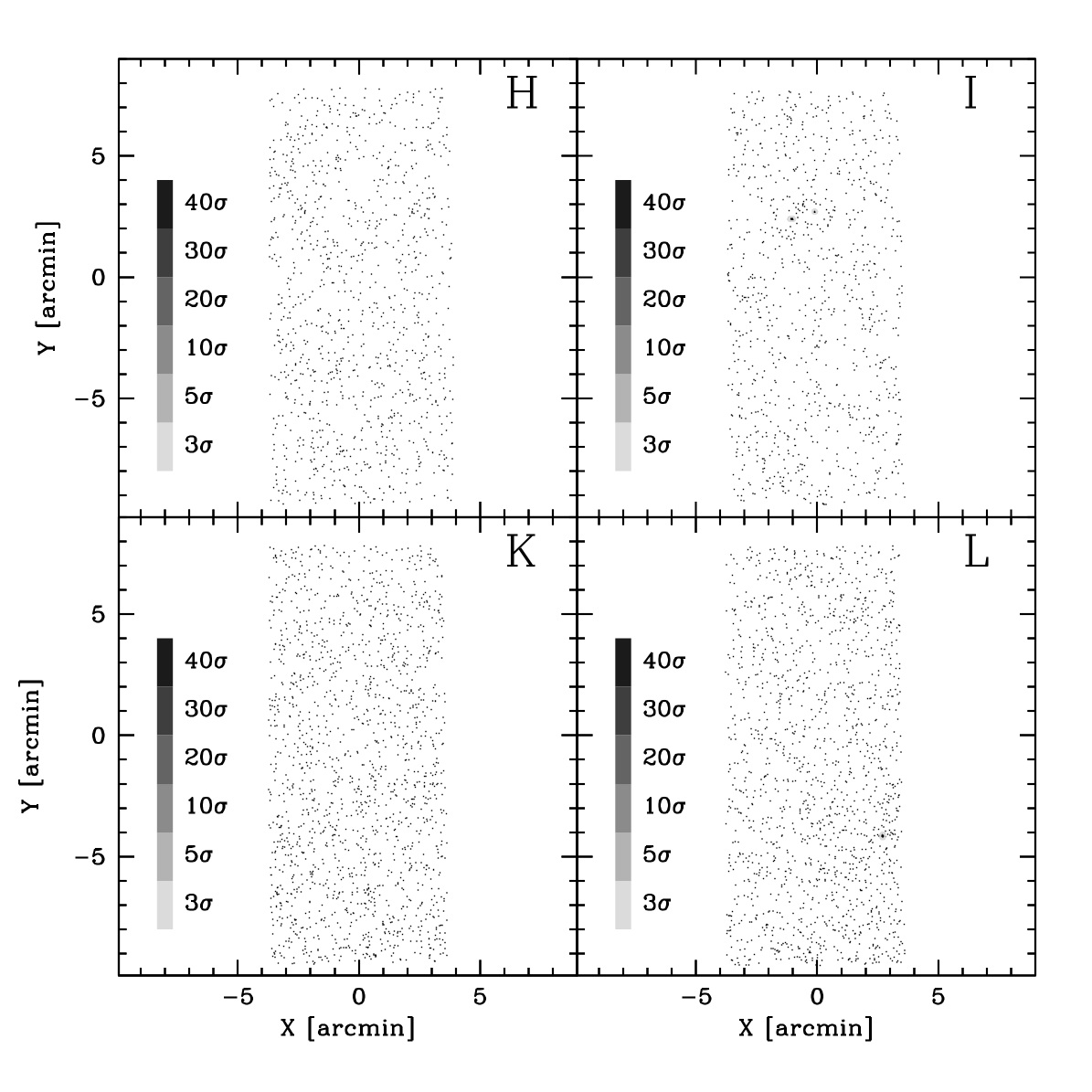}
     \caption{r27 maps and r25 maps for fields H, I, K, and L.
     The arrangement and the meaning of the symbols are the same as in Fig.~\ref{mappe1}.}
        \label{mappe3}
    \end{figure}


   \begin{figure}
   \centering
   \includegraphics[width=\columnwidth]{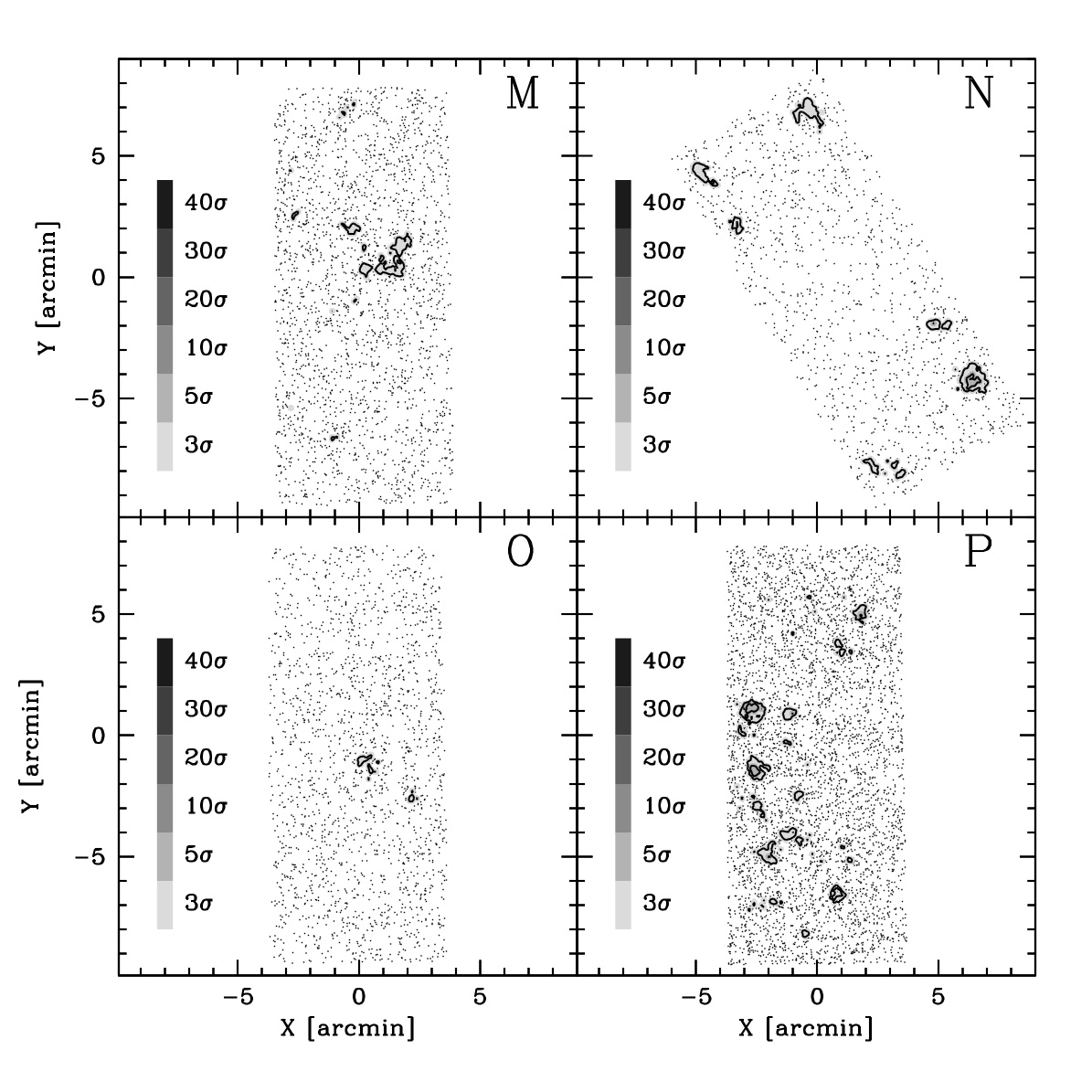}
   \includegraphics[width=\columnwidth]{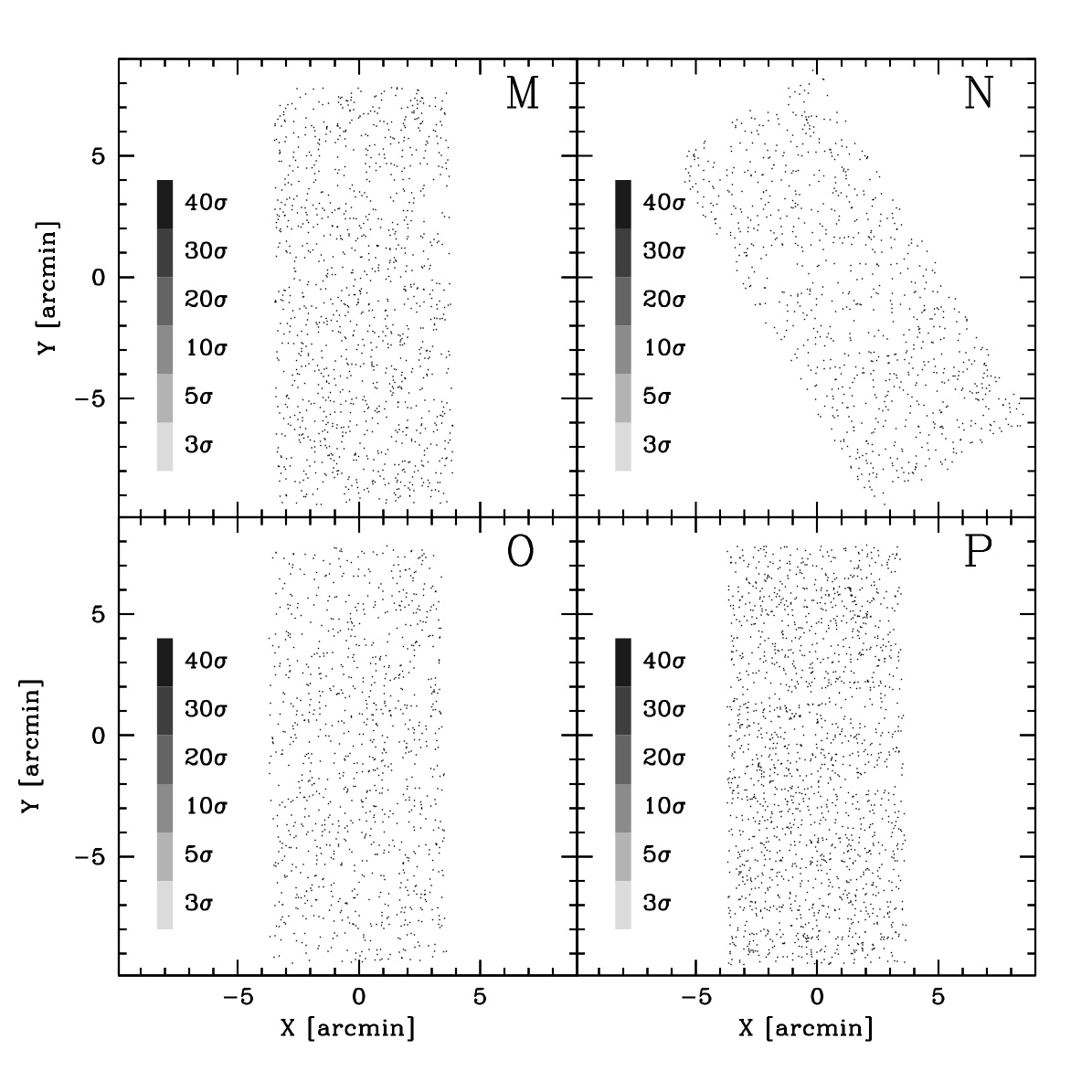}
     \caption{r27 maps and r25 maps fields M, N, O, and P.
     The arrangement and the meaning of the symbols are the same as in Fig.~\ref{mappe1}.}
        \label{mappe4}
    \end{figure}

   \begin{figure}
   \centering
   \includegraphics[width=\columnwidth]{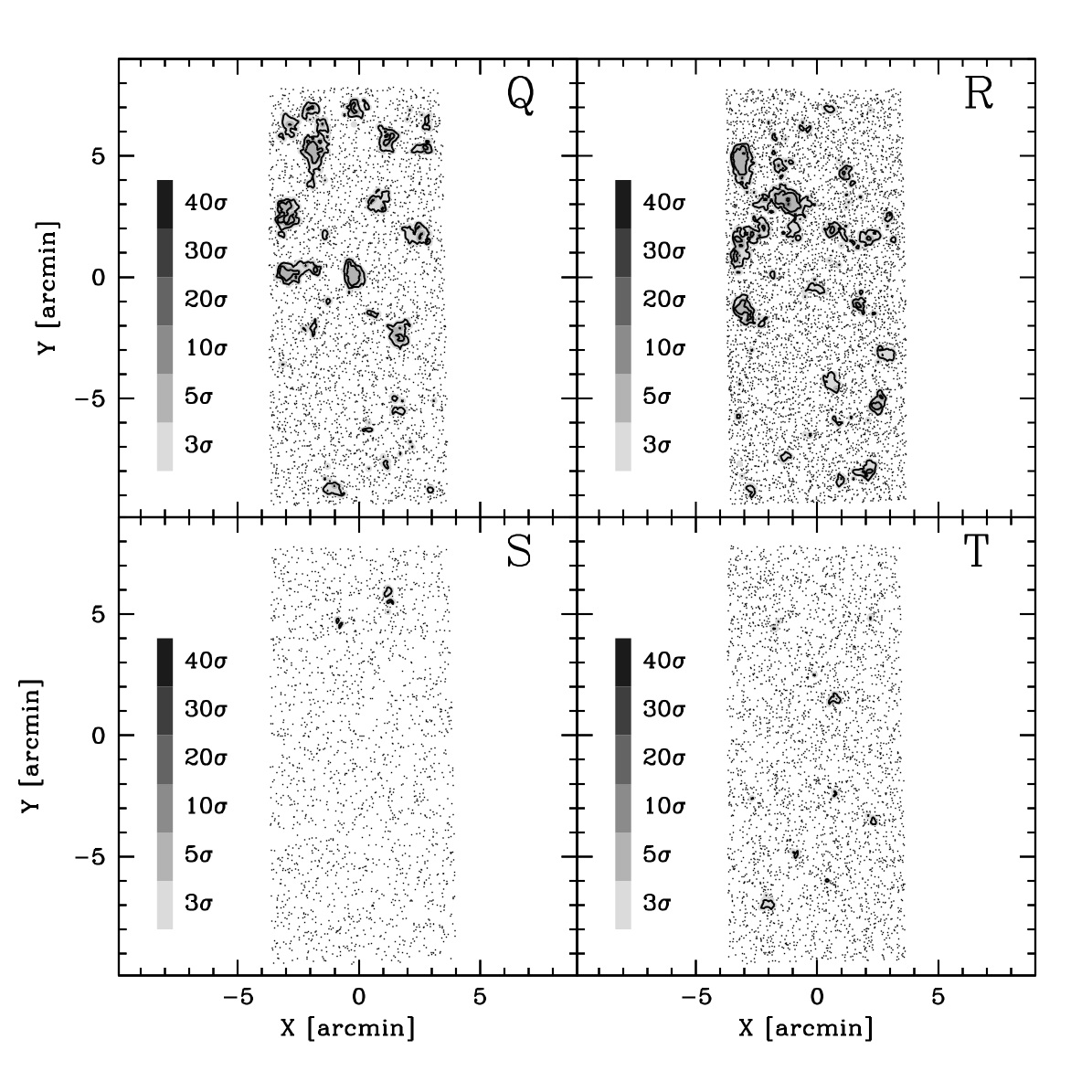}
   \includegraphics[width=\columnwidth]{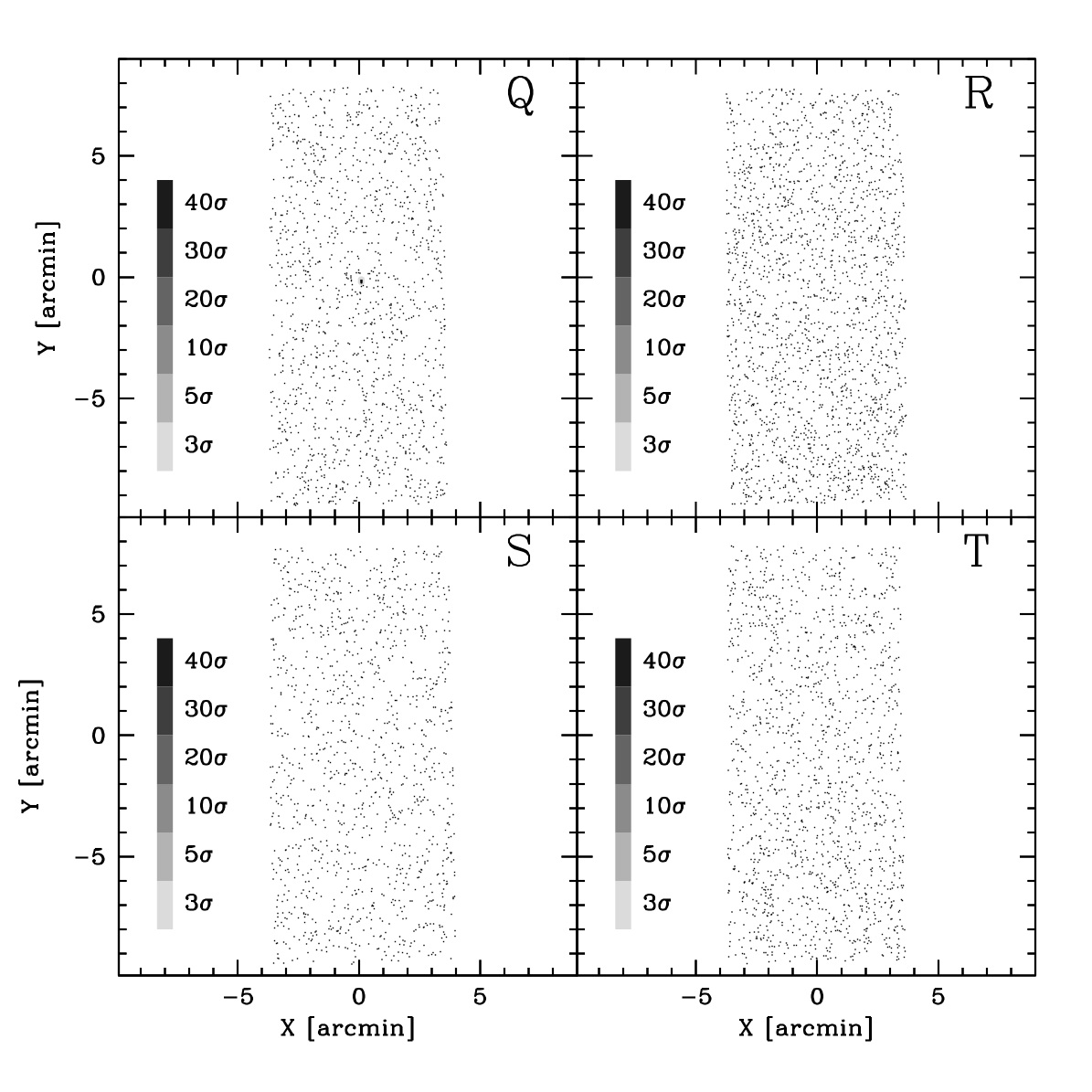}
     \caption{r27 maps and r25 maps for fields Q, R, S, and T.
     The arrangement and the meaning of the symbols are the same as in Fig.~\ref{mappe1}.}
        \label{mappe5}
    \end{figure}

   \begin{figure}
   \centering
   \includegraphics[width=\columnwidth]{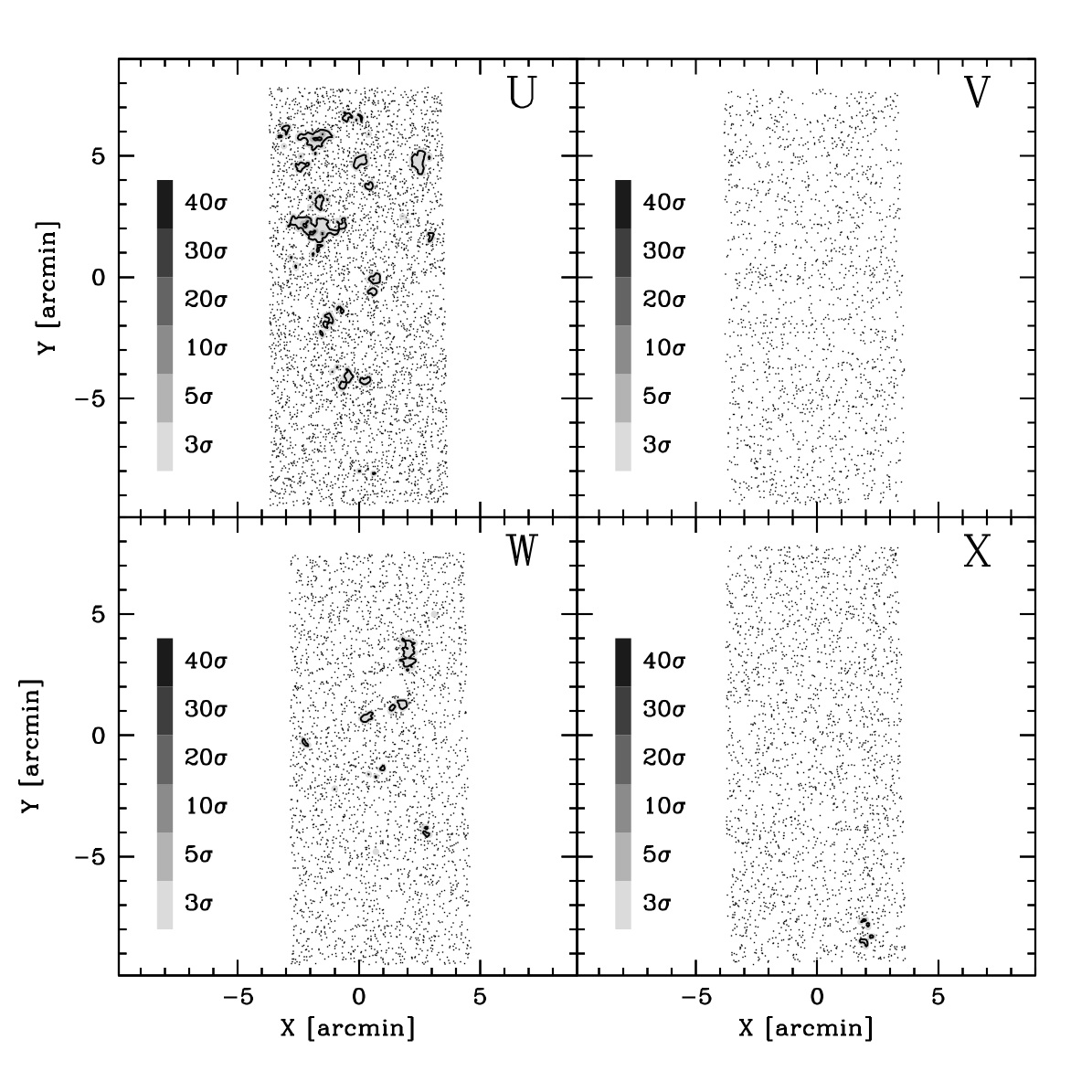}
   \includegraphics[width=\columnwidth]{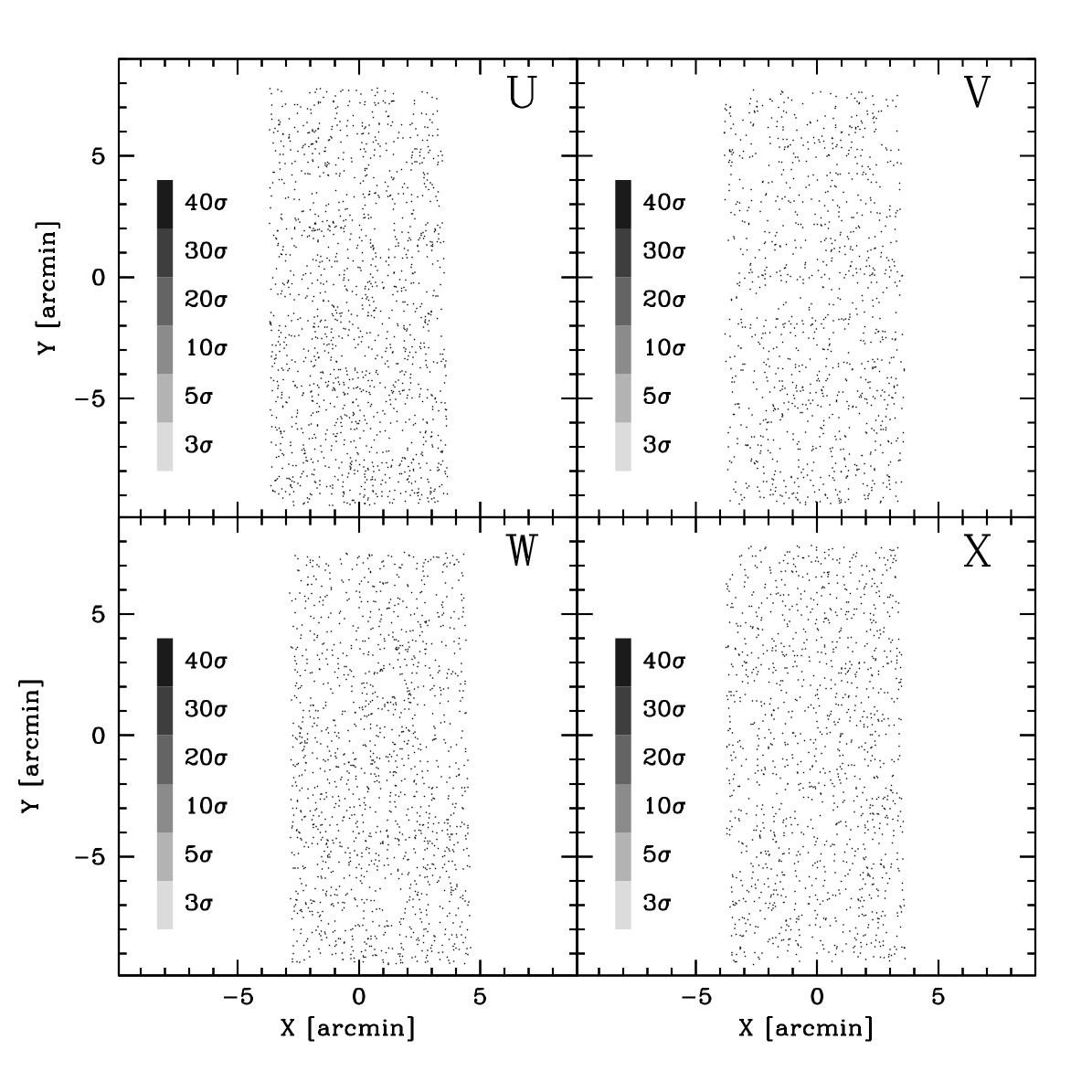}
     \caption{r27 maps and r25 maps the fields U, V, W, X.
     The arrangement and the meaning of the symbols are the same as in Fig.~\ref{mappe1}.}
        \label{mappe6}
    \end{figure}

   \begin{figure}
   \centering
   \includegraphics[width=\columnwidth]{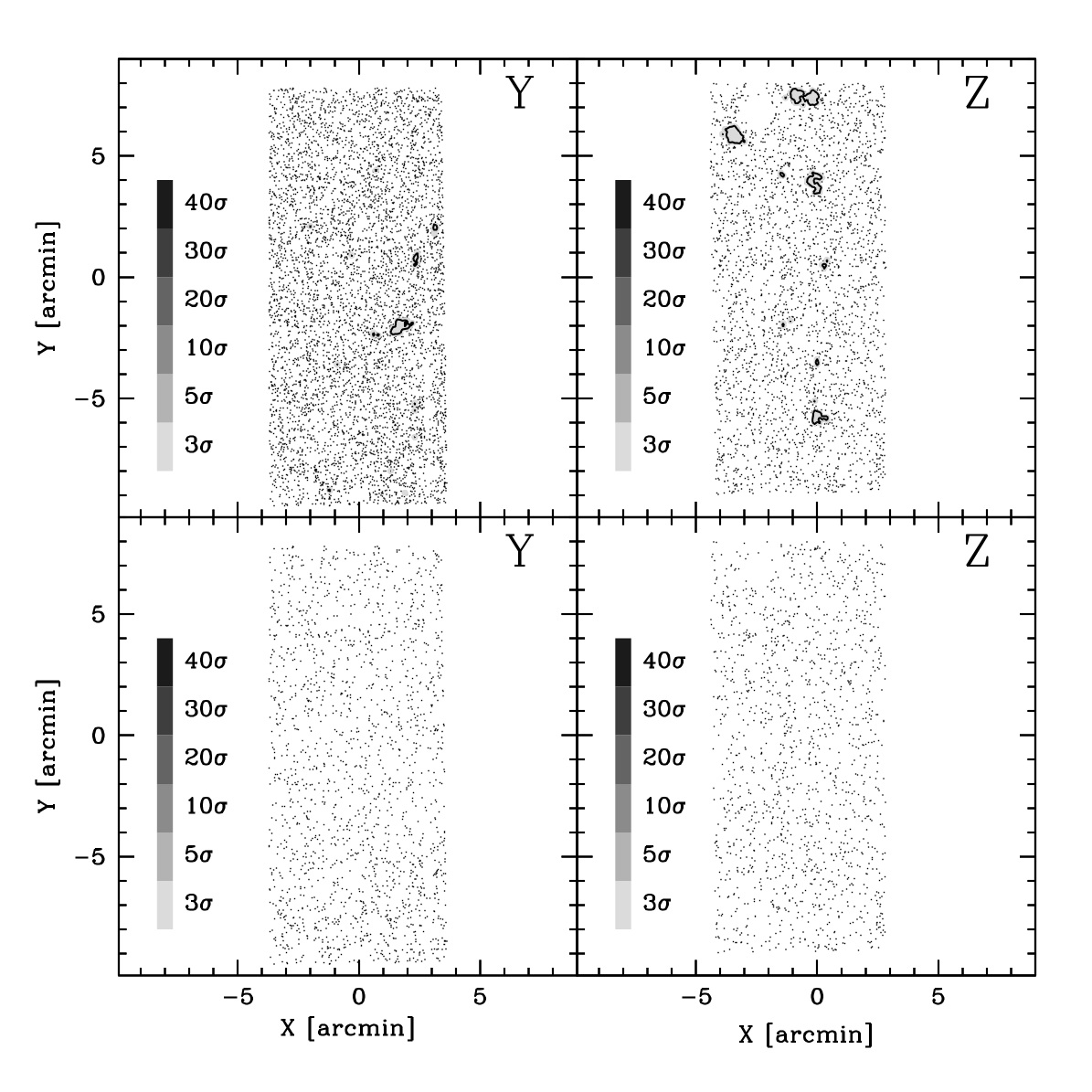}
     \caption{r27 maps (upper panles) and r25 maps (lower panels) the fields Y and Z.
     The meaning of the symbols is the same as in Fig.~\ref{mappe1}}
        \label{mappe7}
    \end{figure}

\clearpage




\end{document}